\documentclass[11pt]{article}
\usepackage[T1]{fontenc}
\usepackage{lmodern}
\usepackage[margin=1in]{geometry}
\usepackage{amsmath,amssymb,amsthm,mathtools}
\usepackage{microtype,booktabs,array,enumitem,xcolor,needspace}
\usepackage{tikz}
\usetikzlibrary{arrows.meta,patterns,calc,positioning,decorations.pathreplacing}
\usepackage[numbers,sort&compress]{natbib}
\usepackage[colorlinks=true,linkcolor=blue!45!black,citecolor=blue!45!black,urlcolor=blue!45!black]{hyperref}
\usepackage[nameinlink,noabbrev]{cleveref}
\hypersetup{pdftitle={Metric Facility Location at the Approximation Threshold},pdfauthor={Mohammad Hajiaghayi, Mohammad Mahdian, Vahab Mirrokni},pdfsubject={Approximation algorithms for facility location}}
\newtheorem{theorem}{Theorem}[section]
\newtheorem{lemma}[theorem]{Lemma}

\newtheorem{corollary}[theorem]{Corollary}
\theoremstyle{definition}
\newtheorem{definition}[theorem]{Definition}
\theoremstyle{remark}

\DeclareMathOperator{\OPT}{OPT}
\DeclareMathOperator{\LP}{LP}
\newcommand{\E}{\mathbb E}
\newcommand{\cost}{\operatorname{cost}}

\newcommand{\ybar}{{\bar y}}
\newcounter{algorithm}
\newenvironment{pseudocode}[1]{%
  \par\addvspace{.8\baselineskip}\noindent\begin{minipage}{\linewidth}
  \refstepcounter{algorithm}\hrule\vspace{5pt}
  \noindent\textbf{Algorithm \thealgorithm. #1}\par\vspace{4pt}
  \small\raggedright
}{\vspace{4pt}\hrule\end{minipage}\par\addvspace{.8\baselineskip}}
\crefname{algorithm}{Algorithm}{Algorithms}
\crefname{theorem}{Theorem}{Theorems}
\crefname{lemma}{Lemma}{Lemmas}
\crefname{section}{Section}{Sections}
\crefname{appendix}{Appendix}{Appendices}
\setlist[enumerate]{leftmargin=1.7em,itemsep=3pt,topsep=4pt}
\setlist[itemize]{leftmargin=1.5em,itemsep=3pt,topsep=4pt}
\definecolor{inkblue}{RGB}{29,66,110}
\definecolor{lightblue}{RGB}{209,226,244}
\definecolor{warmgray}{RGB}{221,222,224}
\newcommand*{\leanrepository}{}
\title{\vspace{-1.5em}Metric Facility Location at the Approximation Threshold}
\author{Mohammad Hajiaghayi\thanks{Google and University of Maryland, College Park.}
  \and
  Mohammad Mahdian\thanks{Google.}
  \and
  Vahab Mirrokni\thanks{Google.}}
\date{}
\begin{document}
\maketitle

\begin{abstract}
The uncapacitated facility location problem (UFL) is a classical
problem in optimization and has been studied extensively in the
approximation-algorithms literature.  A recent result of OpenAI on the
$k$-median problem implies an $(\alpha_*+\varepsilon)$-approximation
algorithm for metric UFL for every $\varepsilon>0$, where
$\alpha_*\approx1.463$ is the inapproximability threshold of Guha and
Khuller.  We improve this algorithm in two ways.
First, we remove $\varepsilon$ from the approximation factor.  We give
polynomial-time algorithms whose factor, measured against the standard
linear programming relaxation, is $\alpha_*-\Omega(1/\sqrt m)$, where
$m$ is the number of facilities, and we show that this is best
possible for the relaxation: its integrality gap on instances with $m$
facilities is $\alpha_*-\Theta(1/\sqrt m)$.
Second, we improve the running time.  A randomized combinatorial
algorithm attains factor $\alpha_*-\Omega(1/\sqrt m)$ in $O(m^2n)$
expected time, where $n$ is the number of clients, and factor
$\alpha_*+\varepsilon$ in $O(m^2+mn/\varepsilon^2)$ expected time.  For
fixed $\varepsilon$ the latter is quadratic in $m+n$, which is optimal
for algorithms that access the metric through distance queries and
return explicit client assignments.  A
deterministic combinatorial algorithm attains factor
$\alpha_*-\Omega(1/\sqrt m)$ in $O((m+n)^4)$ time.
\end{abstract}

\section{Introduction}
\label{sec:introduction}

The uncapacitated facility location problem (UFL) is one of the most
studied optimization problems in the approximation-algorithms
literature.  We are given a set of facilities and a set of clients in
a common metric space.  Each facility has an opening cost.  The goal
is to open a set of facilities and connect every client to an open
facility so as to minimize the total opening cost plus the total
distance from the clients to the facilities they are connected to.
The problem has served as a testing ground for most of the major
techniques of the field: LP rounding, primal--dual algorithms, dual
fitting, local search, cost scaling, and greedy augmentation
\cite{STA97,CS03,JV01,JMM03,CG05,MYZ06,BA10,Li13}.

Shmoys, Tardos, and Aardal gave the first constant-factor
approximation \cite{STA97}.  A long line of work then improved the
factor, using LP rounding \cite{CS03,Sviridenko2002}, primal--dual and
greedy algorithms \cite{JV01,JMS02,JMM03}, and cost scaling combined
with greedy augmentation \cite{CG05,MYZ06}.  Byrka and Aardal obtained
factor $1.5$ \cite{BA10}, and Li obtained factor $1.488$ \cite{Li13}.
On the negative side, Guha and Khuller \cite{GK98,GK99} proved that no
polynomial-time algorithm has a factor below
\begin{equation}
 \alpha_*=1.463\ldots,
 \qquad\text{the solution of }\alpha_*=1+2e^{-\alpha_*},
 \label{eq:threshold-definition}
\end{equation}
under a standard complexity assumption, and their construction also
shows that the standard LP relaxation has integrality gap arbitrarily
close to $\alpha_*$.

Recently, OpenAI gave an improved approximation algorithm for the
metric $k$-median problem \cite{OAI26}.  It builds on the graph-based
iterative rounding framework of Byrka et al.\
\cite{BGH26Clustering,BGH26} and analyzes it with a new potential
function, through a finite comparison inequality.  As observed by
Byrka et al.\ \cite[Section~1.3 and footnote~1]{BGH26}, this result
implies an $(\alpha_*+\varepsilon)$-approximation algorithm for UFL
for every $\varepsilon>0$: the rounding preserves the probability with
which each facility is opened, and therefore applies to arbitrary
opening costs.  This essentially closes the gap between the upper and
lower bounds.  It leaves open whether the factor $\alpha_*$ itself can
be attained, and how fast.

\paragraph{Our contributions.}
We answer both questions:

\emph{The exact threshold.}
We give a deterministic polynomial-time algorithm whose solution costs
at most $\alpha_*$ times the optimum of the standard LP relaxation,
with no additive $\varepsilon$.  In fact the factor $\alpha_*$ is
never attained on a finite instance: we give algorithms whose factor
against the LP is $\alpha_*-\Omega(1/\sqrt m)$ on instances with $m$
facilities.  We show that this dependence on $m$ is tight for the
relaxation.  The integrality gap of the standard LP on instances with
$m$ facilities is $\alpha_*-\Theta(1/\sqrt m)$; the matching lower
bound comes from a finite analysis of the Guha--Khuller instances.

\emph{Fast combinatorial algorithms.}
The algorithm implied by the $k$-median result rounds an optimal
solution of a linear program.  We give a randomized
algorithm that uses no linear-programming solver and attains factor
$\alpha_*-\Omega(1/\sqrt m)$ in $O(m^2n)$ expected time, where $n$ is
the number of clients.  If an additive $\varepsilon$ is allowed, its
expected running time is $O(m^2+mn/\varepsilon^2)$, which is quadratic
in $m+n$ for every fixed $\varepsilon$.  This is optimal when the
metric is accessed through distance queries and the algorithm returns
explicit client assignments: by a classical lower bound, every fixed
approximation factor requires a quadratic number of queries
\cite{Thorup03,BCIS05}.  We also give a deterministic combinatorial
algorithm with factor $\alpha_*-\Omega(1/\sqrt m)$ and running time
$O((m+n)^4)$.

\emph{A modular presentation.}
Each of our algorithms consists of three independent steps: computing
a fractional solution, rounding it, and augmenting the rounded
solution.  We give two ways to perform the first step, three rounding
algorithms, and two augmentation procedures.  Each has a
self-contained guarantee, and the results above are obtained by
combining them.

\subsection{Results}

Throughout, $m\ge1$ is the number of facilities and $n\ge1$ the number
of clients.  For an instance $I$, $\LP(I)$ is the optimum of the
standard LP relaxation \eqref{eq:standard-lp} and $\OPT(I)$ the
optimum of the problem.  Running times count arithmetic operations and
comparisons on rational numbers, distance queries (including distances
between facilities), and, for randomized algorithms, fair random bits.
Every output includes an assignment of each client to a nearest open
facility.  These assignments can be computed in $O(mn)$ time, within
all the running-time bounds below.

\begin{theorem}[The threshold from one LP solution]
\label{thm:main-simple}
There is a deterministic polynomial-time algorithm for metric
facility location with rational input that solves the standard LP once
and returns a solution $S$ with
\[
                  \cost(S)\le\alpha_*\LP(I)\le\alpha_*\OPT(I).
\]
\end{theorem}

Previously, an $(\alpha_*+\varepsilon)$-approximation was known for
every fixed $\varepsilon>0$ \cite{OAI26,BGH26}.

\begin{theorem}[Combinatorial algorithms]
\label{thm:main-upper}
There is a randomized combinatorial algorithm that returns a solution
$S$ with
\[
     \E[\cost(S)]\le\left(\alpha_*-\frac1{400000\sqrt m}\right)\LP(I)
\]
in $O(m^2n)$ expected time.  For every rational $0<\varepsilon\le1$,
a different choice of its accuracy parameter gives
\[
                 \E[\cost(S)]\le(\alpha_*+\varepsilon)\LP(I)
\]
in $O(m^2+mn/\varepsilon^2)$ expected time.  There is also a
deterministic combinatorial algorithm that returns a solution of cost
at most $\bigl(\alpha_*-1/(400000\sqrt m)\bigr)\LP(I)$ in
$O((m+n)^4)$ time.
\end{theorem}

Here \emph{combinatorial} means that the fractional solution is
computed by a direct iterative procedure and not by a general-purpose
linear-programming solver.  Table~\ref{tab:algorithms} summarizes the
algorithms.

\begin{table}[t]
\centering
\small
\begin{tabular}{@{}p{3.3cm}p{2.5cm}p{3.2cm}p{4.2cm}@{}}
\toprule
Algorithm & Factor & Running time & Components\\
\midrule
Deterministic, LP-based
 & $\alpha_*$ & Polynomial
 & LP solver, Rounding~1, greedy augmentation\\[3pt]
Randomized, combinatorial
 & $\alpha_*-\Omega(m^{-1/2})$ & $O(m^2n)$ expected
 & Combinatorial fractional solution, Rounding~2, random augmentation\\[3pt]
Randomized, combinatorial
 & $\alpha_*+\varepsilon$
 & $O(m^2+mn/\varepsilon^2)$ expected
 & As in the preceding row\\[3pt]
Deterministic, combinatorial
 & $\alpha_*-\Omega(m^{-1/2})$ & $O((m+n)^4)$
 & Combinatorial fractional solution, Rounding~3, derandomized augmentation\\
\bottomrule
\end{tabular}
\caption{The algorithms of this paper.  All factors are measured
against the standard LP; $m$ is the number of facilities and $n$ the
number of clients.  The components are defined in
\cref{sec:preliminaries}.}
\label{tab:algorithms}
\end{table}

\begin{theorem}[The finite integrality gap]
\label{thm:main-finite-gap}
Let $G(m)$ be the supremum of $\OPT(I)/\LP(I)$ over rational
metric instances with positive LP value and at most $m$ facilities.
For every $m\ge64$,
\[
 \alpha_*-\frac{72}{\sqrt m}
       \le G(m)
       \le \alpha_*-\frac{1}{300000\sqrt m}.
\]
Consequently, $\alpha_*-G(m)=\Theta(m^{-1/2})$.
\end{theorem}

The upper bound is a consequence of our rounding and augmentation
analysis.  The lower bound follows from a finite analysis of the
classical instances of Guha and Khuller \cite{GK98,GK99}; the instances
are theirs, and only the quantitative estimate is new.  We give it in
\cref{sec:matching-lp-lower-bound}.  In particular, the integrality gap
of the standard LP is exactly $\alpha_*$, as a supremum that is not
attained.

\begin{corollary}[Quadratic time is necessary]
\label{cor:main-quadratic-necessary}
For every fixed $\rho\ge1$, any randomized algorithm that returns
open facilities and explicit client assignments with expected
approximation factor $\rho$ for arbitrary metric facility location
requires $\Omega((m+n)^2)$ expected distance queries on some
instances.  This holds with equal opening costs and rational
distances.
\end{corollary}

This follows from the classical quadratic lower bound for estimating
the optimum within any fixed constant factor \cite[Theorem~2]{BCIS05};
see also \cite{Thorup03}.  An explicitly assigned feasible solution
can be evaluated with $O(n)$ additional distance queries.  Markov's
inequality, constant-factor truncation of the query count, and a
sufficiently large constant gap in the same construction extend the
lower bound to expected guarantees.  The
bound concerns arbitrary metrics accessed through distance queries; it
says nothing about graph or geometric representations.

\paragraph{Techniques.}
Our algorithms use the comparison inequality that underlies the
$k$-median result of OpenAI, but in a simpler rounding scheme that is
specific to facility location.  We open facilities one at a time.
In the associated randomized process, when a facility opens, the
fractional opening values of nearby
facilities are reduced so that the expected opening cost is preserved
for every facility separately.  The connection cost is controlled by a
potential function with a single scalar parameter shared by all
clients.  No batches of simultaneous openings or cardinality
constraints are needed.  The rounding analysis gives a bound on total
cost in which the coefficients of fractional opening and connection
cost are $1$ and $1+2/e$, respectively, up to the numerical and
variance terms specified below; the classical
combination of cost scaling and augmentation
\cite{CG05,MYZ06} then balances the two factors at $\alpha_*$.

Two further ideas give the improvements over $\alpha_*+\varepsilon$.
First, in the very first opening all clients are in the same state,
and there the comparison inequality can be strengthened to give a
strictly positive gain.  This gain, although small, pays for the
rounding of all transcendental quantities to rational numbers, which
gives the factor $\alpha_*$ exactly.  Second, we replace greedy
augmentation by a randomized augmentation that samples additional
facilities according to the fractional solution.  It is faster, and
it gains an amount of order $1/\sqrt m$ in the factor.  This gain
absorbs the error of an approximately optimal fractional solution,
which we compute by a fast combinatorial algorithm, and the error of a
simpler randomized implementation of the rounding.

\subsection{Related work}

\paragraph{Hardness.}
Guha and Khuller proved their bound assuming
$\mathrm{NP}\not\subseteq\mathrm{DTIME}(n^{O(\log\log n)})$
\cite{GK99}.  The strengthening to $\mathrm P\ne\mathrm{NP}$ is
recorded with proof by Vygen \cite[Theorems~4.12--4.13]{Vygen05} and
also follows from the set-cover hardness of Dinur and Steurer
\cite{DinurSteurer14}.  The integrality-gap statements in this paper
are unconditional.

\paragraph{Rounding and $k$-median.}
Lagrangian multiplier preserving guarantees connect facility location
and $k$-median
\cite{JV01,JMM03,LiSvensson2016,CohenAddadEtAl2023LMP,
CohenAddadEtAl2025KMedian}.
Recent work uses iterative randomized rounding
\cite{BGH26Clustering,BGH26}.
Our analytic starting point is OpenAI's finite comparison theorem
\cite[Theorem~4.1]{OAI26}.  We include a self-contained proof of it,
and prove a stronger inequality for the first iteration.

\paragraph{Fast algorithms.}
Fast algorithms for general metrics were studied by Thorup
\cite{Thorup03} and Mahdian, Ye, and Zhang \cite{MYZ06}.
Thorup obtained factor $1.62$ in $O((m+n)^2\log(m+n))$ time in the
distance-query model, as well as faster algorithms for graph
metrics \cite{Thorup05}.  B\u{a}doiu, Czumaj, Indyk, and Sohler
\cite{BCIS05} studied sublinear algorithms and the obstruction posed
by arbitrary metrics.

\paragraph{Fractional solutions.}
Our fractional computation builds on Young's algorithm for
facility-location LPs \cite{Young15Nearly}, whose running time is
$O(mn\log(n+1)/\varepsilon^2)$.
We use the radii of Mettu and Plaxton \cite{MettuPlaxton03} to
assign client penalties.  They give a constant-factor estimate of
the LP value and permit a feasibility repair without increasing
the objective.  A weighted stopping rule then removes the factor
$\log(n+1)$.  Algorithms for pure packing and covering LPs
\cite{KoufogiannakisYoung2014} do not directly give this bound,
because facility location has shared opening variables.

\paragraph{Derandomization.}
Our deterministic combinatorial algorithm uses the method of
conditional expectations, as used for facility-location rounding by
Chudak and Shmoys \cite[Section~4]{CS03}.  Related methods that
preserve marginals appear in
\cite{GandhiKhullerParthasarathySrinivasan2006}.

\paragraph{Organization.}
\Cref{sec:preliminaries} defines the problem, describes the three
steps common to all our algorithms, and states the classical
augmentation theorem.  \Cref{sec:one-opening} introduces the potential
function and analyzes the opening of one facility; this analysis is
shared by the three rounding algorithms of
\cref{sec:rounding-one,sec:quadratic-rounding,sec:weighted-deterministic-rounding}.
\Cref{sec:combinatorial-fractional} gives the combinatorial algorithm
for fractional solutions and \cref{sec:augmentation} the random
augmentation.  \Cref{sec:combining} combines the components and proves
the theorems above.  \Cref{sec:general-comparison-proof} proves the
comparison inequality.  The appendices contain the analysis of greedy
augmentation, the rational-arithmetic details, the integrality-gap
instances, the verification of scalar inequalities, and a flow
formulation of the relaxation.

\section{Preliminaries and the structure of the algorithms}
\label{sec:preliminaries}

\paragraph{The problem.}
Let $\mathcal F$ and $\mathcal C$ be the sets of facilities and clients,
with $|\mathcal F|=m$ and $|\mathcal C|=n$.
They lie in a common metric space with distance function $d$ and may
share locations.
Facility $i$ has opening cost $f_i\ge0$.  A nonempty set $S$ of open
facilities has cost
\[
       \cost(S)=\sum_{i\in S}f_i+
                     \sum_{j\in\mathcal C}\min_{i\in S}d(i,j).
\]
We call the two sums the opening cost and the connection cost of $S$.
We assume $m,n\ge1$, write $(u)_+=\max\{u,0\}$, and use natural
logarithms.

\paragraph{The linear programming relaxation.}
The standard relaxation has opening variables $y_i$ and assignment
variables $x_{ij}$:
\begin{equation}
 \min\left\{\sum_i f_i y_i+\sum_{i,j}d(i,j)x_{ij}:
       \sum_i x_{ij}=1,\quad 0\le x_{ij}\le y_i\le1\right\}.
 \label{eq:standard-lp}
\end{equation}
A \emph{feasible fractional solution} is a feasible solution $(x,y)$ of
\eqref{eq:standard-lp}.  Its opening cost is $F=\sum_i f_iy_i$, the
connection cost of client $j$ is $C_j=\sum_i x_{ij}d(i,j)$, and its
connection cost is $C=\sum_jC_j$.

\paragraph{Three steps.}
Every algorithm in this paper has the following form.
\begin{enumerate}
\item \emph{Fractional solution.}  Compute a feasible fractional
solution $(x,y)$ that is optimal or nearly optimal for
\eqref{eq:standard-lp}.
\item \emph{Rounding.}  Round $(x,y)$ to a set $S_0$ of open
facilities with a bound on total cost or on a prescribed weighted sum
of opening and client connection costs, relative to the corresponding
fractional costs.  The basic connection factor is $\beta=1+2/e$;
the numerical and variance terms are specified below.
The randomized guarantee holds in expectation.  In one of our
algorithms this step is run with the opening costs multiplied by a
scaling factor.
\item \emph{Augmentation.}  Open additional facilities.  This
increases the opening cost and decreases the connection cost, and
balances the two factors at $\alpha_*$.
\end{enumerate}
The three steps are independent.  Each rounding algorithm applies to
an arbitrary feasible fractional solution, however it was computed,
and each augmentation procedure applies to an arbitrary initial set
$S_0$.

\paragraph{Cost scaling and greedy augmentation.}
The classical way to perform the third step is greedy augmentation
combined with cost scaling \cite{CG05,MYZ06}; we follow the
formulation of Mahdian, Ye, and Zhang \cite{MYZ06}.  Starting from a
nonempty set $S_0$, first open every facility of opening cost zero.
For the current set $S$ of open facilities, let
\[
 D_j=\min_{i\in S}d(i,j),\qquad
 G_i=\sum_j(D_j-d(i,j))_+.
\]
Opening $i$ saves $G_i$ in connection cost.  While some facility has
$G_i>f_i$, open one maximizing the ratio $G_i/f_i$.  The procedure
uses only the instance and $S_0$.

\begin{theorem}[Cost scaling and greedy augmentation \cite{CG05,MYZ06,BA10}]
\label{thm:greedy-augmentation}
Let $(x,y)$ be a feasible fractional solution with opening cost $F$
and connection cost $C$, and let $s\ge1$.  Let $S_0$ be a nonempty set
of facilities with opening cost $F_0$ and connection cost $C_0$ such
that
\begin{equation}
               sF_0+C_0\le\gamma_f\,sF+\gamma_c\,C
 \label{eq:scaled-bifactor}
\end{equation}
for some $\gamma_f,\gamma_c\ge0$.  Then greedy augmentation started
from $S_0$ returns a set $S$ with
\[
 \cost(S)\le(\gamma_f+\log s)\,F
              +\left(1+\frac{\gamma_c-1}{s}\right)C.
\]
\end{theorem}

The left side of \eqref{eq:scaled-bifactor} is the cost of $S_0$ in
the instance whose opening costs are multiplied by $s$.  Thus
\eqref{eq:scaled-bifactor} holds whenever $S_0$ is produced, on that
scaled instance, by an algorithm whose cost is at most $\gamma_f$
times the opening cost plus $\gamma_c$ times the connection cost of
$(x,y)$.  The comparison must be with the same solution $(x,y)$ that
appears in the conclusion.  This is the case when the algorithm rounds
the given solution $(x,y)$ and its guarantee does not depend on the
opening costs, as for our rounding algorithms; it is also the case for
algorithms whose guarantee holds against every feasible solution, as in
\cite{MYZ06}.

The statement in \cite{MYZ06} has an additional $\varepsilon$ in the
first factor.  The exact transformation is also stated by Byrka and
Aardal \cite[Lemma~2.1]{BA10}.  We claim no novelty for it; for
completeness, \cref{sec:greedy-augmentation} proves the formulation
above for an arbitrary feasible fractional solution.  With
$(\gamma_f,\gamma_c)=(1,\beta)$ and $s=e^{\alpha_*-1}$, both factors in
the conclusion equal $\alpha_*$.

\paragraph{The components.}
The rest of the paper provides the following components.  Our rounding
algorithms are based on OpenAI's analysis \cite{OAI26} of the
iterative rounding framework of Byrka et al.\
\cite{BGH26Clustering,BGH26}.
\begin{itemize}
\item \emph{Rounding~1} (\cref{sec:rounding-one}) is deterministic.
Its connection factor is strictly smaller than $\beta$, by an amount
that depends on the fractional solution.
\item \emph{Rounding~2} (\cref{sec:quadratic-rounding}) is randomized
and runs in $O(m^2+mn)$ expected time.
\item \emph{Rounding~3} (\cref{sec:weighted-deterministic-rounding}) is
a deterministic version of Rounding~2 for a weighted sum of connection
costs.
\item A \emph{combinatorial algorithm for the fractional solution}
(\cref{sec:combinatorial-fractional}) replaces the LP solver in the
first step.
\item \emph{Random augmentation} (\cref{sec:augmentation}) is an
alternative to greedy augmentation in the third step.  Unlike greedy
augmentation, it takes the fractional solution as an additional input.
It runs in $O(m^2+mn)$ expected time and improves the resulting factor
by an amount of order $1/\sqrt m$.  We also give its derandomization.
\end{itemize}
\Cref{sec:combining} combines these components.

\paragraph{Fixed parameters.}
The following values are used by Rounding~2, Rounding~3, and random
augmentation:
\begin{equation}
 L=\lceil\sqrt m\rceil,\qquad K=1000,\qquad
 \gamma=\frac1{10^6L}.
 \label{eq:final-parameters}
\end{equation}
Here $m$ is the number of facilities of the input instance, even if
co-located facilities are subsequently merged.

\section{The potential function and the analysis of one opening}
\label{sec:one-opening}

The three rounding algorithms open one facility at a time and share
the same state, the same update, and the same potential function.
This section describes them and bounds the change of the potential
when one facility opens.  The potential is defined in terms of a
function $A$; the analysis itself dictates the choice of $A$, which we
make in \cref{sec:choice-of-A}.

\subsection{The state and the potential}
\label{sec:state}

A rounding algorithm starts from a feasible fractional solution
$(x,y)$.  It maintains a set $O$ of open facilities, initially empty,
and a current opening variable $z_i\in[0,1]$ for every unopened
facility, initially $z_i=y_i$.  Let $\mathcal U$ be the set of unopened
facilities with $z_i>0$, and put
\[
                     Y=\sum_{i\in\mathcal U}z_i .
\]
Initially $Y\ge1$ by feasibility.  While $Y\ge1$, an iteration opens
one facility $h\in\mathcal U$ and decreases the variables of the other
facilities in $\mathcal U$.  When $Y<1$, at most one further facility
is opened and the algorithm stops.

For each client $j$, the analysis follows a set $S_j$ of facilities
and assignment variables $u_{ij}$, $i\in S_j$, with
$0\le u_{ij}\le z_i$.  Initially $S_j=\{i:x_{ij}>0\}$ and
$u_{ij}=x_{ij}$.  Fix an order of the initial set by increasing
distance from $j$.  At all times $S_j$ is a prefix, in this order, of
the facilities of the initial set that are still in $\mathcal U$; it
may be shortened after every opening, and a discarded facility never
returns.  Each $u_{ij}$ decreases in the same proportion as $z_i$.
Write
\[
 y_j=\sum_{i\in S_j}u_{ij},\qquad
 V_j=\sum_{i\in S_j}u_{ij}\,d(i,j).
\]
Thus $y_j=1$ and $V_j=C_j$ initially, and both can only decrease.
Finally, let $R_j=\max\{d(i,j):x_{ij}>0\}$, and let $b_j$ be an
auxiliary distance, initially $5R_j$, which is replaced by
$\min\{b_j,d(h,j)\}$ whenever a facility $h$ opens.  We will see
(\cref{lem:nearest-open-facility}) that when $Y<1$, $b_j$ is the
distance from $j$ to the nearest open facility.  If at that point one
more facility is opened, facility $i$ with probability $z_i$ and none
with probability $1-Y$, the expected connection cost of client $j$ is
at most
\begin{equation}
                         V_j+(1-y_j)\,b_j .
 \label{eq:terminal-bound}
\end{equation}

\paragraph{The potential.}
Let $A:[0,1]\to\mathbb R$ be an increasing, concave, differentiable
function with $A(0)=1$ and $A'\le1$, and write
\[
   P(v)=\tfrac12A'(v),\qquad \beta=A(1),\qquad
   a(v,y)=A(v)+2P(v)(y-v).
\]
Thus $y\mapsto a(v,y)$ is the tangent to $A$ at $v$.  For a parameter
$\ybar\in[0,1]$, the potential of client $j$ is
\begin{equation}
        \phi_{j,\ybar}=a(\ybar,y_j)\,V_j+(1-y_j)\,b_j .
 \label{eq:client-potential}
\end{equation}
Intuitively, $A(v)$ is the factor that we are prepared to pay on the
remaining fractional connection cost of a client that still holds a
fraction $v$ of its assignment.  At the end the factor must be at
least one, since \eqref{eq:terminal-bound} is what the client actually
pays; at the beginning it is $\beta$, the connection factor of the
rounding.  The following two facts are all that is needed about $A$
for now.
\begin{itemize}
\item A tangent to a concave function lies above it, so
$a(\ybar,y)\ge A(y)\ge1$ for all $\ybar,y\in[0,1]$.  Hence the
potential \eqref{eq:client-potential} is at least the terminal bound
\eqref{eq:terminal-bound}, for every value of the parameter.
\item Initially $y_j=1$ for every client.  With $\ybar=1$ the sum of
the client potentials is $\beta C$.
\end{itemize}
Consequently, if the opening variables and the sum of the client
potentials do not increase in expectation during the process, the
final solution has expected opening cost at most $F$ and expected
connection cost at most $\beta C$.

\subsection{Opening one facility}
\label{sec:coefficients}

Fix a number $\widehat P\in(0,1/2]$; it will be a rational lower
estimate of $P(\ybar)$.
For each facility $i\in\mathcal U$, choose coefficients $q_{hi}$ indexed
by facilities $h\in\mathcal U$.
If $\sum_{h\in\mathcal U:d(i,h)=0}z_h<1$, choose the least
$r_i>0$ satisfying
\begin{equation}
 q_{hi}=\min\left\{1,
          \frac{(1-d(i,h)/r_i)_+}{\widehat P}\right\},
                   \qquad\sum_{h\in\mathcal U}z_hq_{hi}=1.
 \label{eq:rounding-coefficients}
\end{equation}
Otherwise set $r_i=0$.  Set $q_{ii}=1$, then consider the other
facilities $h\in\mathcal U$ with $d(i,h)=0$ in a fixed order.  At each
facility choose the largest $q_{hi}\in[0,1]$ for which the partial sum
does not exceed one.  Stop when $\sum_hz_hq_{hi}=1$.  All remaining coefficients
are zero.  In either case,
\begin{equation}
 q_{ii}=1,\qquad
 \begin{cases}
 r_i(1-\widehat Pq_{hi})\le d(i,h),&q_{hi}<1,\\
 d(i,h)\le r_i(1-\widehat Pq_{hi}),&q_{hi}>0.
 \end{cases}
 \label{eq:rounding-coefficient-inequalities}
\end{equation}
The positive radius exists: the sum in \eqref{eq:rounding-coefficients} is continuous
as a function of $r_i$, starts below one, and reaches $Y\ge1$.
In words, an opening within distance $(1-\widehat P)r_i$ of $i$
removes all of the remaining value of $i$, an opening beyond distance
$r_i$ removes none, and in between the removed fraction is linear in
the distance.

Opening facility $h\in\mathcal U$ makes the update
\begin{equation}
 O\leftarrow O\cup\{h\},
             \qquad z_i\leftarrow z_i(1-q_{hi})\quad(i\in\mathcal U),
 \label{eq:facility-update}
\end{equation}
so that $h$ itself leaves $\mathcal U$, because $q_{hh}=1$.  The
assignment variables $u_{ij}$ are multiplied by the same factors.
If $h$ is chosen with probability $z_h/Y$, the normalization in
\eqref{eq:rounding-coefficients} shows that facility $i$ opens with
probability $z_i/Y$ and that the expected decrease of $z_i$ is
\begin{equation}
     \frac1Y\sum_{h\in\mathcal U}z_h\,z_iq_{hi}=\frac{z_i}{Y}.
 \label{eq:expected-opening-change}
\end{equation}
Thus, for every facility separately, the probability of opening equals
the expected decrease of its variable.  This preserves the expected
opening cost for arbitrary opening costs.

\begin{lemma}[The auxiliary distance]
\label{lem:nearest-open-facility}
Suppose that in every iteration the variable of a facility $i$ changes
only if $q_{hi}>0$ for the facility $h$ opened in that iteration.
Once a facility with $x_{ij}>0$ is opened or its variable decreases,
$b_j$ equals the distance from client $j$ to the nearest open facility.
In particular, this holds for every client when $Y<1$.
\end{lemma}
\begin{proof}
Let $S_j^0=\{i:x_{ij}>0\}$.
Before the first such iteration, every facility in $S_j^0$ is unopened,
and the sum of their variables $z_i$ is at least one: the initial
assignment variables sum to one and satisfy $x_{ij}\le y_i$.
Any two facilities of $S_j^0$ are within distance $2R_j$.
For any $i\in S_j^0$, these facilities therefore suffice to satisfy the
normalization in \eqref{eq:rounding-coefficients} by radius
$2R_j/(1-\widehat P)\le4R_j$.
Thus a facility $h$ whose opening decreases $z_i$ for some $i\in S_j^0$
is within distance $5R_j$ of client $j$.  If $h\in S_j^0$, its distance
to $j$ is at most $R_j$.
In either case $h$ is opened and remains open.  Since $b_j$ is the
minimum of $5R_j$ and the distances to all facilities opened so far,
it thereafter equals the distance to the nearest open facility.
When $R_j=0$, the rule for $r_i=0$ gives the same conclusion.
If no facility in $S_j^0$ had been opened or had its variable decreased,
these facilities would still contribute at least one to $Y$, which is
impossible when $Y<1$.
\end{proof}

\subsection{The change of one client's potential}
\label{sec:one-client}

We now bound the change of the potential of one client when one
facility opens.  The bound is expressed through the following
quantity, which the comparison inequality of \cref{sec:choice-of-A}
controls.

\begin{definition}[Comparison expression]
\label{def:comparison-expression}
Fix $\ybar\in[0,1]$ and $y\in(0,1]$, and abbreviate
\[
     P=P(\ybar),\qquad \mu_0=A(\ybar)-1-2\ybar P(\ybar).
\]
Let $\pi$ be a probability distribution on a finite index set; all
expectations below are over $\pi$, and two indices in one expectation
are sampled independently.  Let $D_i\ge0$ with $\mathbb E_iD_i=1$,
$\mathcal R_i\ge0$, $q_{ti}\in[0,1]$, and $s_i\in[0,1]$.  Put
$\rho_i=\mathbb E_tq_{ti}$ and $C_i=1-Ps_i$, and define
\begin{align}
 \mathcal E_{\ybar,y}:=\mathbb E_i\bigl[&
 C_i(1-\rho_i)\mathcal R_i
 +\rho_i\{\mu_0D_i+(2-s_i)Py(D_i+1)\}-C_i(D_i+1)\notag\\
 &-\mathbb E_t(1-q_{ti})
       (D_t-(1+\mu_0+2Py)D_i)_+\bigr].
 \label{eq:comparison-expression}
\end{align}
The data are \emph{admissible} if
\begin{equation}
 \mathcal R_i(1-Pq_{ti})\le D_i+D_t
                         \qquad\text{whenever }q_{ti}<1.
 \label{eq:admissible}
\end{equation}
\end{definition}

Concavity of $A$ gives $\mu_0\ge0$: the tangent at $\ybar$ lies above
$A(0)=1$.

\begin{lemma}[One opening, one client]
\label{lem:one-client-comparison}
Fix a state with $Y\ge1$, a parameter $\ybar\in[0,1]$, and
$\widehat P\in(0,1/2]$ with $0\le P(\ybar)-\widehat P\le\delta$.  Let
the coefficients $q_{hi}$ and radii $r_i$ be given by
\eqref{eq:rounding-coefficients}.  Consider one client, and omit its
index: it has a set $S$, assignment variables $u_i\le z_i$ for
$i\in S$, distances $d_i$, and values $y$, $V$, $b$ and
$\phi=a(\ybar,y)V+(1-y)b$.  For $h\in\mathcal U$, apply the update
\eqref{eq:facility-update}, replace $b$ by $b_h=\min\{b,d_h\}$, and let
$\phi_h$ be the minimum of the resulting potential over all prefixes of
the remaining set.  Write $a=a(\ybar,y)$ and $P=P(\ybar)$.
\begin{enumerate}
\item If $y=0$ or $V=0$, then $\sum_hz_h(\phi_h-\phi)\le0$.
\item If $V>0$, then
\begin{equation}
 \sum_{h\in\mathcal U}z_h(\phi_h-\phi)
  \le\{1+3y-(1+y)a-2Py\}V+yV\,\mathcal E_{\ybar,y}+2\delta\,yV,
 \label{eq:client-change-bound}
\end{equation}
where $\mathcal E_{\ybar,y}$ is the comparison expression for the
distribution $\pi_i=u_i/y$ on $S$, the coefficients $q_{ti}$ with
$t,i\in S$, and
\[
   D_i=\frac{y\,d_i}{V},\qquad \mathcal R_i=\frac{y\,r_i}{V},\qquad
   \rho_i=\mathbb E_tq_{ti},
\]
and where $s_i$ is the average of $q_{hi}$ over $h\in\mathcal U$ with
weights $(z_h-u_h)q_{hi}$, or zero if all these weights vanish.
These data are admissible.
\end{enumerate}
\end{lemma}

\begin{proof}
Every sum over $h$ below ranges over $\mathcal U$.  Set $u_i=0$ outside
$S$.  If $y=0$, the potential is $b$; the empty prefix has potential
$b_h\le b$ after any opening.  Assume $y>0$.
Expectations below use $\pi$, whereas sums over facilities $h$ use
the weights $z_h$.

\paragraph{Two prefixes.}
For facility $h$, put
\[
          \Delta_h=\sum_{i\in S}u_iq_{hi},\qquad
          U_h=\sum_{i\in S}u_id_iq_{hi}.
\]
These are the decreases in $y$ and $V$ caused by
\eqref{eq:facility-update}.
The remaining assignment variable at $i$ is $u_i(1-q_{hi})$.
Facility $h$ is removed from $S$ when it is opened, so this formula
also holds for $i=h$, because $q_{hh}=1$.
We compare $\phi_h$ with two prefixes.
The first contains all remaining facilities of $S$; its assignment variables
sum to $y-\Delta_h$, and its fractional connection cost is $V-U_h$.
Expanding the potential and using $b_h\le b$ gives
\begin{equation}
 \phi_h-\phi\le\Delta_hb_h-aU_h-2P\Delta_h(V-U_h).
 \label{eq:all-facilities-bound}
\end{equation}
The second prefix contains exactly the remaining facilities satisfying
$ad_i<d_h$.  Its tangent coefficient is at most $a$, because $P\ge0$,
and $b_h\le d_h$, so
\begin{equation}
 \phi_h\le d_h-
       \sum_{i\in S}u_i(1-q_{hi})(d_h-ad_i)_+.
 \label{eq:nearer-facilities-bound}
\end{equation}

\paragraph{Writing each weight as a sum.}
Both bounds apply to every choice of facility $h$.  Write
$z_h=u_h+(z_h-u_h)$; the two summands are nonnegative because
$u_h\le z_h$.  Apply
\eqref{eq:nearer-facilities-bound} with weight $u_h$, and
\eqref{eq:all-facilities-bound} with weight
$z_h-u_h$.

The definition of $\rho_i$ gives $\sum_hu_hq_{hi}=y\rho_i$.
Subtracting this equality from $\sum_hz_hq_{hi}=1$ gives the first identity below;
the second is the definition of $s_i$.
\begin{equation}
 \sum_h(z_h-u_h)q_{hi}=1-y\rho_i,
 \qquad
 \sum_h(z_h-u_h)q_{hi}^2=s_i(1-y\rho_i).
 \label{eq:weighted-coefficient-sums}
\end{equation}
The same average $s_i$ appears in both the bound on $b_h$ and the quadratic
term, so the two parts of the calculation use a single parameter.

For every facility $h$ with $q_{hi}>0$, \eqref{eq:rounding-coefficient-inequalities}
and the triangle inequality give
$b_h\le d_h\le d_i+r_i(1-\widehat Pq_{hi})$; also $b_h\le b$.
Averaging these bounds using \eqref{eq:weighted-coefficient-sums} gives
\begin{align}
 \sum_h(z_h-u_h)q_{hi}b_h
 &\le(1-y\rho_i)\min\{b,\ d_i+(1-\widehat P s_i)r_i\}\notag\\
 &\le(1-y)b+y(1-\rho_i)\{d_i+(1-\widehat P s_i)r_i\}.
 \label{eq:auxiliary-variable-bound}
\end{align}
The last line bounds the minimum by a weighted average, with weights
$1-y$ and $y(1-\rho_i)$, which are nonnegative and sum to $1-y\rho_i$.
The term involving $b$ will cancel a term from \eqref{eq:nearer-facilities-bound}.

The only quadratic term is controlled directly:
\[
 2\Delta_hU_h
 =\sum_{i,t\in S}u_iu_td_t\,2q_{hi}q_{ht}
 \le\sum_{i\in S}u_iq_{hi}^2(V+yd_i),
\]
using $2q_{hi}q_{ht}\le q_{hi}^2+q_{ht}^2$, including the diagonal.
Therefore
\begin{equation}
 2P\sum_h(z_h-u_h)\Delta_hU_h
       \le P\sum_{i\in S}u_i s_i(1-y\rho_i)(V+yd_i).
 \label{eq:core-quadratic-bound}
\end{equation}

\paragraph{The case $V=0$.}
If $V=0$, all facilities in $S$ are at distance zero from the client.
For each $i\in S$ with $r_i>0$, we have $q_{ti}=1$ for every $t\in S$,
so $\rho_i=1$.  The other case is $r_i=0$.  Thus
$(1-\rho_i)r_i=0$ in either case.  All distance terms in the weighted
bounds vanish.  The bound in \eqref{eq:auxiliary-variable-bound} contributes
$y(1-y)b$, which cancels $-y(1-y)b$ from \eqref{eq:nearer-facilities-bound}.  Hence
$\sum_hz_h(\phi_h-\phi)\le0$.

\paragraph{Normalization.}
Assume henceforth that $V>0$ and normalize by the mean distance:
\[
 \mu=V/y,\qquad D_i=d_i/\mu,\qquad
 \mathcal R_i=r_i/\mu.
\]
Then $\mathbb E_iD_i=1$.
Writing $u_i=y\pi_i$ and $d_i=\mu D_i$ shows that the comparison
terms have common scale $y^2\mu=yV$.
Because $\widehat P\le P$, \eqref{eq:rounding-coefficient-inequalities} and the
triangle inequality give admissibility:
\[
        \mathcal R_i(1-Pq_{ti})
        \le\frac{r_i(1-\widehat Pq_{ti})}{\mu}
        \le D_i+D_t\qquad(q_{ti}<1).
\]

\paragraph{Collecting the terms.}
The terms involving $b$ also cancel when $V>0$.
Inequality~\eqref{eq:nearer-facilities-bound} contributes the scalar terms $V-yaV$;
\eqref{eq:all-facilities-bound} contributes $-aV-2PyV$; and
\eqref{eq:auxiliary-variable-bound} contributes $yV$.  Their sum is
$\{1+y-(1+y)a-2Py\}V$.
After division by $yV$, the remaining $\rho_i$ and quadratic terms
combine by the identity
\[
 \begin{split}
 &\rho_i\{(a-1)D_i+2Py\}
                         +Ps_i(1-y\rho_i)(D_i+1)\\
 &\qquad=\rho_i\{\mu_0D_i+(2-s_i)Py(D_i+1)\}
                              +Ps_i(D_i+1),
 \end{split}
\]
using $a-1=\mu_0+2Py$.  The coefficient $a-1$ combines
$a\rho_iD_i$ from \eqref{eq:all-facilities-bound} with
$-\rho_iD_i$ from \eqref{eq:auxiliary-variable-bound}.
Write $Ps_i=1-C_i$, where $C_i=1-Ps_i$, and use
$\mathbb E_i(D_i+1)=2$.
This leaves exactly the middle terms of
\eqref{eq:comparison-expression} and adds $2yV$ to the scalar sum.
The terms involving $\mathcal R_i$ and the expressions $(\cdot)_+$
give its first and last terms.
The sole discrepancy
is that the radius coefficient is $1-\widehat P s_i$ instead of
$1-Ps_i$.  Therefore
\begin{align*}
 \sum_hz_h(\phi_h-\phi)\le{}&
    \{1+3y-(1+y)a-2Py\}V+yV\mathcal E_{\ybar,y}\\
 &+(P-\widehat P)yV\,
                    \mathbb E_i[s_i(1-\rho_i)\mathcal R_i].
\end{align*}
Admissibility implies
$(1-q_{ti})\mathcal R_i\le D_i+D_t$ when $q_{ti}<1$;
the same inequality is trivial when $q_{ti}=1$.  Thus
$(1-\rho_i)\mathcal R_i\le D_i+1$, whose expectation is two.
Since $0\le s_i\le1$ and $P-\widehat P\le\delta$, the last term is at
most $2\delta yV$.
\end{proof}

\subsection{The choice of \texorpdfstring{$A$}{A} and the comparison inequality}
\label{sec:choice-of-A}

The function $A$ has so far been arbitrary.  The first term of
\eqref{eq:client-change-bound} now determines it.  Substituting
$a=A(\ybar)+2P(\ybar)(y-\ybar)$ and expanding in powers of $y-\ybar$
gives the identity
\begin{equation}
 \begin{split}
 1+3y-(1+y)a-2Py
  ={}&\{1+3\ybar-(1+\ybar)A(\ybar)-2\ybar P(\ybar)\}\\
     &+K_0(\ybar)(y-\ybar)-2P(\ybar)(y-\ybar)^2,
 \end{split}
 \label{eq:bracket-expansion}
\end{equation}
where
\begin{equation}
            K_0(\ybar)=3-A(\ybar)-(4+2\ybar)P(\ybar).
 \label{eq:K0}
\end{equation}
The quadratic term is favourable.  The linear term has no sign, but it
disappears on summing over clients if $\ybar$ is the mean of the $y_j$
with weights $V_j$; this is how Rounding~1 chooses the parameter.  The
constant term is the same for all clients, and the potential can be
nonincreasing only if it is not positive.  We choose $A$ so that it
vanishes:
\begin{equation}
            (1+v)A(v)+v\,A'(v)=1+3v\qquad(0\le v\le1).
 \label{eq:core-coefficient-identities}
\end{equation}
The solutions of this linear differential equation are
$3-2/v+c\,e^{-v}/v$, and exactly one of them, with $c=2$, is bounded
at zero.  Hence
\begin{equation}
 A(v)=1+2\int_0^1(1-e^{-tv})\,dt,
 \qquad P(v)=\tfrac12A'(v)=\int_0^1te^{-tv}\,dt,
 \qquad \beta=A(1)=1+\frac2e.
 \label{eq:analytic-coefficients}
\end{equation}
From now on $A$ and $P$ denote these functions.  They have the
properties required in \cref{sec:state}: $A$ is increasing and
concave, $A(0)=1$, and $1/4<P(v)\le1/2$ on $[0,1]$.  Differentiating
\eqref{eq:core-coefficient-identities} gives
\begin{equation}
                     K_0(v)=2v\,P'(v)\le0 .
 \label{eq:K0-derivative}
\end{equation}
With this choice, \eqref{eq:client-change-bound} reads
\begin{equation}
 \sum_{h\in\mathcal U}z_h(\phi_h-\phi)
  \le K_0(\ybar)(y-\ybar)V-2P(\ybar)(y-\ybar)^2V
          +yV\,\mathcal E_{\ybar,y}+2\delta\,yV.
 \label{eq:client-change-bound-A}
\end{equation}
It remains to control the comparison expression.  The following
theorem, due to OpenAI~\cite[Theorem~4.1]{OAI26}, shows that it is
paid for by the quadratic term.  We give a self-contained proof in
\cref{sec:general-comparison-proof}.

\begin{theorem}[OpenAI finite comparison theorem]
\label{thm:core-finite-comparison}
For all $\ybar\in[0,1]$, $y\in(0,1]$, and admissible data,
\begin{equation}
 \mathcal E_{\ybar,y}\le\frac{2P(\ybar)\,(y-\ybar)^2}{y}.
 \label{eq:core-finite-comparison}
\end{equation}
\end{theorem}

At the initial state, where $\ybar=y=1$, we prove more.  This
strengthening is new, and is the source of the strict improvement in
Rounding~1.

\begin{lemma}[An improved inequality for the first iteration]
\label{lem:first-iteration-improvement}
There is an absolute rational constant $c\in(0,1]$ such that, for all
admissible data,
\begin{equation}
              \mathcal E_{1,1}
                 \le-c\,\mathbb E_i[\rho_i^2D_i].
 \label{eq:first-iteration-improvement}
\end{equation}
One can take $c=1/100$.
\end{lemma}

Combining \eqref{eq:client-change-bound-A} with
\cref{thm:core-finite-comparison} gives the form used by Rounding~2
and Rounding~3.

\begin{corollary}[One opening, one client]
\label{cor:one-client-residual}
In the setting of \cref{lem:one-client-comparison}, for every client
and every parameter $\ybar\in[0,1]$,
\[
 \frac1Y\sum_{h\in\mathcal U}z_h(\phi_h-\phi)
 \le\frac{K_0(\ybar)}Y(y-\ybar)V+\frac{2\delta V}{Y}.
\]
\end{corollary}

\section{Rounding~1: deterministic rounding with a strict improvement}
\label{sec:rounding-one}

Rounding~1 evaluates the client potentials explicitly.  It takes the
parameter $\ybar$ to be a weighted mean over the clients and opens, in
each iteration, the facility that minimizes an upper bound on the new
potential.  Its connection factor is strictly below $\beta$; the gain
comes from the first iteration and pays for rational arithmetic.

\begin{theorem}[Rounding~1]
\label{thm:rounding-bound}
Let a feasible rational fractional solution
have opening cost $F$, connection cost $C$, and $N$ facilities with
positive opening variables.  Let $\xi$ be its smallest positive assignment variable,
let $c$ be the constant of \cref{lem:first-iteration-improvement}, and put
\[
                           \eta=\frac{c\,\xi^2}{2N}.
\]
A deterministic algorithm returns a set of open facilities of
cost at most
\[
                               F+(\beta-\eta)C.
\]
Its running time is polynomial in the rational encoding length of the
instance and the supplied fractional solution.  The guarantee holds
for arbitrary supplied nonnegative opening costs; the fractional
solution need not be optimal for those opening costs.
\end{theorem}

\subsection{Setup}

\paragraph{Zero connection cost.}
If a feasible fractional solution has $C=0$, every client is assigned
only to facilities at distance zero. Each zero-distance equivalence class
containing a client therefore has a sum of opening variables at least
one. Open a cheapest facility with a positive opening variable in
each such class. Its opening cost is at most the fractional opening
cost in that class, so this
construction costs at most $F$ and has no connection cost.
We assume $C>0$ from now on.

\paragraph{The state.}
Discard facilities with $y_i=0$.  On the remaining $N\le m$ facilities
write $p_i=y_i$ and $p_{ij}=x_{ij}$ for the supplied values, which
stay fixed, and put
\[
 \theta_{ij}=p_{ij}/p_i\in[0,1],\qquad
                    \xi=\min\{p_{ij}:p_{ij}>0\}.
\]
The state is that of \cref{sec:state}, with $u_{ij}=\theta_{ij}z_i$ for
$i\in S_j$.  Order $S_j^0=\{i:p_{ij}>0\}$ by increasing distance from
$j$, breaking ties by a fixed facility order; $S_j$ always consists of
the first $k$ facilities of $S_j^0$, for some $k\ge0$, after omitting
open facilities and facilities with $z_i=0$.  Write
$V_j^0=\sum_i p_{ij}d(j,i)$ for the initial connection cost of client
$j$, so $\sum_jV_j^0=C$ and $V_j\le V_j^0$ throughout.
A facility at distance $R_j$ from $j$ has an initial assignment
variable of value at least $\xi$, so
\begin{equation}
                    \xi R_j\le V_j^0,\qquad
                              \xi\sum_jR_j\le C.
 \label{eq:core-support-radius}
\end{equation}
For the opening cost, write
\[
    F(z)=\sum_{i\in O}f_i+\sum_{i\in\mathcal U}f_iz_i .
\]

\paragraph{The grid.}
Fix $\tau$, the reciprocal of an integer, such that every
$p_i$ is a multiple of $\tau$ and
\begin{equation}
                  0<\tau<1/4,\qquad
                         \frac{8N^2\tau}{\xi}\le\eta.
 \label{eq:rounding-error-condition}
\end{equation}
\Cref{sec:rational-precision} constructs such a $\tau$.  After each update
\eqref{eq:facility-update}, round the variable of every unopened
facility down to a multiple of $\tau$.  All stored values of $z_i$ are
then multiples of $\tau$.  If $q_{hi}=0$, neither the update nor the
rounding changes $z_i$, so \cref{lem:nearest-open-facility} applies.

\subsection{The potential at the weighted mean}

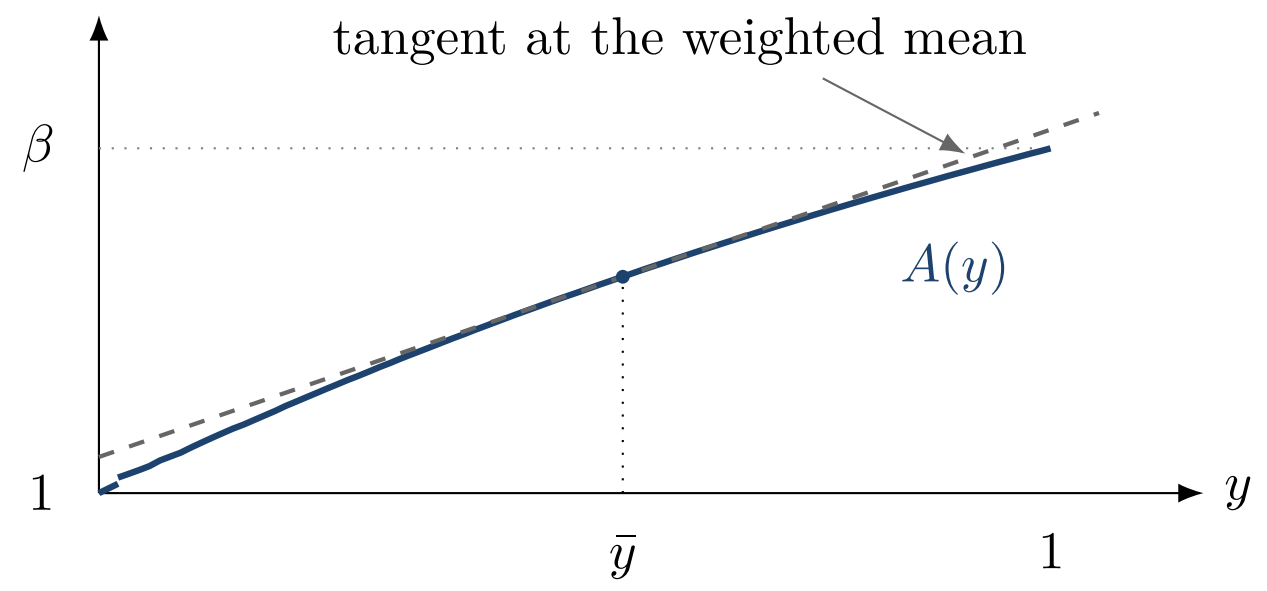
\begin{figure}[t]
\centering
\begin{tikzpicture}[x=6.3cm,y=3.1cm,font=\small,>=Latex]
 \draw[->] (0,0)--(1.16,0) node[right] {$y$};
 \draw[->] (0,0)--(0,1.02);
 \node[anchor=east] at (-.025,0) {$1$};
 \node[anchor=east] at (-.025,.735759) {$\beta$};
 \draw[dotted,gray] (0,.735759)--(1,.735759);
 \draw[inkblue,very thick] (0,0)--(.02,.019867);
 \draw[inkblue,very thick,domain=.02:1,samples=90]
      plot (\x,{2-2*(1-exp(-\x))/\x});
 \draw[dashed,gray!80!black,thick,domain=0:1.05,samples=2]
      plot (\x,{.461635674111+.699026736597*(\x-.55)});
 \draw[dotted] (.55,0)--(.55,.461635674111);
 \fill[inkblue] (.55,.461635674111) circle[radius=1.3pt];
 \node[below=4pt] at (.55,0) {$\bar y$};
 \node[below=4pt] at (1,0) {$1$};
 \node[anchor=west,text=inkblue] at (.82,.48) {$A(y)$};
 \node[align=center] at (.61,.965)
       {tangent at the weighted mean};
 \draw[->,gray!80!black] (.76,.885)--(.91,.724);
\end{tikzpicture}
\caption{The potential uses a tangent to the concave function $A$.
After summing with weights $V_j$, the minimizing tangent is at
$\ybar=\sum_jy_jV_j/\sum_jV_j$.  The vertical coordinate is the
coefficient itself; its value at the horizontal axis is one.
The same tangent is used for every client during an iteration.}
\label{fig:mean-tangent}
\end{figure}

Define the weighted mean
\[
 \ybar=\frac{\sum_jy_jV_j}{\sum_jV_j},
 \qquad\text{with }\ybar=0\text{ if }\sum_jV_j=0.
\]
By concavity of $A$, the tangent at $\ybar$ minimizes the sum of the
client potentials \eqref{eq:client-potential}; see
Figure~\ref{fig:mean-tangent}:
\begin{equation}
 \Phi:=A(\ybar)\sum_jV_j+\sum_j(1-y_j)b_j
                 =\min_{0\le v\le1}\sum_j \phi_{j,v}.
 \label{eq:potential-function}
\end{equation}
The potential function of Rounding~1 is $F(z)+\Phi$.  Initially
$\ybar=1$ and $\Phi=\beta C$.  Because the tangent at the weighted
mean is the minimizing one, we may analyze one opening with the old
tangent fixed; recomputing the mean afterwards can only decrease
$\Phi$.

We evaluate only one scalar function in each iteration.  Choose a
rational lower estimate satisfying
\begin{equation}
               0\le P(\ybar)-\widehat P\le\tau/2,
                        \qquad 0<\widehat P\le1/2,
 \label{eq:scalar-precision}
\end{equation}
and recover
\begin{equation}
 \widehat A=\frac{1+3\ybar-2\ybar\widehat P}{1+\ybar},\qquad
             a^+(v)=\widehat A+2\widehat P(v-\ybar)+\tau v.
 \label{eq:rational-coefficients}
\end{equation}
The identity \eqref{eq:core-coefficient-identities} implies
\begin{equation}
                    0\le a^+(v)-a(\ybar,v)\le2\tau
                              \qquad(0\le v\le1).
 \label{eq:tangent-error}
\end{equation}
\Cref{sec:rational-precision} proves this bound and shows how to
compute $\widehat P$.

\subsection{The algorithm}

For each facility $h\in\mathcal U$, start from a copy of the current
variables and sets.  Make update~\eqref{eq:facility-update}, round
the variables of unopened facilities down to multiples of $\tau$, and set
$b_j^h=\min\{b_j,d(j,h)\}$.
For each client, remove facilities in $O$ and facilities with $z_i^h=0$
from its current set $S_j$, retaining the distance order.
For each $k$ from zero to the number of remaining facilities, let $T$
consist of the first $k$ facilities in this order.  Write $y(T)$ and
$V(T)$ for the corresponding sums, computed with assignment variables
$\theta_{ij}z_i^h$.  Choose a set $T$ minimizing
\begin{equation}
             a^+(y(T))V(T)+(1-y(T))b_j^h,
 \label{eq:client-upper-bound}
\end{equation}
and denote this minimum by $H_j^h$.
Consider $k$ in increasing order and retain the first minimizer.
The new value of $\Phi$ is at most the sum of these minima,
\begin{equation}
                            \Phi^h\le\sum_j H_j^h,
 \label{eq:new-potential-bound}
\end{equation}
because $H_j^h$ bounds $\phi_{j,\ybar}$ on the chosen set and
$\Phi^h$ is a minimum over the parameter.  For each $h\in\mathcal U$,
define
\begin{equation}
             \Gamma(h)=F(z^h)+\sum_jH_j^h ,
 \label{eq:facility-upper-bound}
\end{equation}
an upper bound on the new value of the potential function.  The
algorithm opens a facility minimizing $\Gamma(h)$.

\begin{lemma}[The final step]
\label{lem:final-solution}
When $Y<1$, a minimum-cost solution among $O$ and
$O\cup\{i\}$ for $i\in\mathcal U$ has cost at most
\[
                 F(z)+\sum_j\{V_j+(1-y_j)b_j\}.
\]
\end{lemma}
\begin{proof}
Choose a facility $i\in\mathcal U$ with probability $z_i$, or choose
none with probability $1-Y$.  The expected opening cost is $F(z)$.
By \cref{lem:nearest-open-facility}, each $b_j$ is the distance to an open facility.
If $i\in S_j$ is
chosen, client $j$ can connect for at most
\[
 \min\{d(j,i),b_j\}
       \le\theta_{ij}d(j,i)+(1-\theta_{ij})b_j.
\]
Otherwise it connects to a facility in $O$ at distance $b_j$.
Averaging gives connection cost at most
$V_j+(1-y_j)b_j$.  There are at most $N+1$ choices, and their
minimum cost is no larger than their average.
\end{proof}

\begin{pseudocode}{Rounding a fractional solution}
\label{alg:rounding}
\textbf{Input:} Opening variables $0<p_i\le1$, assignment variables $0\le p_{ij}\le p_i$
with $\sum_i p_{ij}=1$ for a nonempty client set, supplied opening costs $f_i$,
and distances.\\
\textbf{Output:} A set of open facilities of cost at most $F+(\beta-\eta)C$,
where $F$ is computed with the supplied opening costs.
\begin{enumerate}
\item If $C=0$, use the zero-distance construction and return.
Otherwise compute $\theta_{ij}$, $R_j$, $\xi$, and
$\eta=c\xi^2/(2N)$, and choose $\tau$ as in
Section~\ref{sec:rational-precision}.
Set $z_i=p_i$, $O=\varnothing$, and $b_j=5R_j$.
Initially $\mathcal U=\{1,\ldots,N\}$ and $Y=\sum_i p_i$.
Set $S_j=S_j^0$ for every client, ordered by connection cost.
\item While $Y\ge1$:
\begin{enumerate}[label=(\alph*),leftmargin=1.6em]
\item Compute $y_j,V_j$ and the weighted mean $\ybar$.
Evaluate $\widehat P$ satisfying \eqref{eq:scalar-precision} and
form $a^+$ by \eqref{eq:rational-coefficients}.
Compute the coefficients $q_{hi}$ by \eqref{eq:rounding-coefficients} for $h,i\in\mathcal U$,
using the rule for $r_i=0$ when necessary.
\item For each facility $h\in\mathcal U$, start from a separate copy of
the current variables and sets, and apply \eqref{eq:facility-update}.
Round the variables of unopened facilities down to multiples of $\tau$ and set
$b_j^h=\min\{b_j,d(j,h)\}$.
\item For each client $j$ and choice of $h$, remove open facilities and
facilities with $z_i^h=0$ from $S_j$.  For each $k$ in increasing order,
consider the first $k$ remaining facilities in the distance order, including
the empty set.  Retain the first set minimizing \eqref{eq:client-upper-bound}.
After all sets have been chosen, compute $\Gamma(h)$ by \eqref{eq:facility-upper-bound}.
\item Choose $h$ minimizing $\Gamma(h)$, using the fixed facility order
to break ties.  Adopt the corresponding variables $z_i,b_j$, the set $O$,
and the sets $S_j$.  Update $\mathcal U$ and $Y$.
\end{enumerate}
\item Among $O$ and the sets $O\cup\{i\}$ for $i\in\mathcal U$,
return one of minimum total cost.
Use the fixed facility order to break ties, with no additional opening first.
\end{enumerate}
\end{pseudocode}

\subsection{Analysis}

\begin{lemma}[Averaging over the choice of facility]
\label{lem:expected-potential-bound}
At the beginning of every iteration with $Y\ge1$, the quantities $H_j^h$ satisfy
\begin{equation}
 \frac1Y\sum_{h\in\mathcal U}z_h\sum_jH_j^h
           \le\Phi+3\tau C+\frac{5N\tau C}{\xi}.
 \label{eq:average-upper-bound}
\end{equation}
At the initial state, the right side decreases by an additional
$2\eta C$.
\end{lemma}

\begin{proof}
Fix the state at the beginning of an iteration and the weighted mean
$\ybar$.  Write $\phi_j=\phi_{j,\ybar}$, so $\sum_j\phi_j=\Phi$.  For
$h\in\mathcal U$, let $\phi_j^h$ be the quantity of
\cref{lem:one-client-comparison} for client $j$: the minimum over
prefixes of the exact potential, with the exact tangent $a(\ybar,\cdot)$,
after the update \eqref{eq:facility-update} without rounding down.

\paragraph{The numerical error.}
We claim that
\begin{equation}
     H_j^h\le\phi_j^h+e_j,\qquad e_j=2\tau V_j+5N\tau R_j .
 \label{eq:numerical-error}
\end{equation}
Consider a permitted set after rounding down, and let $v,D$ be its sums
of assignment variables and fractional connection costs.
The algorithm bounds its contribution by
$a^+(v)D+(1-v)b_j^h$.
By \eqref{eq:tangent-error}, this estimate is an upper bound
with error at most $2\tau D$.  Thus the minimum $H_j^h$ is at most
the exact potential of any permitted set after rounding down,
plus $2\tau V_j$.
Now compare with a permitted set before rounding down.
After facilities whose variables become zero are omitted, this set still
consists of the first $k$ facilities in the remaining order, for some $k$.
Each assignment variable is a fixed ratio at most one times its opening
variable, so the sum of the assignment variables decreases by at most $N\tau$.
The fractional connection cost also decreases.
The coefficient $a(\ybar,v)$ decreases when $v$ decreases and remains at least one,
so only the term $(1-v)b_j^h$ can increase.
This increase is at most $N\tau b_j^h\le5N\tau R_j$.
This proves \eqref{eq:numerical-error}.

\paragraph{One client.}
Apply \cref{lem:one-client-comparison} with $\delta=\tau/2$ and the
identity \eqref{eq:client-change-bound-A}.  For a client with $V_j>0$,
\begin{equation}
 \begin{split}
 \sum_{h\in\mathcal U}z_h(H_j^h-\phi_j)
  \le{}&K_0(\ybar)(y_j-\ybar)V_j-2P(\ybar)(y_j-\ybar)^2V_j\\
       &+y_jV_j\,\mathcal E^{(j)}+\tau y_jV_j+Ye_j,
 \end{split}
 \label{eq:rounding-one-client}
\end{equation}
where $\mathcal E^{(j)}$ is the comparison expression of client $j$.
For a client with $y_j=0$ or $V_j=0$ the left side is at most $Ye_j$.

\paragraph{Summing over clients.}
Since $\ybar$ is the weighted mean, $\sum_j(y_j-\ybar)V_j=0$, and the
terms with $K_0$ cancel.
\Cref{thm:core-finite-comparison} gives
$y_jV_j\,\mathcal E^{(j)}\le2P(\ybar)(y_j-\ybar)^2V_j$, which cancels
the quadratic terms.
Since $y_j\le1\le Y$, $\sum_jV_j\le C$, and
$\sum_jR_j\le C/\xi$ by \eqref{eq:core-support-radius}, the remaining
error satisfies
\[
 \frac1Y\sum_j\{\tau y_jV_j+Ye_j\}
                       \le3\tau C+5N\tau C/\xi.
\]
This proves \eqref{eq:average-upper-bound}.

\paragraph{The first iteration.}
Initially $\ybar=y_j=1$ for every client.  The quadratic terms
vanish, and for the comparison expression we use
\cref{lem:first-iteration-improvement} instead of
\cref{thm:core-finite-comparison}.  For client $j$ we have
$\pi_i=p_{ij}$ because $y_j=1$, and
$\rho_{ij}=\sum_t p_{tj}q_{ti}\ge p_{ij}\ge\xi$ because $q_{ii}=1$.
Hence the comparison terms contribute at most
\[
  -c\sum_j\sum_{i\in S_j^0}p_{ij}\,d(j,i)\,\rho_{ij}^2
                                      \le-c\,\xi^2C.
\]
Dividing by the initial
sum $Y\le N$ gives an additional decrease of at least
$c\,\xi^2C/N=2\eta C$.
\end{proof}

\begin{proof}[Proof of \cref{thm:rounding-bound}]
Consider the choices $h\in\mathcal U$ with probabilities $z_h/Y$.
By \eqref{eq:expected-opening-change}, the expected value of $F(z)$
does not change under the update \eqref{eq:facility-update}, and
rounding down only decreases it.
By \cref{lem:expected-potential-bound}, the average value of $\Gamma(h)$ is at
most
\[
          F(z)+\Phi+3\tau C+\frac{5N\tau C}{\xi}
                 \le F(z)+\Phi+\frac{8N\tau C}{\xi},
\]
where $N/\xi\ge1$.  At the first step there is an additional
decrease of $2\eta C$.  The minimum value of $\Gamma(h)$ is no larger
than this average, and it bounds the new value of the potential function.

Initially $Y\ge1$, so the first iteration is executed.
Every iteration opens a new facility, and open facilities remain open.
The loop therefore ends with $Y<1$ after at most $N$ iterations.
Starting from potential function value $F+\beta C$ and using
\eqref{eq:rounding-error-condition}, the final state satisfies
\begin{equation}
 F(z)+\Phi
       \le F+\beta C-2\eta C+\frac{8N^2\tau C}{\xi}
       \le F+(\beta-\eta)C.
 \label{eq:final-potential-bound}
\end{equation}
At this point each $b_j$ is the distance to the nearest open facility, and
\cref{lem:final-solution} gives a final solution of no larger cost.

For running time, $\tau$ is fixed throughout the at most $N$ iterations.
Every assignment variable is a fixed rational ratio $\theta_{ij}$ times
a multiple of $\tau$.  Each set $S_j$ is determined by an initial part
of its original order after omitting open facilities and facilities with
$z_i=0$.  Stored quantities have polynomial encoding length.
The coefficients $q_{hi}$ are computed by sorting rational breakpoints,
each set $S_j$ is chosen by a linear scan, and each iteration evaluates
$P$ only once.
\Cref{sec:running-time} gives the detailed running-time bound.
\end{proof}

\section{Rounding~2: randomized rounding in quadratic expected time}
\label{sec:quadratic-rounding}

Rounding~2 never computes a client potential: the sets and potentials
of \cref{sec:state} are used only in its analysis.  The implementation
uses the opening variables and the distances between facilities.  The
two ingredients are an update of the parameter $\ybar$ that works for
every client separately, and rejection sampling for the deletion
decisions.

The algorithm applies to any feasible fractional solution $(x,y)$,
after zero-distance classes of facilities have been merged as described
below.  Write
\[
 T_j=\sum_i x_{ij}^2d(i,j),
\]
recall $L,K,\gamma$ from \eqref{eq:final-parameters}, and put $M=Km$.

\begin{theorem}[Rounding~2]
\label{thm:quadratic-rounding}
Given the preprocessed feasible fractional solution, a finite randomized
algorithm returns a nonempty set of open facilities such that
\[
 \mathbb E[F_0]\le F,
 \qquad
 \mathbb E[D_j]\le\beta C_j+\frac{T_j}{K}+\gamma C_j
                         \quad\text{for every client }j,
 \qquad \beta=1+2/e.
\]
Here $F_0$ is the opening cost and $D_j$ is the distance from client $j$
to the returned set.  The algorithm uses $O(m^2+mn)$ expected arithmetic
operations and fair random bits.
\end{theorem}

\paragraph{Preprocessing.}
To merge a zero-distance class, use its cheapest opening cost, sum its
client assignments, and replace its total opening by the smaller of
that total and one.  Each summed client assignment is at most one, so
the merged solution remains feasible.  Neither cost increases.
All classes are found by scanning the facility-distance table in
$O(m^2)$ operations, and merging their assignments takes $O(mn)$
operations.  The assignment values are otherwise left unchanged.
The proof uses only $x_{ij}\le y_i$; it imposes no restriction on
the number of partial assignments of a client.  The distance order
used for prefixes is needed only in the analysis, so the
randomized implementation does not sort client distances.

\subsection{The rounding process}

Replace each facility $i$ with positive opening by $K$ co-located
copies, each having opening mass $y_i/K$ and assignment $x_{ij}/K$.
There are at most $M$ copies.  A surviving copy keeps
these values throughout the process.  At original location $i$, let
$z_i$ be the sum of the masses of the surviving copies, and put
$Y=\sum_i z_i$.  Initially $Y\ge1$ by fractional feasibility.
Maintain a set $O$ of open original facilities, initially empty.

Only a rational lower approximation to $P$ is used by the
implementation.  Set
\[
                              \delta_P=\frac{\gamma}{2}.
\]
Start with $\ybar=1$.  Before each opening iteration replace $\ybar$ by
\begin{equation}
                         \ybar'=\frac{\ybar Y}{Y+1}.
 \label{eq:quadratic-rational-schedule}
\end{equation}
Compute a rational lower estimate $\widehat P$ such that
\begin{equation}
 0\le P(\ybar')-\widehat P\le\delta_P,
 \qquad \frac14\le\widehat P\le\frac12.
 \label{eq:quadratic-P-precision}
\end{equation}
For the active locations, with their current masses $z_i$, compute the
coefficients $q_{hi}$ and radii $r_i$ of
\eqref{eq:rounding-coefficients}.  Zero-distance classes have been
merged, so $r_i=0$ occurs only if the mass at $i$ is one; then
$q_{ii}=1$ and the other coefficients $q_{hi}$ are zero.
Coefficients between copies at the same original location are always
one.

Choose original facility $h$ with probability $z_h/Y$ and add it to
$O$.  Independently for every surviving copy at original location $i$,
delete that copy with probability $q_{hi}$.  Copies at $h$ are deleted
with certainty.  Every decision in this iteration uses the old weight
vector; the new opening masses are committed only after all decisions
have been made.  Repeat while $Y\ge1$.

When $Y<1$, choose one remaining original facility $i$ with probability
$z_i$, or choose none with probability $1-Y$, and add the chosen
facility to $O$.  Connect every client to a nearest facility in $O$.
This last randomized choice is needed for the separate per-client
guarantees in the theorem.

\begin{pseudocode}{Randomized rounding from a fractional solution}
\label{alg:quadratic-rounding}
\begin{enumerate}
\item Form the $K$ copies of each positive-opening facility.  Set $O=\varnothing$
and $\ybar=1$.
\item While $Y\ge1$:
\begin{enumerate}
\item Update $\ybar$ by \eqref{eq:quadratic-rational-schedule} and compute
$\widehat P$ satisfying \eqref{eq:quadratic-P-precision}.
\item Choose $h$ with probability $z_h/Y$ and add $h$ to $O$.
\item Using the current masses, independently delete each copy at $i$
with probability $q_{hi}$ from \eqref{eq:rounding-coefficients}.
The rejection-sampling implementation below avoids computing most radii.
\item Commit all deletions, update the opening masses, and set
$Y=\sum_i z_i$.
\end{enumerate}
\item Choose one remaining facility $i$ with probability $z_i$, or none
with probability $1-Y$.  Add it to $O$ and return $O$.
\end{enumerate}
\end{pseudocode}

\subsection{Updating the common parameter}

In the analysis, each client has the state of \cref{sec:state}, with
the surviving copies in the role of the facilities: it retains a
prefix of its surviving assigned copies in order of increasing
distance, $u_i$ are their assignments, and $y$, $V$, $b$ are as
before.  Selecting an original facility is equivalently viewed as
selecting one of its surviving copies in proportion to mass.  In
addition put $W=\sum_i u_i^2d_i$.  The potential of the client is
\[
                         a(\ybar,y)V+(1-y)b+W.
\]
After each deletion outcome, the retained prefix may be shortened to
minimize this expression among prefixes of its current surviving
support.

By \cref{cor:one-client-residual}, at a fixed parameter $\ybar$ the
expected change of the base potential $a(\ybar,y)V+(1-y)b$, under the
fractional update \eqref{eq:facility-update} and before numerical
errors, is at most the residual
\begin{equation}
 \frac{K_0(\ybar)}Y(y-\ybar)V,
 \qquad
 K_0(\ybar)=3-A(\ybar)-(4+2\ybar)P(\ybar)=2\ybar P'(\ybar).
 \label{eq:quadratic-single-client-residual}
\end{equation}

\begin{lemma}[A common parameter for every client]
\label{lem:quadratic-schedule}
Replacing $\ybar$ by $v=\ybar Y/(Y+1)$ before an opening iteration makes the
combined scalar change and the residual in
\eqref{eq:quadratic-single-client-residual} nonpositive for every
client separately.
\end{lemma}

\begin{proof}
Assume first that $\ybar>0$.  Put $\lambda=1/Y=(\ybar-v)/v$ and
$a_0(t)=A(t)-2tP(t)$.  Dividing the combined bound by $V$ gives
$B_0+B_1y$, where
\[
 B_1=2(P(v)-P(\ybar))+\lambda K_0(v),
 \qquad
 B_0=a_0(v)-a_0(\ybar)-\lambda K_0(v)v.
\]
Convexity of $P$ gives $B_1\le0$.  Set
$g(t)=-tP'(t)$.  Direct differentiation yields
\[
 g'(t)=\int_0^1s^2e^{-st}(1-st)\,ds\ge0
                                      \qquad(0\le t\le1).
\]
Since $a_0'(t)=2g(t)$ and $K_0(v)=-2g(v)$,
\[
 B_0=-2\int_v^\ybar g(t)\,dt+2(\ybar-v)g(v)\le0.
\]
The case $\ybar=0$ follows directly from $K_0(0)=0$; when $V=0$, the
zero-distance case of \cref{lem:one-client-comparison} applies.
\end{proof}

The update is exact rational arithmetic.  The surviving copy masses
are fixed input rationals.  If their common denominator is $D$, every
total mass has the form $Y=a/D$, so $\ybar$ is a product of at most $m$
ratios $a/(a+D)$.  Its rational representation therefore has polynomial
size.  No rounding of the common parameter is needed.

\subsection{The approximation guarantee}

\paragraph{The variance of independent deletions.}
The additional term $W$ accounts for the variance from independent
deletions.  Only the full-prefix estimate for the base potential in
the comparison proof is quadratic in the retained assignments.
In this calculation, use $a$ for a copy and
$h$ for a current opening location, and let $U_h$ be the client's
total retained assignment at $h$.  Conditional on opening $h$,
independent deletion of copies changes the full-prefix expectation from
the value at the mean assignments by
\[
                       2P\sum_a u_a^2d_aq_{ha}(1-q_{ha}).
\]
For the full prefix, the expected change in $W$ is
\[
                         -\sum_a u_a^2d_aq_{ha}.
\]
Thus, for each fixed $h$, adding $W$ changes the bound by
\[
             \sum_a u_a^2d_aq_{ha}\{2P(1-q_{ha})-1\}\le0,
\]
because $2P\le1$.  The shorter-prefix estimate for the base potential
is linear and has no variance term, and its $W$ can only decrease.
The proof of \cref{lem:one-client-comparison} therefore applies to the
augmented potential.  Its two estimates still have respective weights
$z_h-U_h$ and $U_h$, exactly the sums of the copy weights at $h$.
The additional initial potential is
\begin{equation}
 \sum_i K\left(\frac{x_{ij}}K\right)^2d(i,j)
                              =\frac{T_j}K.
 \label{eq:quadratic-variance-charge}
\end{equation}

\paragraph{The numerical error.}
The discrepancy from using $\widehat P$ instead of $P$ in the radii is
at most $2\delta_PV/Y$ per client and per opening, by the same radius
term in the comparison calculation.  Normalization deletes every
retained copy with conditional probability $1/Y$.  Since shortening
the surviving prefix can only decrease $V$, one has
\[
 \mathbb E[V_{t+1}\mid\mathcal F_t]\le(1-1/Y_t)V_t.
\]
Summing these expected decreases gives
\[
                         \mathbb E\sum_t\frac{V_t}{Y_t}\le C_j.
\]
The total scalar-evaluation error is therefore at most
$2\delta_PC_j=\gamma C_j$.

\paragraph{The connection cost.}
Initially $y=1$ and the client potential is $\beta C_j+T_j/K$.  The parameter
update, the augmented-potential comparison, and the numerical error estimate
therefore bound its final expectation by
$\beta C_j+T_j/K+\gamma C_j$.

When $Y<1$, the auxiliary value $b$ is the distance to a facility
already in $O$.  This is \cref{lem:nearest-open-facility}, applied to
the original locations with their masses: the mass at a location
decreases only if one of its copies is deleted, which requires
$q_{hi}>0$.

In the randomized final choice, the probability of choosing original
location $i$ is $z_i$, and the client's retained assignment at $i$ is
at most $z_i$.  Thus, conditional on the final state,
\[
                   \mathbb E[D_j\mid\mathcal F]\le V+(1-y)b.
\]
The tangent coefficient $a(\ybar,y)$ is at least one, so the final
potential dominates this bound.  Also $O$ is nonempty: initial
feasibility gives $Y_0\ge1$, and at least one opening iteration occurs.
This proves the claimed connection guarantee.

\paragraph{The opening cost.}
Opening cost is controlled separately.  For an unopened original
facility $i$, the probability of being opened in the next iteration is
$z_i/Y$.  Its expected surviving unopened mass is $z_i-z_i/Y$.
Therefore its indicator of being open plus its residual unopened mass
is a martingale.  The final randomized choice preserves this quantity
in expectation.  All copies at a location are deleted when that
location opens, so it is opened at most once and has its initial
opening marginal.  The expected opening cost is $F$ after preprocessing;
merging zero-distance classes can only reduce the original value of $F$.

\subsection{Computing each radius in linear time}

For fixed rational $P\in(0,1/2]$ and target $c$, define
\[
 G_i(r)=\sum_hz_h\min\{1,(1-d(i,h)/r)_+/P\}.
\]
Assume that the mass at distance zero is less than $c\le Y$.
The smallest radius satisfying $G_i(r)=c$ is found in
$O(m)$ arithmetic operations without sorting.  To see this, multiply
by $Pr$ and write
\[
 H_i(r)=\sum_hz_h(r-d(i,h))_+
 -(1-P)\sum_hz_h\left(r-\frac{d(i,h)}{1-P}\right)_+-cPr.
\]
For positive $r$, its sign is the sign of $G_i(r)-c$.
First absorb all zero-distance positive-part terms into the affine term
$P(z_0-c)r$, where $z_0$ is the mass at distance zero.  The remaining
breakpoints are strictly positive, and the affine slope at zero is
negative because $z_0<c$.  This convention excludes the spurious
zero of $H_i$ at $r=0$.
Maintain an interval containing
the least root, the affine contribution of breakpoints already known
to lie below that interval, and the unresolved breakpoint list.  Select
the median unresolved breakpoint and evaluate $H_i$ there.  If the
value is negative, discard lower breakpoints and add their now-active
positive-part terms to the affine contribution.  Otherwise discard upper breakpoints,
whose positive-part terms are zero on the remaining interval.  Process equal
breakpoints together.  A zero value retains the tested breakpoint as
the upper endpoint; this also handles a flat interval on which $G_i=c$
and the case $c=Y$.  After the list is exhausted, solve the remaining
affine equation in the maintained interval, including a root at the
upper endpoint.  Its lower endpoint has $G_i<c$ (interpreted as the
right limit if it is zero), so the spurious root zero is never returned.
Each pass
discards at least half the unresolved list, so the sum of the work over
all passes is $O(m)$.  Signed coefficients of positive-part terms cause no difficulty:
the direction of each search step uses the monotonicity of $G_i$.

\subsection{Rejection sampling and the number of radius updates}

Suppose $Y\ge2$.  At every active original facility $i$, choose a
stored radius $R_i$ with
\begin{equation}
 b_{hi}=\min\{1,2(1-d(i,h)/R_i)_+\},\qquad
                         \sum_hz_hb_{hi}=\frac32.
 \label{eq:quadratic-reference-radius}
\end{equation}
The target is strictly larger than the mass at distance zero, which is
at most one, and smaller than $Y$.  The preceding linear-time procedure
therefore computes $R_i$ and all its coefficients in $O(m)$ operations.
Keep the radius fixed and maintain $v_i=\sum_hz_hb_{hi}$.  A decrease
$\Delta_h$ in opening mass $z_h$ decreases every active $v_i$ by
$\Delta_hb_{hi}$.  A single mass change thus takes $O(m)$ operations.
After all deletions in an iteration have been committed, recompute a
stored radius only when the new total mass satisfies $Y\ge2$,
$v_i<1$, and $i$ remains active.  If the new total mass is below two,
discard all stored radii.  After recomputation its value is again $3/2$.

For a valid stored radius, put
\[
                         u_{hi}=\min\{1,4(1-d(i,h)/R_i)_+\}.
\]
Because $\widehat P\le1/2$, the true coefficient function at $R_i$ dominates
$b_{hi}$; its normalization sum is at least $v_i\ge1$.  Thus
$r_i\le R_i$.  Since $\widehat P\ge1/4$, it follows that
\begin{equation}
 q_{hi}\le u_{hi}\le2b_{hi},
 \qquad \sum_hz_hu_{hi}\le2v_i\le3.
 \label{eq:quadratic-upper-probabilities}
\end{equation}

\begin{figure}[t]
\centering
\begin{tikzpicture}[x=5.2cm,y=2.6cm,font=\small,>=Latex]
  \draw[->] (0,0) -- (1.15,0) node[below right] {$d(i,h)/R_i$};
  \draw[->] (0,0) -- (0,1.2) node[above] {probability};
  \draw (-.015,1) -- (.015,1) node[left=5pt] {$1$};
  \draw (1,-.025) -- (1,.025) node[below=5pt] {$1$};
  \node[below left=3pt] at (0,0) {$0$};
  \draw[dashed,gray!75!black,line width=.9pt]
     (0,1) -- (.5,1) -- (1,0);
  \draw[inkblue,line width=1.3pt]
     (0,1) -- (.6,1) -- (.8,0);
  \draw[black,line width=.9pt]
     (0,1) -- (.75,1) -- (1,0);
  \draw[black,line width=.9pt] (1.28,1.04) -- (1.43,1.04);
  \node[anchor=west] at (1.47,1.04) {$u_{hi}$: proposal probability};
  \draw[inkblue,line width=1.3pt] (1.28,.77) -- (1.43,.77);
  \node[anchor=west] at (1.47,.77) {$q_{hi}$: deletion probability};
  \draw[dashed,gray!75!black,line width=.9pt] (1.28,.50) -- (1.43,.50);
  \node[anchor=west] at (1.47,.50) {$b_{hi}$: coefficient defined using $R_i$};
  \node[anchor=west] at (1.28,.12) {$q_{hi}\le u_{hi}\le2b_{hi}$};
\end{tikzpicture}
\caption{The stored radius gives a proposal probability that
dominates the true deletion probability.  The displayed curves
illustrate $r_i=4R_i/5$ and $\widehat P=1/4$; the proof applies to
every $r_i\le R_i$ and $1/4\le\widehat P\le1/2$.
Only a successful proposal requires the radius $r_i$.
The maintained bound $\sum_hz_hb_{hi}\le3/2$ implies
$\sum_hz_hu_{hi}\le3$.}
\label{fig:proposal-probabilities}
\end{figure}
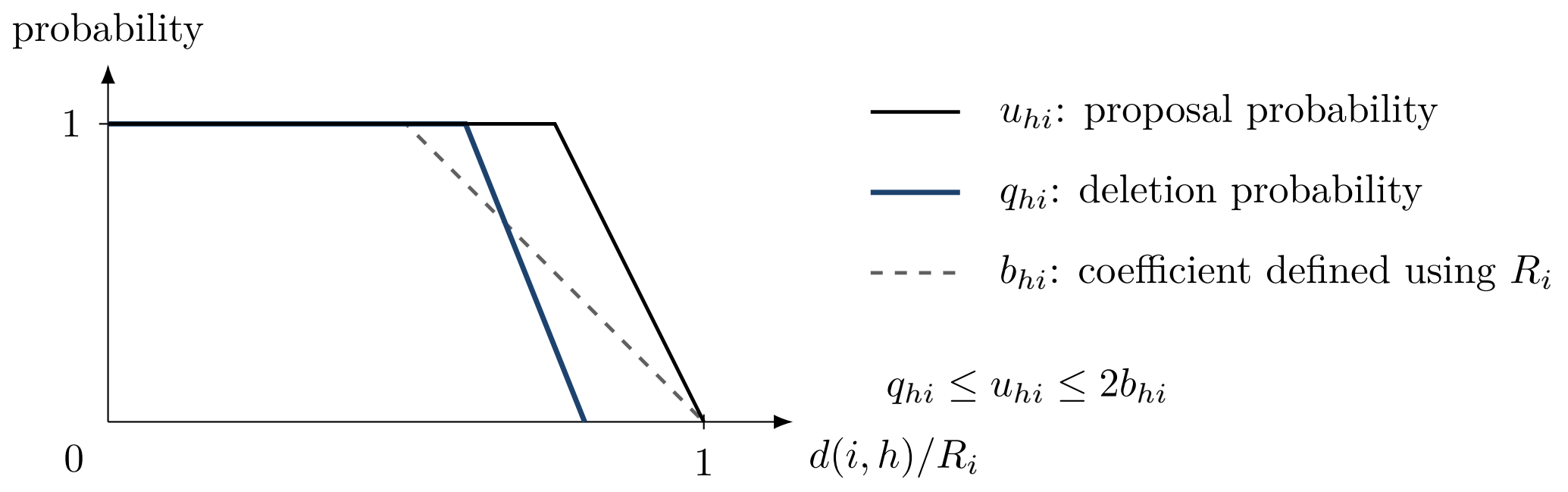

For a copy at $i$, first propose deletion with probability $u_{hi}$.
Only when a proposal occurs, compute the true radius $r_i$ in $O(m)$
time and accept the deletion with conditional probability $q_{hi}/u_{hi}$.
The resulting deletion probability is exactly $q_{hi}$.  All proposals
and acceptance coins use independent fresh bits.  Once a true radius
has been computed, it can be reused for all copies at the same original
location in that opening iteration.  The sums $v_i$ and opening masses
are updated after all deletion decisions, so every probability uses the
same old state.

\begin{lemma}
\label{lem:quadratic-proposals}
The expected total number of deletion proposals while $Y\ge2$ is at
most $3M$.  The expected number of recomputations of stored radii is at
most $3M$, in addition to the initial construction at each original
location.
\end{lemma}

\begin{proof}
At a state with $M_t$ surviving copies, a fixed copy at $i$ receives a
proposal with conditional probability at most
$\sum_hz_hu_{hi}/Y\le3/Y$.  Thus the expected number of proposals in
the iteration is at most $3M_t/Y$, three times the expected number of
copy deletions.  Summing over all iterations proves the first bound.

Fix the coefficients $b_{hi}$ at the beginning of an iteration.
Every opening mass has expected decrease $z_h/Y$.  Hence the expected
decrease in $v_i$ is $v_i/Y\le3/(2Y)$.  There are at most
$M_t$ active original locations, so the total expected loss of sums $v_i$ is at most $3M_t/(2Y)$.  Summed over the execution, its expectation
is at most $3M/2$.  Every recomputation requires a loss greater than $1/2$
since the last initialization of that stored radius.  These losses
are disjoint for each location.  Removing a location after its last
copy dies, or discarding stored radii when $Y<2$, only leaves some
losses uncharged.  The expected number of recomputations is consequently at
most $3M$.
\end{proof}

If $1\le Y<2$, compute true radii directly for all active locations.
The expected number of copy deletions in such an iteration is at least
$M_t/2$.  Therefore the expected total number of visits to surviving copies
over these iterations is at most $2M$, and the direct normalization
computations cost $O(mM)$ in expectation.  This also covers $Y_0<2$.

\subsection{Running time}

Exact sampling from fair random bits and the evaluation of
$\widehat P$ are described in \cref{sec:sampling-details}: a Bernoulli
variable with a rational parameter uses $O(1)$ expected operations and
bits, sampling an opening location uses $O(m)$, and all evaluations of
$\widehat P$ together use $O(m^2)$ operations.

Finally, there are at most $m$ opening iterations and at most $M=Km$
copy deletions.  Scanning copies and generating preliminary trials costs
$O(mM)$ expected operations and bits.  Choosing opening locations costs
$O(m^2)$.  Initial stored radii cost $O(m^2)$.  By
\cref{lem:quadratic-proposals}, the expected total cost of successful
proposals and recomputations of stored radii is $O(mM)$.  Updating every reference
sum after each copy deletion also costs $O(mM)$.  The last iterations
have the same expected bound.  Connecting clients to the final open
set costs $O(mn)$.  Since $K$ is an absolute constant,
the total is $O(m^2+mn)$, as claimed.

\section{Rounding~3: deterministic rounding for weighted connection cost}
\label{sec:weighted-deterministic-rounding}

The potential of Rounding~2, with the variance term $W$, also gives a
deterministic rounding algorithm.  This version controls any prescribed nonnegative
linear combination of client connection costs.  Its use is to
derandomize an objective obtained after linearizing the augmentation
bound.  Rounding~2 remains the faster algorithm.

\begin{lemma}[Rounding~3]
\label{lem:weighted-deterministic-rounding}
Let $w_j\ge0$ be rational client weights, and let $\varepsilon>0$ be
rational.  From any feasible fractional solution, after the
zero-distance merging used in
\cref{thm:quadratic-rounding}, a deterministic polynomial-time algorithm
returns a nonempty open set with opening cost $F_0$ and client distances
$D_j$ such that
\[
 F_0+\sum_jw_jD_j
 \le F+\beta\sum_jw_jC_j+
                  \frac1K\sum_jw_jT_j+\varepsilon\sum_jw_jC_j.
\]
The algorithm uses only rational arithmetic.  It does not require a
lower bound on positive opening or assignment variables, or a
restriction on partial assignments.
\end{lemma}

\paragraph{The potential.}
Use the same $K$ copies, current opening variables, normalization
coefficients, and sets of retained client assignments as in the randomized
algorithm.  In this section the client supports are maintained explicitly.
Sort each client's fractional support once by distance, breaking ties
by facility index and copy index; subsequent prefixes use the inherited
order of the surviving copies.
For client $j$, keep
\[
 y_j=\sum_a u_{aj},\qquad V_j=\sum_a u_{aj}d(a,j),\qquad
 W_j=\sum_a u_{aj}^2d(a,j),
\]
where the sum is over its currently retained copies.  Let $b_j$ be its
auxiliary distance.  The true potential is
\[
 \Phi=F(O)+\sum_{i\notin O}f_i z_i+
       \sum_jw_j\{a(\ybar,y_j)V_j+(1-y_j)b_j+W_j\}.
\]
The analysis uses this potential; the algorithm compares rational
upper bounds for conditional expectations of its value.  These bounds
serve as pessimistic estimators, as in \cite[Section~4]{CS03}.

Let $M=Km$ and choose
\[
\delta=\min\left\{\frac{\varepsilon}{32M},\frac1{1000}\right\}.
\]
Before each iteration use the same parameter update
$\ybar\leftarrow \ybar Y/(Y+1)$, evaluated exactly.
Compute rational bounds
\[
 A(\ybar)\le A^+\le A(\ybar)+\delta,
 \qquad P(\ybar)-\delta\le\widehat P\le P(\ybar),
 \qquad \frac14\le\widehat P\le\frac12.
\]
Odd partial sums of the alternating Taylor series give the upper bound
for $A$ and the lower bound for $P$, with degree $O(\log(1/\delta))$;
take the maximum of the latter with $1/4$.  All these quantities are
rational and have polynomial-size representations.  Explicitly,
\[
 A(\ybar)=1+2\sum_{k\ge1}\frac{(-1)^{k+1}\ybar^k}{k!(k+1)}.
\]
An odd degree $d$ with $2/((d+1)!(d+2))\le\delta$ suffices for both
Taylor bounds; it is found by increasing the odd degree and testing
this rational remainder bound.  Set
\[
 a^+(y)=A^++2\widehat P(y-\ybar)+2\delta=a_0^++a_1^+y.
\]
Then
\begin{equation}
 a(\ybar,y)\le a^+(y)\le a(\ybar,y)+5\delta,
 \qquad 0<a_1^+=2\widehat P\le1
                                    \quad(0\le y\le1).
 \label{eq:weighted-upper-tangent}
\end{equation}
No scalar identity is asserted for these approximations.  All
comparison inequalities are applied to the true potential, and the
excess in \eqref{eq:weighted-upper-tangent} is charged separately.

\paragraph{Expected value for a fixed prefix.}
Fix an opening location $h$ and a prefix of client $j$'s current support.
For a copy $a$ in this prefix, let $s_a=1-q_{ha}$ be its survival
probability.  Put $b_j^h=\min\{b_j,d(h,j)\}$ and define
\[
 \bar y=\sum_a u_{aj}s_a,\quad
 \bar V=\sum_a u_{aj}d(a,j)s_a,\quad
 R=\sum_a u_{aj}^2d(a,j)s_a(1-s_a),\quad
 \bar W=\sum_a u_{aj}^2d(a,j)s_a.
\]
Independence gives $\mathbb E[y'V']=\bar y\bar V+R$.  Thus the
expected rational upper potential of this prefix is exactly
\begin{equation}
        a_0^+\bar V+a_1^+(\bar y\bar V+R)
                   +(1-\bar y)b_j^h+\bar W.
 \label{eq:weighted-prefix-score}
\end{equation}
All four sums are computed for every prefix by one scan in distance
order.  For each client choose the prefix minimizing
\eqref{eq:weighted-prefix-score}.  It is a prefix of the current
retained support; no previously discarded assignment is restored.

Add the weighted client estimates to the expected opening-cost part,
\[
             F(O)+f_h+\sum_a f_a z_a(1-q_{ha}),
\]
where $z_a$ is the fixed mass of a surviving copy, $f_a$ is its
facility's opening cost, and $q_{ha}=1$ for all copies at $h$.
This gives a rational estimate $S_h$ for opening $h$.
Choose a location minimizing $S_h$ and fix its chosen client prefixes.

\paragraph{Fixing the deletion outcomes.}
Keep all $q_{ha}$ fixed for this iteration.  Reveal the deletion
decisions one copy at a time.  For a copy whose probability is strictly
between zero and one, evaluate the conditional rational estimate for
deleting it and for retaining it, and choose the smaller estimate.  Forced
decisions are made as prescribed.  In the formula above, a decided
copy simply has $s_a=0$ or $s_a=1$; the undecided copies keep their
original survival probabilities.  Hence the same four sums maintain
the conditional estimate.  Each chosen outcome has positive probability
under the product distribution and does not increase the conditional
estimate.  Only after all decisions are fixed are the original
opening masses and retained client supports updated.

\begin{pseudocode}{Deterministic rounding for weighted connection cost}
\label{alg:weighted-deterministic-rounding}
\begin{enumerate}
\item Form the $K$ copies.  For each client retain its entire assigned
support in distance order, and set $b_j$ to five times its largest
initial support distance.  Initialize $O=\varnothing$ and $\ybar=1$.
\item While $Y\ge1$:
\begin{enumerate}
\item Update the scalar and its rational approximations.  Compute the
normalization coefficients from the current opening masses.
\item For every candidate opening $h$, compute the minimizing expected
prefix estimate for each client using \eqref{eq:weighted-prefix-score},
and add the expected opening cost.
\item Choose the candidate with minimum total estimate and fix its
minimizing client prefixes.
\item Fix each deletion decision by choosing the smaller conditional
estimate.  Apply all decisions and update the retained supports.
\end{enumerate}
\item When $Y<1$, return the minimum-weighted-cost solution among $O$
and $O\cup\{i\}$ over remaining original facilities $i$.
\end{enumerate}
\end{pseudocode}

\begin{proof}[Proof of \cref{lem:weighted-deterministic-rounding}]
The augmented-potential argument in the randomized rounding theorem
applies to each client separately.  For a fixed opening location, the
expected full-prefix variance is paid by the decrease of $W_j$, and
the shorter-prefix estimate is linear while $W_j$ only decreases.
Both prefixes are among the ones considered by the algorithm.  Thus,
after multiplication by the nonnegative weights $w_j$, their comparison
bounds may be summed.

Consider one iteration and write $C_w=\sum_jw_jC_j$.  The exact rational
parameter update makes the combined scalar and comparison change nonpositive.
Using $\widehat P$ in the normalization adds at most
$2\delta C_w/Y\le2\delta C_w$.  For every deletion outcome, replacing
the true tangent by $a^+$ adds at most $5\delta C_w$, because the
retained connection cost does not exceed its initial value.  Consequently
\[
                    \sum_h\frac{z_h}{Y}S_h\le\Phi+7\delta C_w.
\]
Choosing the minimum $S_h$ cannot exceed this weighted average.  Fixing
the deletion decisions cannot increase that chosen conditional estimate.
The true potential of the resulting state is at most its rational upper
bound.  Therefore each actual iteration satisfies
\[
                            \Phi'\le\Phi+7\delta C_w.
\]
There are at most $m\le M$ iterations.  Initially
\[
                  \Phi=F+\beta C_w+\frac1K\sum_jw_jT_j.
\]
The accumulated numerical error is at most
$7M\delta C_w\le\varepsilon C_w$.

At termination, the distance argument from the randomized proof applies
pathwise.  The randomized choice of one remaining original facility
with probabilities $z_i$, or none with probability $1-Y$, has expected
weighted total cost at most
\[
                 F(O)+\sum_i f_i z_i+
                    \sum_jw_j\{V_j+(1-y_j)b_j\}\le\Phi.
\]
The minimum candidate examined in the final step is no more expensive
than this average.  This proves the asserted bound.
\end{proof}

\paragraph{Running time.}
The algorithm is polynomial and requires no exponential prefix
enumeration.  A direct implementation computes all radii in
$O(m^2)$ operations per iteration.  It computes each candidate's
bounds for client prefixes in $O(nM)$ operations.  For a chosen candidate,
maintaining the four sums while fixing a copy's outcome takes $O(n)$
operations, so all conditional decisions cost $O(nM)$ per iteration.
There are at most $m$ iterations.  Including the initial distance
orders and scalar evaluations, one obtains
\[
 O\bigl(m^3+m^3n+mn\log(m+1)+m\log(m/\varepsilon+2)\bigr)
\]
arithmetic operations.  For constant $K$ and inverse-polynomial
$\varepsilon$, this is $O((m+n)^4)$.

\section{A combinatorial algorithm for the fractional solution}
\label{sec:combinatorial-fractional}

We compute the fractional solution by increasing one facility opening
and a set of its assignments at a time.  The algorithm uses neither a
linear-programming oracle nor a flow oracle.  It adapts Young's
algorithm for fractional facility location~\cite{Young15Nearly}.
The difference is in the client weights and the stopping rule.  Explicit
client penalties let us stop before every client receives a full unit
of assignment.  We then complete the remaining assignments without
increasing the penalized objective.  This gives a running time without
a factor of $\log n$.

There are $m\ge1$ facilities and $n\ge1$ clients.  We use $\mathrm{LP}$
for the optimum of the standard fractional relaxation.  An equivalent
dual formulation maximizes $\sum_j\alpha_j$ over nonnegative dual
variables satisfying
\[
                \sum_j(\alpha_j-d(i,j))_+\le f_i
                  \qquad\text{for every facility }i.
\]

\subsection{Client penalties and feasibility repair}

We use the facility radii of Mettu and Plaxton~\cite{MettuPlaxton03}.
The notation of this section is local.  For $f_i>0$, let $\hat r_i$ be the unique radius satisfying
\[
                         \sum_j(\hat r_i-d(i,j))_+=f_i.
\]
For $f_i=0$, put $\hat r_i=0$.  Define
\[
                  \ell_j=\min_i\{d(i,j)+2\hat r_i\},
                  \qquad B=\sum_j\ell_j,
\]
and, for every client $j$, store a facility $i(j)$ attaining the minimum.
A client may temporarily
assign unserved demand to a private dummy facility at distance $\ell_j$.
The dummy facility has no opening cost.  For fixed opening variables
$0\le y_i\le1$, write $C_\ell(y)$ for the cheapest total connection cost
when these dummy assignments are allowed.

\begin{lemma}[Client penalties]
\label{lem:initializer-penalties}
The values $\ell_j$ and facilities $i(j)$ can be computed in $O(mn)$ time.
They satisfy
\[
               \mathrm{LP}\le B\le3\,\mathrm{LP},
\]
and $\alpha_j=\ell_j/3$ is dual feasible.  Moreover, from any
$y\in[0,1]^m$ one can construct an ordinary feasible fractional
solution of cost at most $\sum_i f_i y_i+C_\ell(y)$ in $O(mn)$
time.
\end{lemma}

\begin{proof}
Each radius is obtained by linear selection on one row of client
distances.  Maintain the count and the sum of distances already
known to lie below the radius.  At a median unresolved distance
$v$, compare $\sum_j(v-d(i,j))_+$ with $f_i$, then discard the
half of the unresolved distances on the other side of the root.
When discarding distances below the root, add them to the maintained
count and sum.  Process ties together.  Once all distances are
resolved, the radius is $(f_i+\text{sum})/\text{count}$; the count
is positive because $f_i>0$.  The total work is $O(n)$.
Computing all the minima defining $\ell_j$ then takes one scan of the
facility--client pairs.  The function $\ell$ is
$1$-Lipschitz on clients, since it is the minimum of the
$1$-Lipschitz functions $d(i,j)+2\hat r_i$.

We first prove the repair claim.  Open all zero-cost facilities
fully.  Group clients with the same stored facility $i(j)$.
For each client maintain
\[
                         c_j=\sum_{h:d(j,h)\le \ell_j}y_h.
\]
Process each nonempty group once.  In the group associated with
facility $i$, choose a client $j$ minimizing the current value $c_j$.
If $c_j\ge1$, proceed to the next group.  Otherwise put
$\eta=1-c_j$.  Facility $i$ has positive opening cost;
otherwise it was already opened fully.  Thus $\hat r_i>0$ and
$d(i,j)<\ell_j$.  In particular, $\eta\le1-y_i$.  Increase $y_i$
by $\eta$.

For any client $k$ with $d(i,k)<\hat r_i$, put $a=\ell_j-d(j,k)$.
The triangle inequality and the Lipschitz property give
\[
                         \ell_k\ge a\ge2\hat r_i-d(i,k)>d(i,k).
\]
Every original facility at distance less than $a$ from $k$ lies in
the closed ball counted by $c_j$.  Therefore at least $\eta$
units of a cheapest assignment with penalties for $k$ have distance at
least $a$.  The added mass at $i$ can replace these units, saving
at least $2\eta(\hat r_i-d(i,k))$.  These reassignments can be made
for every such client simultaneously because each assignment
has its own bound $x_{ik}\le y_i$.  The total saving is at least
$2\eta f_i$, while the opening cost increases by only $\eta f_i$.
Thus the objective including penalties does not increase.

Every client $k$ with $i(k)=i$ has $d(i,k)\le \ell_k$,
so its coverage also increases by $\eta$.  Because $c_k\ge c_j$
before the increase, the whole group is now covered.  No coverage
can decrease later.  There are at most $\min\{m,n\}$ nonempty
groups.  Afterwards every
client has a cheapest assignment with penalties using only original facilities;
at equal distance we prefer an original facility to a dummy.  Initializing
the $c_j$ takes $O(mn)$ time, all group minima together take
$O(n)$ time, and each increase updates the coverages in $O(n)$
time.  Weighted selection computes final cheapest unit
assignments in $O(m)$ time per client.  This proves the claimed
running time and cost bound, including the case of zero penalties.
Applying the repair to $y=0$ gives $\mathrm{LP}\le B$.

For any feasible solution of the standard relaxation with costs $F$ and
$C$,
\[
 f_i y_i
 =y_i\sum_j(\hat r_i-d(i,j))_+
 \ge \hat r_i\sum_jx_{ij}-\sum_jd(i,j)x_{ij}.
\]
Consequently
\[
 B\le\sum_{ij}x_{ij}\{d(i,j)+2\hat r_i\}\le2F+3C.
\]
Using an optimal fractional solution proves $B\le3\,\mathrm{LP}$.
Finally $\ell_j\le d(i,j)+2\hat r_i$ implies
\[
 (\ell_j/3-d(i,j))_+\le\tfrac23(\hat r_i-d(i,j))_+.
\]
Summing over $j$ proves dual feasibility of $\ell/3$.
\end{proof}

\subsection{The weighted algorithm}

Fix rational $0<\delta\le1/10$ and put
\[
              U=\lceil\delta^{-2}\rceil,
              \qquad q=1-\delta,
              \qquad {W_{\min}}=Bq^U.
\]
If $B=0$, the repair in
Lemma~\ref{lem:initializer-penalties} already gives a zero-cost
solution.  Otherwise initialize
\[
 X_{ij}=Y_i=\kappa_j=0,\qquad w_j=\ell_j,\qquad\lambda=1/3.
\]
The count $\kappa_j$ always equals $\sum_iX_{ij}$.  Its weight is
$w_j=\ell_jq^{\kappa_j}$ until $\kappa_j=U$, and is zero thereafter.
Initially retain the feasible dual vector $\widehat\alpha=\ell/3$
and its value $D=B/3$.

\begin{pseudocode}{A combinatorial fractional solution}
\label{alg:capped-combinatorial-initializer}
\begin{enumerate}
\item While $W=\sum_jw_j>{W_{\min}}$, perform the following pass.
\begin{enumerate}
\item For each facility $i$, form
\[
 S_i=\{j:\kappa_j<U\text{ and }d(i,j)<(1+\delta)\lambda w_j\}.
\]
While $S_i\ne\varnothing$ and
\begin{equation}
 f_i+\sum_{j\in S_i}d(i,j)
       \le(1+\delta)\lambda\sum_{j\in S_i}w_j,
 \label{eq:capped-star-test}
\end{equation}
increase $Y_i$ by one.  For every $j\in S_i$, increase $X_{ij}$
and $\kappa_j$ by one; multiply $w_j$ by $q$ if $\kappa_j<U$, and set
$w_j=0$ otherwise.  Update $S_i$.  Stop immediately if
$\sum_jw_j\le{W_{\min}}$.
\item After a completed pass that has not stopped the algorithm,
replace $\lambda$ by $(1+\delta)\lambda$.  The vector
$\alpha_j=\lambda w_j$ is dual feasible.  If its value exceeds
$D$, retain it as $\widehat\alpha$ and update $D$.
\end{enumerate}
\item Set $x_{ij}=X_{ij}/U$ and $y_i=\min\{Y_i/U,1\}$.
Assign the remaining $1-\kappa_j/U$ units of client $j$ to its dummy
facility.  Apply Lemma~\ref{lem:initializer-penalties} to return a
feasible solution of the standard relaxation, together with the retained
dual vector $\widehat\alpha$.
\end{enumerate}
\end{pseudocode}

Zero-penalty clients have weight zero throughout.  Their remaining
dummy demand has zero cost and is served by the repair step.
The original and dummy assignments in step~2 are feasible: $\kappa_j\le U$,
$X_{ij}\le Y_i$, and $X_{ij}\le U$.

The array $X$ is needed only to describe the assignment history in the
proof.  The implementation stores $Y_i$, $\kappa_j$, and $w_j$, and updates
the latter two quantities for the clients in the current set $S_i$.
The repair step uses only the final opening variables, the penalties,
and the stored facilities $i(j)$.  If explicit fractional assignments are requested,
compute a cheapest unit of assignment for one client at a time by
weighted selection and write that client's assignments to the output.
Thus the algorithm need not store the history array $X$.

\begin{theorem}[Combinatorial fractional approximation]
\label{thm:capped-combinatorial-initializer}
Algorithm~\ref{alg:capped-combinatorial-initializer} returns an
exactly feasible fractional solution of cost $Q$ and a feasible
dual solution of value $D$ such that
\[
                    Q\le(1+6\delta)D
                         \le(1+6\delta)\,\mathrm{LP}.
\]
Its running time is
\[
                  O(mn/\delta^2).
\]
It uses $O(m+n)$ working registers in addition to the metric input
and any explicit fractional-assignment output.
\end{theorem}

\begin{proof}
Throughout the algorithm,
\begin{equation}
                  \lambda W\le D\le\mathrm{LP},
                  \qquad B\le3D.
 \label{eq:capped-initializer-invariant}
\end{equation}
These statements hold initially.  During a pass the weights decrease,
preserving the first inequality.  At the end of a pass every facility
either has an empty set $S_i$ or has failed~\eqref{eq:capped-star-test}.
A facility exhausted earlier
in the pass remains exhausted because weights only decrease.  Hence
the vector $\alpha=\lambda w$ after the scaling at the end of a pass satisfies
all dual constraints.  Retaining the better dual vector proves the
invariant.  The retained vector need not satisfy $\alpha_j\le \ell_j$;
only dual feasibility for the standard relaxation is used here.

Index the successful increments by $t=1,\ldots,T$ and put
$a_t=\sum_{j\in S_t}w_j/W_t$, with weights taken just before the
increment.  Its increase in the unnormalized opening and connection cost
is at most
\[
                    (1+\delta)\lambda_tW_t a_t
                         \le(1+\delta)D a_t,
\]
where $D$ can be taken to be its final retained value.  Before newly
completed clients are removed, the weight total becomes
\[
                         \widetilde W_t=W_t(1-\delta a_t).
\]
Therefore
\[
                  \delta a_t\le\log W_t-\log\widetilde W_t.
\]
Removing completed clients only decreases the total further.  For the
last logarithm, keep the weights before this removal.  Immediately
before the last increment $W_T>{W_{\min}}$, and consequently
\[
                  \widetilde W_T\ge qW_T>q{W_{\min}}=Bq^{U+1}.
\]
Since $W_1=B$, telescoping gives
\[
                      \sum_ta_t\le\frac{(U+1)(-\log q)}{\delta}.
\]
After division by $U$, the opening and connection cost is thus at most
\begin{align*}
 (1+\delta)D\left(1+\frac1U\right)
                      \frac{-\log(1-\delta)}{\delta}
 &\le (1+\delta)^2(1+\delta^2)D\\
 &\le(1+3\delta)D.
\end{align*}
We used $1/U\le\delta^2$,
$-\log(1-\delta)/\delta\le1+\delta$, and $\delta\le1/10$.
Capping the opening variables at one cannot increase this cost.

It remains to pay for the dummy assignments.  For an unfinished
client, let $s=U-\kappa_j\ge1$.  Since $q^{-s}\ge1+\delta s$,
we have $s q^s\le1/\delta$.  Hence its remaining penalty satisfies
\[
 \ell_j\left(1-\frac{\kappa_j}{U}\right)
    =w_jq^{-U}\frac{s q^s}{U}
    \le\frac{w_j}{\delta Uq^U}.
\]
At termination the sum of these penalties is at most
\[
             \frac{W}{\delta Uq^U}
                   \le\frac B{\delta U}\le\delta B\le3\delta D.
\]
Completed clients have no remaining demand.  The repair lemma
does not increase the total cost including dummy assignments, proving the cost
bound.

For the running time, initialize each facility's set $S_i$ and the
sums in~\eqref{eq:capped-star-test} by one scan of the clients.
During its run, $S_i$ only shrinks.  Updating its members after a
successful increment takes time proportional to the assignment
variables just increased.  Maintain $W$ by the same weight changes,
so testing the stopping condition takes constant additional time.
The total number of assignment increases is at most $nU$.
Before termination, $W>Bq^U$; by
\eqref{eq:capped-initializer-invariant} and $\mathrm{LP}\le B$,
\[
                              \lambda<q^{-U}.
\]
The scalar starts at $1/3$ and grows by $1+\delta$ after every
completed pass.  The number of passes is therefore
\[
 O\left(\frac{1+U[-\log(1-\delta)]}{\log(1+\delta)}\right)
                            =O(U+1/\delta)=O(U).
\]
This also proves termination: only finitely many successful increments
are possible, and otherwise another completed pass would eventually
violate the displayed bound on $\lambda$.

The passes cost $O(mnU)$, retaining the dual vectors costs $O(nU)$,
and penalty construction and repair cost $O(mn)$.
Computing $q^U$ takes $O(\log(U+1))$ rational multiplications by
repeated squaring, within the same bound.  The integer counts,
weights, and updates require no evaluations of logarithms.
\end{proof}

\section{Random augmentation}
\label{sec:augmentation}

Random augmentation is an alternative to greedy augmentation.  It
opens additional facilities at random, with probabilities determined
by a feasible fractional solution, which it therefore takes as an
input in addition to the initial set $S_0$.  It runs in $O(m^2+mn)$
expected time and gains an amount of order $m^{-1/2}$ over the bound
of \cref{thm:greedy-augmentation}.  We state it for an initial set
whose connection factor is about $\beta$, which is what our rounding
algorithms provide.

\paragraph{The fractional solution.}
We use any feasible fractional solution on the original facilities:
\[
 \sum_i x_{ij}=1,\qquad 0\le x_{ij}\le y_i\le1.
\]
Facilities with zero opening amount are omitted from sampling.  No
reassignment or splitting of facilities is needed.

Fix such a solution.  Write
\[
 \begin{aligned}
 F&=\sum_i f_i y_i,& C_j&=\sum_i x_{ij}d(i,j),& C&=\sum_j C_j,\\
 T_j&=\sum_i x_{ij}^2d(i,j),& T&=\sum_jT_j,\\
 U_j&=\sum_i x_{ij}y_i d(i,j),& U&=\sum_jU_j.
 \end{aligned}
\]
Since $x_{ij}\le y_i$, we have $T_j\le U_j$.
Let $L=\lceil\sqrt m\rceil$ as in \eqref{eq:final-parameters}, and
partition the facilities into
\[
 \mathcal H=\{i:y_i\ge1/L\},\qquad
 \mathcal L=\{i:y_i<1/L\}.
\]
The total opening amount in $\mathcal L$ is at most
$m/L\le L$.

\paragraph{The two augmentation procedures.}
Fix a rational $t\in[2/5,1/2]$.  All the following choices are
independent of the initial integral solution.  Facilities with $y_i=0$
are omitted from sampling.
\begin{enumerate}
\item In the first procedure, open each facility $i$ independently
      with probability $ty_i$.
\item In the second procedure, process $\mathcal H$ and $\mathcal L$
      independently.  For a group $\mathcal G$ of total amount
      $Y_{\mathcal G}>0$, put
      \[
       k_{\mathcal G}=\lfloor tY_{\mathcal G}\rfloor,
       \qquad \theta_{\mathcal G}=tY_{\mathcal G}-k_{\mathcal G}.
      \]
      Make $k_{\mathcal G}$ draws, and one additional draw with
      probability $\theta_{\mathcal G}$.  Each draw independently
      chooses $i\in\mathcal G$ with probability $y_i/Y_{\mathcal G}$;
      open every facility drawn.  A group of total amount zero is skipped.
\end{enumerate}
The implemented procedure chooses the first alternative with probability
$3/4$ and the second with probability $1/4$, and adds the chosen
facilities to the initial solution.  Clients use their nearest open
facility.  Either alternative adds expected opening cost at most $tF$.

\begin{lemma}[Failure probability within one group]
\label{lem:group-failure}
Suppose a group $\mathcal G$ has total opening amount $Y>0$, of which
an amount $a\in[0,Y]$ lies at distance at most $r$ from a client.
The probability that no draw selects a facility at distance at most
$r$ is at most
\[
 \exp\!\left(-ta-\frac{t^2a^2}{2\max\{1,Y\}}\right).
\]
\end{lemma}
\begin{proof}
Write $k=\lfloor tY\rfloor$ and $\theta=tY-k$.  Each draw selects
a facility at distance at most $r$ with probability $a/Y$.
Thus the probability
in question is exactly
\[
 (1-a/Y)^k(1-\theta a/Y).
\]
We use $(1-a/Y)^0=1$, including when $a=Y$.
If this probability is zero, the result holds.  When $k=0$, the first
factor is absent, so the argument also covers $a=Y$ and $k=0$.
For the remaining nonzero factors,
$\log(1-u)\le-u-u^2/2$ gives
\[
 \log\bigl((1-a/Y)^k(1-\theta a/Y)\bigr)
 \le-ta-\frac{(k+\theta^2)a^2}{2Y^2}.
\]
Cauchy--Schwarz gives $k+\theta^2\ge(tY)^2/(k+1)$.
For $Y<1$ one has $k=0$; for $1\le Y<2$ one again has $k=0$;
and for $Y\ge2$ one has $k+1\le Y/2+1\le Y$.
Thus $k+1\le\max\{1,Y\}$ in all cases, proving the claim.
\end{proof}

For a fixed client and distance $r$, define
\[
 \Lambda(r)=\sum_{i:d(i,j)\le r}x_{ij},\qquad
 B_y(r)=\sum_{i:d(i,j)\le r}x_{ij}y_i,
\]
and let $a_{\mathcal H}(r),a_{\mathcal L}(r)$ be the two groups'
contributions to $\Lambda(r)$.  Put
\[
 J(r)=\sum_{\mathcal G\in\{\mathcal H,\mathcal L\}:
                          Y_{\mathcal G}>0}
          \frac{a_{\mathcal G}(r)^2}{\max\{1,Y_{\mathcal G}\}}.
\]
Let $p_{\rm ind}(r)$ and $p_{\rm grp}(r)$ be the probabilities that
no facility at distance at most $r$ is selected under the two
alternatives, respectively.  These probabilities use every facility
with $y_i>0$, including facilities with $x_{ij}=0$.

\begin{lemma}[Two complementary improvements]
\label{lem:two-improvements}
For $c_1=e^{-t}t^2/4\ge1/50$, one has
\begin{align*}
 p_{\rm ind}(r)&\le1-(1-e^{-t})\Lambda(r)
             -c_1\bigl(\Lambda(r)(1-\Lambda(r))+B_y(r)\bigr),\\
 p_{\rm grp}(r)&\le1-(1-e^{-t})\Lambda(r)
             -c_1\bigl(\Lambda(r)(1-\Lambda(r))+J(r)\bigr).
\end{align*}
Moreover, whenever $\Lambda(r)\ge1/3$,
\[
                         B_y(r)+J(r)\ge\frac1{9L}.
\]
\end{lemma}
\begin{proof}
Independence and $x_{ij}\le y_i$ give
\[
 p_{\rm ind}(r)
 =\prod_{i:d(i,j)\le r,\,y_i>0}(1-ty_i)
 \le e^{-t\Lambda(r)-t^2B_y(r)/2}.
\]
Indeed, $\sum_{d(i,j)\le r}y_i\ge \Lambda(r)$ and
$\sum_{d(i,j)\le r}y_i^2\ge B_y(r)$.
For each group its opening amount within distance $r$ is at least
its assignment amount $a_{\mathcal G}(r)$.  Applying
Lemma~\ref{lem:group-failure} to the actual opening amounts and then
decreasing them to the assignment amounts gives
$p_{\rm grp}(r)\le e^{-t\Lambda(r)-t^2J(r)/2}$.
Both $B_y(r)$ and $J(r)$ belong to $[0,1]$.
For $a,z\in[0,1]$, the chord bound for the strongly convex function
$e^{-ta}$, followed by $1-e^{-u}\ge u/2$ for $u\in[0,1]$, gives
\[
 e^{-ta-t^2z/2}
 \le1-(1-e^{-t})a
       -\frac{e^{-t}t^2}{2}a(1-a)
       -\frac{e^{-t}t^2}{4}z.
\]
This proves the first two inequalities.  The lower bound $c_1\ge1/50$
follows from $t\ge2/5$ and $e^{-1/2}>3/5$.

Let $a=a_{\mathcal H}(r)$ and $b=a_{\mathcal L}(r)$, so $\Lambda(r)=a+b$.
Every facility in $\mathcal H$ has $y_i\ge1/L$, and hence
\[
 B_y(r)\ge\frac{a}{L},\qquad
 J(r)\ge\frac{b^2}{L}.
\]
The latter inequality uses $\max\{1,Y_{\mathcal L}\}\le L$;
it also holds when the group is empty.  Therefore
\[
 B_y(r)+J(r)\ge\frac{a+b^2}{L}.
\]
Finally,
\[
 \min\{a+u^2:a,u\ge0,\ a+u\ge1/3\}=1/9,
\]
which proves the last claim.
\end{proof}

For $D\ge0$, put
\[
 g_{\rm ind}(D)=\int_0^D p_{\rm ind}(r)\,dr,
 \qquad g_{\rm grp}(D)=\int_0^D p_{\rm grp}(r)\,dr.
\]
Each function is increasing, concave, and $1$-Lipschitz.

\begin{lemma}[Integrated improvement]
\label{lem:integrated-improvement}
For each client,
\begin{align*}
 g_{\rm ind}(\beta C_j)
 &\le\bigl(1+(\beta-1)e^{-t}\bigr)C_j-U_j/150,\\
 \frac{g_{\rm ind}(\beta C_j)+g_{\rm grp}(\beta C_j)}2
 &\le\bigl(1+(\beta-1)e^{-t}\bigr)C_j-C_j/(5400L).
\end{align*}
Consequently the expected connection cost after the implemented mixture is
$g=\tfrac34g_{\rm ind}+\tfrac14g_{\rm grp}$ satisfying
\[
 g(\beta C_j)
 \le\bigl(1+(\beta-1)e^{-t}\bigr)C_j
                    -U_j/300-C_j/(10800L).
\]
\end{lemma}
\begin{proof}
We omit the client subscript.  If $C=0$, all terms vanish.  Otherwise
put $D=\beta C$, $v_i=\min\{d(i,j),D\}$, and
\[
 \begin{gathered}
 C_D=\sum_i x_i v_i,\qquad U_D=\sum_i x_i y_i v_i,\qquad
 S_{xy}=\sum_i x_i y_i,\\
 E=C-C_D,\qquad G=\int_0^D \Lambda(r)(1-\Lambda(r))\,dr,\qquad
 B_D=\int_0^D B_y(r)\,dr.
 \end{gathered}
\]
Then $B_D=DS_{xy}-U_D$.  Symmetrizing gives
\begin{align*}
 U_D-C_DS_{xy}
 &=\sum_{i<h}x_ix_h(y_i-y_h)(v_i-v_h)\\
 &\le\sum_{i<h}x_ix_h|v_i-v_h|=G.
\end{align*}
The inequality uses $0\le y_i,y_h\le1$.
Since $C_D\le C=D/\beta$,
\[
 U_D\le\frac{\beta G+B_D}{\beta-1},\qquad
 U\le E+\frac{\beta}{\beta-1}(G+B_D).
\]
The second inequality also uses $U-U_D\le C-C_D=E$.
Integrating the common chord in Lemma~\ref{lem:two-improvements} gives
\[
 \int_0^D\bigl(1-(1-e^{-t})\Lambda(r)\bigr)\,dr
 =\bigl(1+(\beta-1)e^{-t}\bigr)C-(1-e^{-t})E.
\]
Now $1-e^{-t}\ge1/150$, $c_1\ge1/50$, and
$(\beta-1)/\beta\ge1/3$.  The first inequality follows by charging
$U/150$ to $(1-e^{-t})E+c_1(G+B_D)$.

For the second inequality, average the two bounds of
Lemma~\ref{lem:two-improvements} and retain the term
$\tfrac12c_1(B_y(r)+J(r))$.
For every $r\ge3C/2$, Markov's inequality gives
$\Lambda(r)\ge1-C/r\ge1/3$.  Also $\beta\ge5/3$.
Integrating only over $[3C/2,5C/3]$, which has length $C/6$, gives
an improvement of at least
\[
 \frac{c_1}{2}\,\frac1{9L}\,\frac C6
 \ge\frac C{5400L}.
\]
Finally, the implemented mixture is one half of the first procedure
and one half of the equal mixture, proving the last inequality.
\end{proof}

\begin{theorem}[Random augmentation]
\label{thm:grouped-augmentation}
Put $\gamma=1/(10^6L)$.  Suppose a randomized rounding algorithm gives a nonempty initial
solution $S_0$, with opening cost $F_0$ and client distances $D_j$, such
that
\[
 \mathbb E F_0\le F,\qquad
 \mathbb E D_j\le(\beta+\gamma)C_j+T_j/1000
                                      \quad\text{for every client }j.
\]
Run the augmentation mixture above independently of $S_0$.
Its expected total cost is at most
\[
 (1+t)F+\bigl(1+(\beta-1)e^{-t}\bigr)C
                    -C/(10800L)-7U/3000+\gamma C.
\]
Let $\alpha_*$ be the root of
$\alpha_*-1=2e^{-\alpha_*}$, and put $t_*=\alpha_*-1$.
Choose a rational $t$ such that
\[
 \frac1{86400L}\le t_*-t\le\frac1{43200L}.
\]
Then the expected cost is at most
\[
                \left(\alpha_*-\frac1{86400L}+\gamma\right)(F+C).
\]
The augmentation, including the choice of
$t$, uses $O(m^2+mn)$ expected operations, fair random bits, and
distance accesses.
This bound does not
include finding the fractional solution or running the initial
rounding algorithm.
\end{theorem}
\begin{proof}
For a fixed initial distance $D_j$, the expected connection cost after
augmentation is at most $g_j(D_j)$.  Concavity and the $1$-Lipschitz
property give
\[
 \mathbb E g_j(D_j)
 \le g_j(\mathbb E D_j)
 \le g_j(\beta C_j)+T_j/1000+\gamma C_j.
\]
Lemma~\ref{lem:integrated-improvement} and the expected added opening
cost $tF$ prove the first bound.  Since $T_j\le U_j$,
$U_j/300-T_j/1000\ge7U_j/3000$.

Let $h(t)=1+(\beta-1)e^{-t}$.  Since $h(t_*)=\alpha_*$ and
$|h'(t)|\le\beta-1<1$ for $t\ge0$,
\[
 1+t\le\alpha_*-1/(86400L),\qquad
 h(t)-1/(10800L)\le\alpha_*-1/(14400L).
\]
The prescribed $t$ lies in $[2/5,1/2]$.  Dropping the remaining
nonnegative improvement in $U$ proves the approximation bound.

There are only two groups.  Their total number of draws is at most
$\lfloor tY_{\mathcal H}\rfloor+
 \lfloor tY_{\mathcal L}\rfloor+2=O(m)$.
Each draw can be implemented by a scan of its group, for $O(m^2)$
operations in total; more elaborate sampling is unnecessary.
Each newly drawn facility is compared with the current connection
of every client, for $O(mn)$ operations.  The independent procedure
has the same or smaller bound.  Repeated selections of a facility
incur its opening cost only once.

The rational parameter $t$ is computed in $O(\log^2(L+1))$ operations;
see \cref{sec:sampling-details}.
\end{proof}

\subsection{Derandomized augmentation}
\label{sec:deterministic-augmentation}

The separate expected-distance bounds are useful for the fast randomized
algorithm.  A deterministic algorithm needs only a suitable weighted sum
of distances.  Concavity of the augmentation bound supplies the weights.

For a feasible fractional solution and the rational parameter $t$, let
$g_j(D)$ be the expected connection distance obtained by applying the
augmentation mixture to an initial distance $D$.  This is the actual
augmentation procedure: every selected facility is available to every
client.  As in the augmentation analysis, $g_j$ is increasing, concave, and
$1$-Lipschitz.  It is piecewise linear with rational breakpoints and
rational slopes.  In particular, its value and a supporting slope at
any rational argument can be computed exactly.  Its distance intervals
include every facility with $y_i>0$, even if $x_{ij}=0$.

\begin{lemma}[Supporting affine functions]
\label{lem:augmentation-tangents}
Suppose a deterministic rounding procedure works for arbitrary
nonnegative rational client weights $w_j$ and returns a nonempty set
with opening cost $F_0$ and distances $D_j$ satisfying
\[
 F_0+\sum_jw_jD_j
 \le F+\beta\sum_jw_jC_j+\frac1K\sum_jw_jT_j
                                      +\gamma\sum_jw_jC_j.
\]
Then a deterministic polynomial-time algorithm returns a solution of
cost at most
\[
 (1+t)F+\bigl(1+(\beta-1)e^{-t}\bigr)C
             -\frac{C}{10800L}-\frac{7U}{3000}+2\gamma C.
\]
Here $K=1000$, $L=\lceil\sqrt m\rceil$, and
$\gamma=1/(10^6L)$.
\end{lemma}
\begin{proof}
Choose a rational $b$ with $\beta\le b\le\beta+\gamma$.
Let $w_j\in[0,1]$ be the right derivative of $g_j$ at $bC_j$.
Concavity gives, for every $D\ge0$,
\[
              g_j(D)\le g_j(bC_j)+w_j(D-bC_j).
\]
Apply weighted deterministic rounding with these weights.  The
resulting set satisfies
\begin{align*}
 F_0+tF+\sum_jg_j(D_j)
 &\le (1+t)F+\sum_jg_j(bC_j)
       +\sum_jw_j\bigl(T_j/K+\gamma C_j-(b-\beta)C_j\bigr)\\
 &\le (1+t)F+\sum_jg_j(\beta C_j)+T/K+2\gamma C.
\end{align*}
The second inequality uses $w_j\le1$ and the $1$-Lipschitz property.
The integrated augmentation bound proves the claimed expression,
because $T_j\le U_j$ and
$U_j/300-T_j/1000\ge7U_j/3000$.

It remains to realize the first line by deterministic augmentation.
Charge the opening cost of every selected facility, counting
repeated draws separately.  This only increases the actual opening
cost, and its expectation is at most $tF$.  Conditional on any partial
augmentation, the expected remaining connection cost is a finite sum
over the client's distance intervals.  In independent opening,
an undecided facility $i$ contributes $1-ty_i$ to the survival
probability at radii $r\ge d(i,j)$.  A facility fixed to open contributes
zero, and a facility fixed to stay closed contributes one.
Facilities beyond $r$ also contribute one.  In the grouped procedure
put $c_{\mathcal G}(r)=\sum_{i\in\mathcal G:d(i,j)\le r}y_i$.
The remaining draws contribute
\[
 \left(1-\frac{c_{\mathcal H}(r)}{Y_{\mathcal H}}\right)^{R_{\mathcal H}}
 \left(1-\frac{c_{\mathcal L}(r)}{Y_{\mathcal L}}\right)^{R_{\mathcal L}},
\]
with the remaining numbers of draws and with zero-mass groups omitted.
If an already fixed draw selected a facility at distance at most $r$,
the failure probability is zero.  Otherwise fixed draws contribute
one.  As before, a zeroth power is one even when its base is zero.
All these quantities are rational.

First fix the choice of augmentation procedure by conditional
expectation.  For grouped augmentation, fix the two trial counts, then
fix the facility selected by each draw.  For independent opening, fix
its Bernoulli decisions one by one.  At each step select an outcome of
positive conditional probability whose conditional expected upper bound
does not exceed the current one.  Once all facility choices have been
fixed, the connection cost is determined.  The final feasible solution has
cost at most $F_0+tF+\sum_jg_j(D_j)$, as required.
\end{proof}

\paragraph{Implementation.}
For each client, sort the distances to all positive-opening facilities
and compute prefix opening amounts in the two groups.  The survival
probability on each resulting interval is rational.  Precomputing
all powers up to the largest possible number of draws takes
$O(nm^2)$ operations.  Maintain the failure-probability factors from already
fixed selections on all client distance intervals.  Conditional
expectations then take $O(nm)$ operations per candidate facility.
There are $O(m)$ draws and at most $m$ candidates per draw, so
deterministic augmentation takes $O(nm^3)$ operations.  Independent
opening needs less work.
The same tables give $g_j(bC_j)$ and its supporting slope, with one
additional breakpoint per client.  A certified rational upper estimate
of $\beta$ takes $O(\log(m+1))$ series terms.  Thus the complete
derandomization is polynomial and uses only rational arithmetic.

\section{Combining the components}
\label{sec:combining}

We now combine the components and prove the theorems of the
introduction.

\subsection{The threshold from one LP solution}

Let
\[
        s_*:=e^{\alpha_*-1},\qquad\text{so that}\qquad
        \beta-1=(\alpha_*-1)s_*.
\]

\begin{pseudocode}{LP solver, Rounding~1, and greedy augmentation}
\label{alg:threshold}
\begin{enumerate}
\item Solve the LP \eqref{eq:standard-lp}.  Retain an optimal
      rational solution $(x^*,y^*)$ and its costs $F^*,C^*$.
      If $C^*=0$, open a cheapest facility with $y_i^*>0$ in each
      zero-distance class containing a client, and return.
\item Compute the quantity $\eta$ of \cref{thm:rounding-bound} for
      $(x^*,y^*)$, and construct a rational scaling factor $s$ with
      \begin{equation}
                 1<s<s_*,\qquad s_*-s\le\eta/2.
       \label{eq:scaling-factor-conditions}
      \end{equation}
\item Run Rounding~1 (Algorithm~\ref{alg:rounding}) on $(x^*,y^*)$ with
      opening costs $sf_i$ and the original distances.  Let $S_0$ be the
      returned set.
\item Run greedy augmentation from $S_0$ with the original opening
      costs, and return its set of open facilities.
\end{enumerate}
\end{pseudocode}

The LP is solved once, with the original opening costs.  The scaling
factor depends on $\eta$, and hence on the LP solution; and the
guarantee of Rounding~1 for the costs $sf_i$ is relative to the same
solution $(x^*,y^*)$, as \cref{thm:greedy-augmentation} requires.
\Cref{sec:scaling-factor-choice} constructs $s$ in time polynomial in
the encoding length of $\eta$.

\begin{proof}[Proof of \cref{thm:main-simple}]
If $C^*=0$, the zero-distance construction costs at most $F^*$.
Otherwise, \cref{thm:rounding-bound} applied to $(x^*,y^*)$ with
opening costs $sf_i$ gives
\[
                  sF_0+C_0
                  \le sF^*+(\beta-\eta)C^*,
\]
where $F_0,C_0$ are the costs of $S_0$ in the original instance.
\Cref{thm:greedy-augmentation} with
$(\gamma_f,\gamma_c)=(1,\beta-\eta)$ gives
\[
 \cost(S)
 \le(1+\log s)F^*
          +\left(1+\frac{\beta-\eta-1}{s}\right)C^* .
\]
Both coefficients are smaller than $\alpha_*$:
\begin{equation}
 \begin{split}
 1+\log s&<1+\log s_*=\alpha_*,\\
 1+\frac{\beta-\eta-1}{s}
 &=\alpha_*+\frac{(\alpha_*-1)(s_*-s)-\eta}{s}
 \le\alpha_*-\frac{3\eta}{4s}<\alpha_*.
 \end{split}
 \label{eq:approximation-coefficients}
\end{equation}
The second bound uses $\alpha_*-1<1/2$ and
\eqref{eq:scaling-factor-conditions}.  Hence
$\cost(S)\le\alpha_*(F^*+C^*)=\alpha_*\LP$.

A polynomial-time LP algorithm returns an optimal solution of
polynomial encoding length.  Hence $\log(1/\xi)$ and
$\log(1/\eta)$ are polynomially bounded in the input length, and the
construction of $s$ and Rounding~1 run in polynomial time.  Greedy
augmentation makes at most $m$ additions and uses $O(m^2n)$
operations.
\end{proof}

The strict improvement $\eta$ depends on the fractional solution and
is not bounded away from zero across instances.

\subsection{Rounding~2 and random augmentation}

\begin{theorem}[Quadratic rounding and augmentation]
\label{thm:main-quadratic}
Given a feasible fractional solution of cost $F+C$, one can find a
solution in $O(m^2+mn)$ expected time with expected
cost at most
\[
       \left(\alpha_*-\frac{1}{86400L}
                         +\frac{1}{10^6L}\right)(F+C),
             \qquad L=\lceil\sqrt m\rceil.
\]
\end{theorem}
\begin{proof}
Merge zero-distance facility classes as in
\cref{sec:quadratic-rounding}; this does not increase $F$ or $C$.
Apply Rounding~2 to the merged solution.  By
\cref{thm:quadratic-rounding}, with $K=1000$, its output satisfies the
hypothesis of \cref{thm:grouped-augmentation}.  Apply random
augmentation with the merged solution and the parameter $t$ of that
theorem.  Its conclusion is the stated bound, since
$\gamma=1/(10^6L)$.  Both steps take $O(m^2+mn)$ expected time.
\end{proof}

\subsection{The combinatorial algorithms}

\begin{pseudocode}{The combinatorial algorithms}
\label{alg:complete-algorithm}
\textbf{Parameters:} $L,K,\gamma$ as in \eqref{eq:final-parameters};
the accuracy $\delta$ of the fractional solution is
$\delta=\gamma/6$ for the factor below $\alpha_*$, and
$\delta=\varepsilon/12$ for the factor $\alpha_*+\varepsilon$.
\begin{enumerate}
\item Run the combinatorial fractional algorithm
(\cref{alg:capped-combinatorial-initializer}) with accuracy $\delta$.
If the cost of its solution is zero, return an integral solution of
cost zero.
Merge zero-distance facility classes as in
\cref{sec:quadratic-rounding}, and use the resulting fractional
solution in both rounding and augmentation.
\item \emph{Randomized version.}  Apply Rounding~2 and random
augmentation as in \cref{thm:main-quadratic}.
\item \emph{Deterministic version.}  Apply the derandomized
augmentation of \cref{lem:augmentation-tangents}, which calls
Rounding~3 with accuracy $\varepsilon=\gamma$.
\end{enumerate}
\end{pseudocode}

A fractional solution of cost zero has $C=0$, and the zero-distance
construction of \cref{sec:rounding-one} then gives an integral
solution of cost at most $F=0$.

\begin{proof}[Proof of \cref{thm:main-upper}]
Put $\mu=1/(86400L)$, and note that $L\le3\sqrt m$ and
$\alpha_*<3/2$.

\emph{The randomized algorithm.}
With $\delta=\gamma/6$, \cref{thm:capped-combinatorial-initializer}
gives a fractional solution with $F+C\le(1+\gamma)\LP$ in
$O(mn/\delta^2)=O(m^2n)$ time.  By \cref{thm:main-quadratic}, the
expected cost is at most
\[
 (\alpha_*-\mu+\gamma)(1+\gamma)\LP
 \le\bigl(\alpha_*-\mu+(1+\alpha_*)\gamma\bigr)\LP .
\]
Since $(1+\alpha_*)\gamma<5/(2\cdot10^6L)$,
\[
 \mu-(1+\alpha_*)\gamma
   >\left(\frac1{86400}-\frac1{400000}\right)\frac1L
   >\frac1{125000L}\ge\frac1{375000\sqrt m}.
\]
The remaining work takes $O(m^2+mn)$ expected time.

\emph{The factor $\alpha_*+\varepsilon$.}
With $\delta=\varepsilon/12\le1/10$, the fractional solution satisfies
$F+C\le(1+\varepsilon/2)\LP$ and is computed in $O(mn/\varepsilon^2)$
time.  The factor of rounding and augmentation is below $\alpha_*$, so
the expected cost is at most
$\alpha_*(1+\varepsilon/2)\LP<(\alpha_*+\varepsilon)\LP$.  The
remaining work takes $O(m^2+mn)$ expected time.

\emph{The deterministic algorithm.}
By \cref{lem:weighted-deterministic-rounding} with
$\varepsilon=\gamma$, applied to the same merged fractional solution
used in augmentation, Rounding~3 satisfies the hypothesis of
\cref{lem:augmentation-tangents}.  The choice of $t$ in
\cref{thm:grouped-augmentation} bounds both cost coefficients in the
conclusion of \cref{lem:augmentation-tangents}, before the term
$2\gamma C$, by $\alpha_*-\mu$.  Hence the cost is at most
$(\alpha_*-\mu+2\gamma)(F+C)$.  With $F+C\le(1+\gamma)\LP$,
\[
 (\alpha_*-\mu+2\gamma)(1+\gamma)
 \le\alpha_*-\mu+(\alpha_*+2+2\gamma)\gamma
 <\alpha_*-\frac1{125000L}
 \le\alpha_*-\frac1{375000\sqrt m},
\]
because $(\alpha_*+2+2\gamma)\gamma<7/(2\cdot10^6L)$ and
$1/86400-7/(2\cdot10^6)>1/125000$.
The fractional solution takes $O(m^2n)$ time, Rounding~3 with
$\varepsilon=\gamma$ takes $O(m^3n+m^3)$ operations up to lower-order
terms, and the derandomized augmentation takes $O(nm^3)$ operations.
The total is $O((m+n)^4)$.
\end{proof}

\subsection{The integrality gap}

\begin{proof}[Proof of the upper bound in \cref{thm:main-finite-gap}]
Apply \cref{thm:main-quadratic} to an optimal fractional solution.
Some outcome of the algorithm has cost no greater than its expectation.
Since
\[
 \frac1{86400}-\frac1{10^6}>\frac1{100000},
 \qquad \lceil\sqrt m\rceil\le3\sqrt m,
\]
every instance with at most $m$ facilities has integrality gap at most
$\alpha_*-1/(300000\sqrt m)$.
\end{proof}

The lower bound in \cref{thm:main-finite-gap} is proved in
\cref{sec:matching-lp-lower-bound}.

\section{Proof of the comparison inequality}
\label{sec:general-comparison-proof}

We prove \cref{thm:core-finite-comparison} and
\cref{lem:first-iteration-improvement} by the same charging argument.
The theorem is due to OpenAI~\cite[Theorem~4.1]{OAI26}, and we prove it
to keep the paper self-contained; the lemma is new.  The notation of
this section is local: the symbols $B$, $C_s$, $b_s$, $k$, $d_s$,
$H_s$, $L_s$, $K_s$, $G$, $W$, $U$ introduced below are used only here
and in \cref{sec:scalar-verification}.
We separate the common argument from the scalar inequalities for its
two applications.  The argument uses one affine upper bound for each
summand and one convex function for the total charge.

\subsection{Four sufficient inequalities}

Fix $\ybar,y\in[0,1]$, and put
\[
 P=P(\ybar),\quad \mu_0=A(\ybar)-1-2\ybar P\ge0,\qquad a=1+\mu_0+2Py.
\]
We subtract $\varepsilon(r)D+B$ from a summand with distance $D$
and mean $r=\mathbb E_tq_{ti}$.  Here $B$ is constant in $r$, and
\begin{equation}
 B\ge0,\qquad \varepsilon(r)\le B\quad(0\le r\le1).
 \label{eq:general-correction-order}
\end{equation}
For the general comparison, the correction is nonnegative and bounds
the error due to the difference between $\ybar$ and $y$.  At $\ybar=y=1$,
we take $\varepsilon(r)=-cr^2$ and $B=0$; a negative correction gives
the desired improvement.  The argument treats these choices identically.
For $s\in\{0,1\}$, write
\[
 C_s=1-sP,\qquad b_s=(2-s)Py,\qquad
 \kappa_s(q)=\frac{P(1-q)(q-s)}{1-Pq}.
\]
Choose $k\ge0$ satisfying
\begin{equation}
                         P(1+k)^2\le4k,
 \label{eq:kappa-upper-bound}
\end{equation}
and set $k_s=(1-s)k$.  Then $\kappa_s(q)\le k_s$.
Indeed, $q-s\le(1-s)q$, and clearing the positive denominator gives
\[
 k(1-Pq)-Pq(1-q)
   =P\left(q-\frac{1+k}{2}\right)^2
                     +k-\frac{P(1+k)^2}{4}\ge0.
\]
We use this bound on $\kappa_s(q)$ in both applications.
Define
\begin{align}
 L_s(r)&=(C_s-\mu_0-b_s)r+\varepsilon(r),\notag\\
 K_s(r)&=C_s-k_s-b_sr+B,\label{eq:general-affine-functions}\\
 G(r)&=a(1-r).\notag
\end{align}
Inequality~\eqref{eq:scalar-inequality-one} below, together with
Jensen's inequality, gives the same coefficient $G(r)$ of $D$ for
$s=0$ and $s=1$.  We will average $K_s(r)$ over the indices $t$,
so $K_s$ must be affine in $r$; this is why $B$ is
constant.  No affineness of $L_s$ is required.

We will choose nonnegative parameters $d_0,d_1,H_0,H_1$
such that
\begin{equation}
 0\le H_0,H_1\le H_*:=\max(H_0,H_1)\le\min(d_0,d_1).
 \label{eq:parameter-order}
\end{equation}
The following four conditions are sufficient:
\begin{gather}
 L_s(r)\ge\frac{C_sPr(1-r)}{1-Pr}
              \qquad(0\le r\le1),\label{eq:scalar-inequality-one}\\
 K_s(r)\ge d_sG(r)
              \qquad(0\le r\le1),\label{eq:scalar-inequality-two}\\
 K_s(r)L_s(r)+\{(C_sH_s-b_s)r+B\}G(r)\ge0
              \qquad(0<r<1),\label{eq:scalar-inequality-three}\\
 K_s(q)+d_t\{k_s-\kappa_s(q)\}\ge H_t
              \qquad(0\le q\le1,\ t\in\{0,1\}).
 \label{eq:scalar-inequality-four}
\end{gather}
Each inequality holds for both choices of $s$.

\paragraph{Roadmap.}
The comparison is an inequality on average: an individual summand may
be positive.  After subtracting its correction, call the summand
$F_i$.  Inequalities \eqref{eq:scalar-inequality-one} and
\eqref{eq:scalar-inequality-two} show that $F_i>0$ is possible only if
$D_i>d_{s_i}$.  Inequality \eqref{eq:scalar-inequality-three} bounds
such a summand by $\mathbb E_t(H_{s_i}-D_t)_+$.  Inequalities
\eqref{eq:scalar-inequality-four} and \eqref{eq:scalar-inequality-two}
show that the summands with small $D_i$ are negative enough to pay for
this; the two bounds cancel after averaging, by interchanging the two
independent indices.

\subsection{Bounding the summands}

\begin{lemma}[The charging argument]
\label{lem:general-charging}
For admissible data (\cref{def:comparison-expression}),
suppose $s_i\in\{0,1\}$ for every index $i$.
Parameters satisfying the preceding scalar conditions give
\[
              \mathcal E_{\ybar,y}
                 \le B+\mathbb E_i[\varepsilon(\rho_i)D_i].
\]
\end{lemma}

\begin{proof}
Subtract $\varepsilon(\rho_i)D_i+B$ from the summand indexed by $i$
in $\mathcal E_{\ybar,y}$, and call the result $F_i$.
It remains to prove $\mathbb E_iF_i\le0$.

\paragraph{An affine upper bound.}
Fix an index $i$ and omit it from the notation.  Write $D=D_i$, $R=\mathcal R_i$,
$Z=D_t$, $q=q_{ti}$, $r=\mathbb E_tq$, and $s=s_i$.
Use $C,b,k,L,K$ for the quantities with subscript $s$, and put
$\delta(q)=k-\kappa_s(q)\ge0$.
In particular, $\mathbb E_tZ=1$.
The corrected summand is
\begin{align*}
 F={}&C(1-r)R+r\{\mu_0 D+b(D+1)\}-C(D+1)\\
    &-\mathbb E_t(1-q)(Z-aD)_+-\varepsilon(r)D-B.
\end{align*}
The radius inequality $R(1-Pq)\le D+Z$, multiplied by $1-q$,
remains valid at $q=1$.  Apply it and the inequality
$(Z-aD)_+\ge Z-aD$.
The coefficient of $Z$ is
\[
 \frac{C(1-q)}{1-Pq}-(1-q)=\kappa_s(q).
\]
Writing this coefficient as $k-\delta(q)$ and using
$\mathbb E_tZ=1$ changes the constant $-C+br-B$ to $-K(r)$.
The remaining coefficient of $D$ is
\[
 a(1-r)-L(r)+C\{\mathbb E_t\psi(q)-(1-r)\},
 \qquad \psi(q)=\frac{1-q}{1-Pq}.
\]
The function $\psi$ is concave:
$\psi''(q)=-2P(1-P)/(1-Pq)^3<0$.
Jensen's inequality and \eqref{eq:scalar-inequality-one} therefore give
\[
 C\{\mathbb E_t\psi(q)-(1-r)\}
 \le\frac{CPr(1-r)}{1-Pr}\le L(r).
\]
Thus the coefficient of $D$ is at most $a(1-r)=G(r)$, and
\begin{equation}
 F\le V(D):=DG(r)-K(r)-\mathbb E_t[\delta(q)Z].
 \label{eq:affine-upper-bound}
\end{equation}

At $r=1$, this bound is nonpositive because $G(1)=0$ and $K(1)\ge0$.
At $r=0$, every index $t$ with positive probability has $q=0$, so $R\le D+1$.
Also $\varepsilon(0)=L_s(0)\ge0$ by \eqref{eq:scalar-inequality-one}, and
$B\ge0$; the corrected summand is again nonpositive.
Consequently, $F>0$ implies $0<r<1$ and
\begin{equation}
                           D>K(r)/G(r)\ge d_s.
 \label{eq:positive-summand-distance}
\end{equation}
This proves the required implication.

\paragraph{The case $D>d_s$.}
We show that every summand with $D>d_s$ satisfies
\begin{equation}
                         F\le\mathbb E_t(H_s-Z)_+.
 \label{eq:bound-for-large-distance}
\end{equation}
It suffices to consider $F>0$.  Define
\[
                   U=D+\frac{br-B-L(r)D}{Cr}.
\]
By the definition of $L$, this rewrites the corrected summand as
\[
 F=C\{(1-r)R+rU-D-1\}
             -\mathbb E_t(1-q)(Z-aD)_+.
\]
The choice of $U$ makes comparisons with $q<1$ nonpositive; the
third scalar inequality will bound the remaining comparisons, where
$q=1$.  Indeed, since $Cr-L(r)-br+B=\mu_0 r+B-\varepsilon(r)\ge0$, inequality
\eqref{eq:scalar-inequality-one} gives
\[
 U\le(D+1)\left(1-\frac{L(r)}{Cr}\right)
                  \le\frac{(D+1)(1-P)}{1-Pr}.
\]
Since $F>0$ and $(Z-aD)_+\ge0$,
\[
 R>\frac{D+1-rU}{1-r}\ge\frac{U}{1-P}.
\]
The second inequality is the preceding bound on $U$, rearranged.
Thus $U<(1-P)R$.

For an index $t$ with $q<1$, this strict inequality and the radius
condition imply
\[
                    (1-q)R+qU\le R(1-Pq)\le D+Z.
\]
These indices contribute at most zero.  Only indices with $q=1$ can
make a positive contribution, and hence
\[
                          F\le C\mathbb E_t(U-D-Z)_+.
\]
Finally,
\[
 U-D\le\frac{br-B-L(r)K(r)/G(r)}{Cr}\le H_s.
\]
Here we used \eqref{eq:positive-summand-distance}, the nonnegativity of
$L$ from \eqref{eq:scalar-inequality-one}, and \eqref{eq:scalar-inequality-three}, in
that order.
Because $C\le1$, this proves \eqref{eq:bound-for-large-distance}.

\paragraph{The remaining terms.}
Let
\[
 \mathcal T=\{i:D_i>d_{s_i}\},\qquad
 W(D)=\mathbb E_t\!\left[
       \mathbf1_{\{t\in\mathcal T\}}(H_{s_t}-D)_+\right].
\]
For each $i\in\mathcal T$, the preceding bound gives
$F_i\le\mathbb E_t(H_{s_i}-D_t)_+$.
The function $W$ collects these quantities with the indices interchanged.
It is convex and vanishes for $D\ge H_*$.
In particular, $W(D_i)=0$ for every $i\in\mathcal T$.
See Figure~\ref{fig:convex-function}.

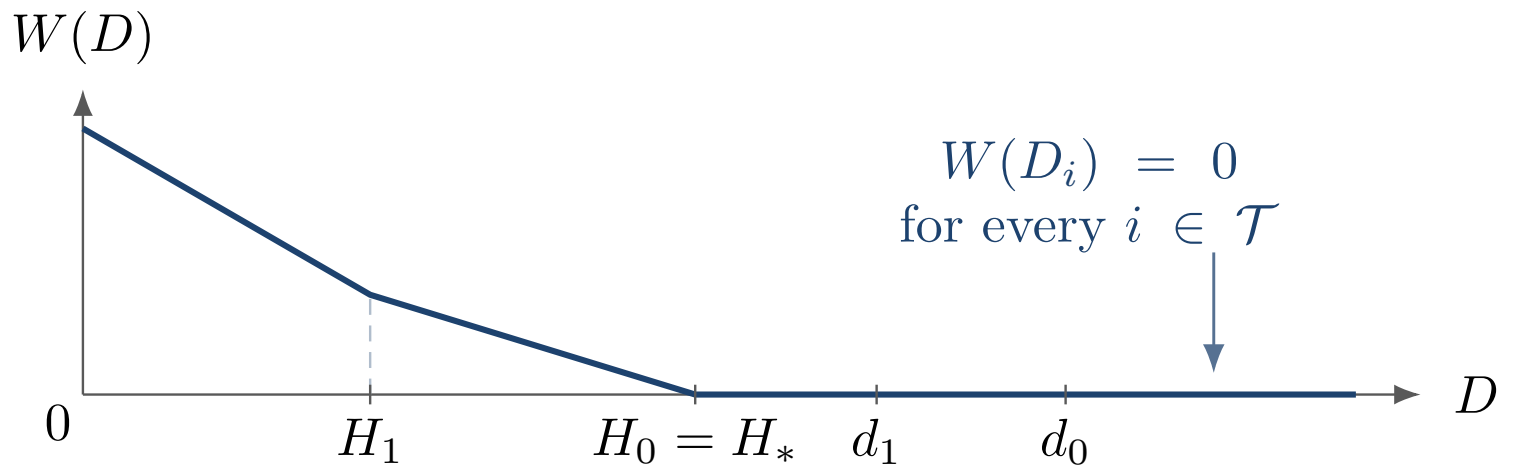
\begin{figure}[ht]
\centering
\begin{tikzpicture}[x=1cm,y=1cm,>=Latex,font=\small]
  \path[use as bounding box] (-.55,-.48) rectangle (9.15,2.25);
  \draw[->,black!65,line width=.45pt] (0,0) -- (8.85,0);
  \draw[->,black!65,line width=.45pt] (0,0) -- (0,2.02);
  \node[anchor=south] at (0,2.05) {$W(D)$};
  \node[anchor=west] at (8.93,0) {$D$};
  \node[anchor=north east,inner sep=2pt] at (0,0) {$0$};

  \draw[densely dashed,inkblue!35,line width=.45pt]
       (1.9,0) -- (1.9,.66);
  \draw[inkblue,line width=1.15pt,line join=round]
       (0,1.76) -- (1.9,.66) -- (4.05,0) -- (8.42,0);

  \foreach \x in {1.9,4.05,5.25,6.5}
    \draw[black!65,line width=.45pt] (\x,.065) -- (\x,-.065);
  \node[anchor=north,inner sep=2.5pt] at (1.9,-.065) {$H_1$};
  \node[anchor=north,inner sep=2.5pt] at (4.05,-.065) {$H_0=H_*$};
  \node[anchor=north,inner sep=2.5pt] at (5.25,-.065) {$d_1$};
  \node[anchor=north,inner sep=2.5pt] at (6.5,-.065) {$d_0$};

  \node[align=center,text=inkblue,text width=3.8cm]
       at (6.65,1.32) {$W(D_i)=0$\\for every $i\in\mathcal T$};
  \draw[->,inkblue!75,line width=.55pt]
       (7.48,.94) -- (7.48,.14);
\end{tikzpicture}
\caption{The function $W(D)$, shown schematically for
$H_1<H_0=H_*<d_1<d_0$, with both values of $s_t$ represented in
$\mathcal T$.  It is convex, and $W(D_i)=0$ for every
$i\in\mathcal T$.  Equal values of $H_0,H_1$, or the absence of one
value of $s_t$ from $\mathcal T$, remove a bend.}
\label{fig:convex-function}
\end{figure}

For a summand with $D\in[0,H_*]$, we prove
\begin{equation}
                                  F\le-W(D).
 \label{eq:bound-for-small-distance}
\end{equation}
Hold the data $(q,Z)$, the value of $s$, and the set $\mathcal T$ fixed.
Then $V+W$ is convex in its argument, so its values at $0$ and $H_*$
suffice.  These endpoint evaluations are purely algebraic;
\eqref{eq:affine-upper-bound} is used only at the given distance $D$.

At $D=H_*$, we have $W(H_*)=0$ and
\[
                         V(H_*)\le H_*G(r)-K(r)\le0.
\]
Here $G(r)\ge0$, $H_*\le d_s$, and $K(r)\ge d_sG(r)$.
At $D=0$, affineness gives $K_s(r)=\mathbb E_tK_s(q)$.
Also $Z>d_{s_t}$ for $t\in\mathcal T$.  Thus
\begin{align*}
 V(0)+W(0)
 &\le-\mathbb E_tK_s(q)
       +\mathbb E_t\!\left[\mathbf1_{\{t\in\mathcal T\}}
                 \{H_{s_t}-d_{s_t}\delta(q)\}\right]\\
 &\le-\mathbb E_t\!\left[
            \mathbf1_{\{t\notin\mathcal T\}}K_s(q)\right]\le0,
\end{align*}
by \eqref{eq:scalar-inequality-four} and $K_s(q)\ge0$.
Convexity proves \eqref{eq:bound-for-small-distance}.
For $i\notin\mathcal T$ with $D_i>H_*$, we have $F_i\le0$ by
\eqref{eq:positive-summand-distance} and $W(D_i)=0$.
Thus \eqref{eq:bound-for-small-distance} holds for every $i\notin\mathcal T$.

\paragraph{Summing the inequalities.}
Sum the bounds for $i\in\mathcal T$ and $i\notin\mathcal T$.
Since $W(D_i)=0$ for $i\in\mathcal T$, the result is
\begin{align*}
 \mathbb E_iF_i
 &\le\mathbb E_i\!\left[
       \mathbf1_{\{i\in\mathcal T\}}
                    \mathbb E_t(H_{s_i}-D_t)_+\right]
                         -\mathbb E_iW(D_i)\\
 &=\mathbb E_tW(D_t)-\mathbb E_iW(D_i)=0.
\end{align*}
This proves the lemma.
\end{proof}

\subsection{Recovering the stated comparison theorem}

The general comparison uses a constant correction in $r$.
The following scalar statement supplies its parameters.

\begin{lemma}[Scalar choices for the general comparison]
\label{lem:general-scalar-choices}
For $0\le \ybar\le1$, $0\le w\le1-\ybar$, $y=\ybar+w$, and $z=Pw^2$,
there are $\lambda\in[0,1]$, $\nu=2-\lambda$, and parameters
$k,d_s,H_s$, with $k_s=(1-s)k$, satisfying all the preceding scalar conditions with
\[
                    \varepsilon(r)=\lambda z,\qquad B=\nu z.
\]
\end{lemma}

The explicit choices appear in \cref{sec:comparison-parameter-values},
and \cref{sec:scalar-verification} verifies their constraints.
All parameters are functions of
$\ybar$ or absolute constants; they do not depend on the individual summands
or on the comparison matrix.

\begin{proof}[Proof of Theorem~\ref{thm:core-finite-comparison}]
For fixed distances, radii, and comparison matrix, the expression
$\mathcal E_{\ybar,y}$ is affine in each $s_i$.
It is therefore enough to prove the claim when every $s_i$ is zero
or one.
The expression is also nondecreasing in $y$: its explicit
$y$-dependent term increases, and the subtracted term involving
$(D_t-(1+\mu_0+2Py)D_i)_+$ decreases.

If $y\le \ybar$, apply Lemmas~\ref{lem:general-scalar-choices}
and~\ref{lem:general-charging} at $w=0$ to obtain
$\mathcal E_{\ybar,y}\le\mathcal E_{\ybar,\ybar}\le0$.
If $y\ge \ybar$, put $w=y-\ybar$.  Since the distances have mean one,
the correction has average $(\lambda+\nu)z=2Pw^2$.
The same lemmas therefore give
\[
          \mathcal E_{\ybar,y}\le2Pw^2
                  \le\frac{2P(y-\ybar)^2}{y},
\]
because $y\le1$.
These are the required bounds in both cases.
\end{proof}

\subsection{The improved inequality at the initial state}
\label{sec:first-iteration-inequality}

At $\ybar=y=1$, the same charging argument gives the additional
term $-c\,\mathbb E_i[\rho_i^2D_i]$ in the comparison bound.  We prove
Lemma~\ref{lem:first-iteration-improvement} by applying
Lemma~\ref{lem:general-charging} with a negative correction term.
We use $d_0=d_1=H_0=H_1=h$ and verify the four inequalities directly.

Abbreviate the initial coefficients by
\[
 P=P(1)=1-2/e,\qquad a=A(1)=2-P,\qquad \mu_0=\mu_0(1)=1-3P.
\]
Without a correction, $L_s(r)=Pr$ for $s=0$ and $s=1$.
Inequality~\eqref{eq:scalar-inequality-one} is stronger when $s=0$; in that case,
the difference between its two sides is
\[
 Pr-\frac{Pr(1-r)}{1-Pr}=\frac{P(1-P)r^2}{1-Pr}.
\]
This quadratic slack suggests the correction $\varepsilon(r)=-cr^2$.
Under $k\le P$, the smaller of $K_0(0),K_1(0)$ is $1-P$.
We therefore choose $h$ so that $ah=1-P$:
\begin{equation}
 h=\frac{1-P}{2-P},\qquad ah=1-P.
 \label{eq:common-parameter}
\end{equation}
Here $0<P<1/3$, so $0<h<1$.  Choose $c,k$ satisfying
\begin{gather}
 0<c\le P/2,\qquad 0<k\le P,\qquad P(1+k)^2\le4k,
 \label{eq:first-iteration-parameter-order}\\
 2P\le h(1+P-c),
 \label{eq:first-iteration-condition-one}\\
 2P\le(1-h)(1-k).
 \label{eq:first-iteration-condition-two}
\end{gather}
The paragraph on explicit constants below verifies these conditions for
absolute rational choices of $c,k$.

For Lemma~\ref{lem:general-charging}, take
\[
 \varepsilon(r)=-cr^2,\qquad B=0,\qquad
 d_0=d_1=H_0=H_1=h,\qquad k_s=(1-s)k.
\]
Then $B\ge0$ and $\varepsilon\le B$.
Affineness in each $s_i$ allows us to restrict to $s_i\in\{0,1\}$;
the inequalities below in fact hold for all $0\le s\le1$.
Put $C=1-sP$ and $b=(2-s)P$.
Since $\mu_0+b=C-P$, the functions in that lemma become
\begin{equation}
 L(r)=r(P-cr),\quad K(r)=C-k_s-br,\quad
 G(r)=a(1-r).
 \label{eq:first-iteration-functions}
\end{equation}
We verify its four conditions directly.

\paragraph{The first inequality.}
Inequality~\eqref{eq:kappa-upper-bound} is included in
\eqref{eq:first-iteration-parameter-order}, so $\kappa_s(q)\le k_s$.
For \eqref{eq:scalar-inequality-one}, the displayed slack dominates $cr^2$
because $c\le P/2\le P(1-P)$.  Hence
\[
 L(r)=Pr-cr^2
 \ge\frac{Pr(1-r)}{1-Pr}
 \ge\frac{CPr(1-r)}{1-Pr}.
\]

\paragraph{The second and third inequalities.}
The definition of $G$ gives the exact identity
\[
 K(r)-hG(r)=(1-s)(P-k)+r(1-P-b)\ge0,
\]
because $k\le P$, $b\le2P$, and $P<1/3$.
This proves \eqref{eq:scalar-inequality-two}.

For \eqref{eq:scalar-inequality-three}, put $T(r)=(Ch-b)r$.  The identity
\[
 KL+TG=L(K-hG)+G(T+hL)
\]
reduces \eqref{eq:scalar-inequality-three} to $T+hL\ge0$, since
$L,G\ge0$ and $K-hG\ge0$.  In fact,
\begin{align*}
 T+hL
 &=r\{h(C+P-cr)-b\}\\
 &=r\bigl[\{h(1+P-c)-2P\}
                  +sP(1-h)+hc(1-r)\bigr]\ge0.
\end{align*}

\paragraph{The fourth inequality.}
Since $h\le1-P$ and $b\ge P$,
\[
 h\kappa_s(q)
 \le\frac{hP(1-q)}{1-P}
 \le b(1-q).
\]
Consequently,
\begin{align*}
 K_s(q)+h\{k_s-\kappa_s(q)\}
 &\ge C-b-(1-h)k_s\\
 &\ge1-2P-(1-h)k\ge h,
\end{align*}
where the last step is \eqref{eq:first-iteration-condition-two}.
This is \eqref{eq:scalar-inequality-four} for both values of $t$.

All hypotheses of Lemma~\ref{lem:general-charging} hold.  Its conclusion
with $\varepsilon(r)=-cr^2$ is exactly
\[
                 \mathcal E_{1,1}
                    \le-c\,\mathbb E_i[\rho_i^2D_i],
\]
which proves the stronger inequality at $\ybar=y=1$.

\paragraph{Explicit constants.}
\label{sec:first-iteration-parameters}
Here $P=P(1)=1-2/e$.  Choose
\[
 c=\frac1{100},\qquad k=\frac2{25},\qquad
                         h=\frac{1-P}{2-P}.
\]
The parameter $h$ satisfies $h(2-P)=1-P$ by definition.
It is used only in the proof; the algorithm needs the
rational constant $c$.

Summing the exponential series through degree three and bounding the
remaining tail geometrically gives
\[
 \frac83<e<\frac83+\frac{1/4!}{1-1/5}=\frac{87}{32},
       \qquad \frac14<P<\frac{23}{87}<\frac{53}{200}.
\]
Thus $0<c<P/2$, $0<h<1$, and $0<k<P$.  To verify
\eqref{eq:kappa-upper-bound}, compute
\[
 4k-P(1+k)^2
    \ge\frac8{25}-\frac{53}{200}\left(\frac{27}{25}\right)^2
    =\frac{1363}{125000}>0.
\]

Substituting $h$ into the two remaining scalar inequalities
leaves quadratic numerators:
\begin{align*}
 (2-P)\{h(1+P-c)-2P\}
      &=1-4P+P^2-c(1-P),\\
 (2-P)\{(1-h)(1-k)-2P\}
      &=1-k-4P+2P^2.
\end{align*}
Both right sides decrease on $[1/4,53/200]$.  At $P=53/200$ they
equal $23/8000$ and $9/20000$, respectively, and hence are positive
throughout the interval.  This verifies
\eqref{eq:first-iteration-parameter-order} and
\eqref{eq:common-parameter}--
\eqref{eq:first-iteration-condition-two}.

\paragraph{AI Use Statement.}
This work resulted from extensive interaction between the authors and OpenAI Astra and Gemini 4 Argon and Carbon models, with the authors providing suggestions and critically checking model responses through iterative exchanges. These models were also used for writing through detailed prompts from the authors. Throughout, the authors directed the process and evaluated the proposed ideas and arguments. Lean files formalizing the main theorem are available on
\href{https://github.com/hajiaghayi/Facility}{GitHub}; this formalization was produced with the Gemini models.

\begingroup
\small
\raggedright
\setlength{\bibsep}{3pt}
\interlinepenalty=10000
\bibliographystyle{plainnat}
\bibliography{facility_location_threshold}
\endgroup

\appendix

\section{Greedy augmentation}
\label{sec:greedy-augmentation}

This appendix proves \cref{thm:greedy-augmentation}.  The argument is
the classical one \cite{CG05,MYZ06,BA10}.  The exact transformation,
without an additive $\varepsilon$, is stated in
\cite[Lemma~2.1]{BA10}; we include the proof for an arbitrary feasible
fractional solution and claim no novelty.

\begin{pseudocode}{Greedy augmentation}
\label{alg:augmentation}
\textbf{Input:} A nonempty set $S_0$ of open facilities, opening costs $f_i$,
and distances.
\begin{enumerate}
\item Set $S=S_0\cup\{i:f_i=0\}$ and
      $D_j=\min_{i\in S}d(i,j)$.
\item Compute $G_i=\sum_j(D_j-d(i,j))_+$ for every $i\notin S$.
      If $G_i\le f_i$ for all such $i$, return $S$.
\item Choose $i\notin S$ of maximum $G_i/f_i$, breaking ties by a
      fixed facility order.  Add $i$ to $S$, replace each $D_j$ by
      $\min\{D_j,d(i,j)\}$, and repeat step~2.
\end{enumerate}
\end{pseudocode}

Each iteration decreases the total cost, and there are at most $m$
iterations.  Recomputing all gains in every iteration takes $O(m^2n)$
operations in total.

\begin{lemma}[Augmentation]
\label{lem:greedy-augmentation}
Let $F_0,C_0$ be the costs of $S_0$.  For every feasible fractional
solution of costs $F^*,C^*$ and every $s\ge1$, the final solution
satisfies
\begin{equation}
 \operatorname{cost}(S)
 \le F_0+\frac{C_0}{s}+\left(1-\frac1s\right)C^*
                         +F^*\log s.
 \label{eq:greedy-augmentation}
\end{equation}
The algorithm does not use this fractional solution or $s$.
\end{lemma}

\begin{proof}
For a current total connection cost $D=\sum_jD_j$, feasibility gives
\begin{equation}
 \sum_i y_i^*G_i
 \ge\sum_{i,j}x_{ij}^*(D_j-d(i,j))_+
 \ge D-C^*.
 \label{eq:augmentation-ratio}
\end{equation}
An open facility has $G_i=0$.  After all facilities of opening cost zero are open,
some unopened facility therefore has ratio $G_i/f_i$ at least
$(D-C^*)/F^*$ whenever $F^*>0$ and $D>C^*$.

Suppose first that $F^*>0$.  Define the increasing concave function
\[
 H(u)=
 \begin{cases}
 u,&u\le F^*,\\
 F^*\bigl(1+\log(u/F^*)\bigr),&u>F^*.
 \end{cases}
\]
Use the potential function given by the current opening cost plus
$C^*+H(D-C^*)$.  If an addition costs $a$ and saves $v$, its excess
$u=D-C^*$ decreases to $u-v$.  When $u>F^*$, the bound on $G_i/f_i$ gives
$a\le F^*v/u$, and concavity gives
\[
                    H(u)-H(u-v)\ge H'(u)v=F^*v/u\ge a.
\]
When $u\le F^*$, the function $H$ is linear on $[u-v,u]$, so its
decrease is $v>a$.  Thus the potential function never increases.
At termination the bound on $G_i/f_i$ forces $D-C^*\le F^*$, where the
potential function equals the total cost.  Opening facilities of opening cost zero preserves
$F_0$ and cannot increase $C_0$, so
\[
                 \operatorname{cost}(S)
                 \le F_0+C^*+H(C_0-C^*).
\]
The tangent to $H$ at $sF^*$ is $u/s+F^*\log s$, which proves
\eqref{eq:greedy-augmentation}.

If $F^*=0$, every facility with a positive assignment variable in the
fractional solution has opening cost zero.  Opening all such facilities reduces the connection cost to
at most $\min\{C_0,C^*\}$, which is at most
$C_0/s+(1-1/s)C^*$.  Later additions only improve total cost.
\end{proof}

\begin{proof}[Proof of \cref{thm:greedy-augmentation}]
Apply \cref{lem:greedy-augmentation} with the fractional solution
$(x,y)$, so $F^*=F$ and $C^*=C$.  By \eqref{eq:scaled-bifactor},
\[
   F_0+\frac{C_0}{s}=\frac{sF_0+C_0}{s}
        \le\gamma_fF+\frac{\gamma_c}{s}\,C .
\]
Substituting in \eqref{eq:greedy-augmentation} gives
$\cost(S)\le(\gamma_f+\log s)F+\bigl(1+(\gamma_c-1)/s\bigr)C$.
\end{proof}

\section{Rational arithmetic and exact sampling}
\label{sec:rational-precision}

This appendix collects the routine numerical details: the precision
of Rounding~1, the rational scaling factor, the running time of
Rounding~1, and the exact sampling used by the randomized algorithms.

\subsection{Precision of Rounding~1}

Assume $C>0$, and let $\xi$ be the least positive value among the
initial assignment variables.  Put $\eta=c\xi^2/(2N)$.
Choose an integer $D$ such that every $Dp_i$ and $D\theta_{ij}$ is an
integer; \cref{sec:running-time} gives an explicit
choice of polynomial encoding length.  Set
\begin{equation}
       \frac1\tau=D\left\lceil\frac{8N^2}{D\eta\xi}\right\rceil.
 \label{eq:precision-choices}
\end{equation}
All initial opening variables are multiples of $\tau$, and every opening
variable remains a multiple of $\tau$.  Thus rounding down leaves a
variable unchanged whenever the preceding update leaves it unchanged.

Since $\tau\le\eta\xi/(8N^2)$ and at most $N$ iterations occur,
the total numerical error is at most
\begin{equation}
                  N\cdot\frac{8N\tau C}{\xi}\le\eta C.
 \label{eq:rounding-error-verified}
\end{equation}
The stronger comparison at the first iteration decreases the bound by
$2\eta C$, leaving the claimed decrease of $\eta C$.  This verifies
\eqref{eq:rounding-error-condition}.
The same $\tau$ bounds the error in evaluating $P$ and the decrease in
each variable due to rounding down.
As $0<\xi\le1$, $N\ge1$, and $c\le1$, we have
$0<\tau\le1/16<1/4$.

\paragraph{A rational approximation to $P$.}
The definition of $q_{hi}$ uses a lower bound on $P$, which preserves
the hypothesis involving the radii in the comparison theorem.  The same evaluation
gives a rational upper bound on the contribution of client $j$ for every
choice of $S_j$.
At the current rational $\ybar\in[0,1]$, use an odd partial sum of
\[
                 P(\ybar)=\sum_{k=0}^{\infty}
                              \frac{(-\ybar)^k}{k!(k+2)}.
\]
Choose the least odd degree $r\ge1$ for which
$1/[(r+1)!(r+3)]\le\tau/2$, and call the partial sum
$\widehat P$.  The alternating-series bound gives
\begin{equation}
             0\le P(\ybar)-\widehat P\le\tau/2,
                    \qquad 0<\widehat P\le1/2.
 \label{eq:rational-lower-estimate}
\end{equation}
For positivity, odd partial sums increase from
$1/2-\ybar/3\ge1/6$.
The degree is $O(\log(1/\tau)+1)$.

Use \eqref{eq:rational-coefficients} to define
\[
 \widehat A=\frac{1+3\ybar-2\ybar\widehat P}{1+\ybar},\qquad
 a^+(y)=\widehat A+2\widehat P(y-\ybar)+\tau y.
\]
To verify the direction and size of the error, put
$\varepsilon=P(\ybar)-\widehat P\in[0,\tau/2]$.  For every
$y\in[0,1]$,
\[
 a^+(y)-a(\ybar,y)
   =2\varepsilon\left(\ybar+\frac{\ybar}{1+\ybar}\right)
                        +(\tau-2\varepsilon)y
   \in[0,3\tau/2].
\]
The two nonnegative coefficients sum to $\tau$, while the two
quantities they multiply are at most $3/2$ and one, respectively.
Thus, for a set $S_j$ with sum of assignment variables $y$,
fractional connection cost $V$, and auxiliary variable $b$, we obtain
the upper bound
\[
        \widehat\phi=a^+(y)V+(1-y)b,
                \qquad 0\le\widehat\phi-\phi\le2\tau V,
\]
where $\phi=a(\ybar,y)V+(1-y)b$ is the exact potential.
Every choice of $S_j$ and facility $h$ uses this same evaluation
at the current common parameter $\ybar$.  The algorithm evaluates $P$ only
once per iteration.

\subsection{A rational scaling factor}
\label{sec:scaling-factor-choice}

We construct the scaling factor in \eqref{eq:scaling-factor-conditions} with
one truncated exponential polynomial.  Put $\delta=\eta/16$ and
choose the least odd integer $r\ge3$ such that
\[
              \frac{2(3/2)^{r+1}}{(r+1)!}\le\delta.
\]
Define
\[
 E_r(t)=\sum_{k=0}^{r}\frac{t^k}{k!},\qquad
 g_r(t)=t-1-2E_r(-t),\qquad g(t)=t-1-2e^{-t}.
\]
The function $g$ has its unique zero at $\alpha_*$.  On
$[5/4,3/2]$, the alternating-series bound gives
\begin{equation}
                         g(t)<g_r(t)\le g(t)+\delta.
 \label{eq:scaling-polynomial-bounds}
\end{equation}
The initial polynomial signs are
\[
 g_r(5/4)\le\tfrac14-2E_3(-5/4)=-31/192<0,
             \qquad g_r(3/2)>g(3/2)>0.
\]
For the first inequality, group the terms added after degree three in
positive pairs.  The second uses
$e^{3/2}>1+3/2+(3/2)^2/2+(3/2)^3/6>4$.

Bisect $[L,U]=[5/4,3/2]$ until $U-L\le\delta$.
At a midpoint $t$, replace $L$ by $t$ if $g_r(t)\le0$, and replace
$U$ by $t$ otherwise.  Exact rational comparisons preserve
$g_r(L)\le0\le g_r(U)$.  Thus $g(L)<0$ and $g(U)\ge-\delta$.
Since $g'(t)>1$, the final endpoints satisfy
\[
                 L<\alpha_*\le U+\delta,
                    \qquad \alpha_*-L\le2\delta.
\]
Set
\begin{equation}
                             s=E_r(L-1).
 \label{eq:rational-scaling-factor}
\end{equation}
Since $1/4\le L-1<1/2$, the positive Taylor tail is at most twice
its first term.  The same choice of $r$ therefore gives
\[
                  1<s<e^{L-1},\qquad e^{L-1}-s\le\delta.
\]
Writing $s_*=e^{\alpha_*-1}<2$, the mean value theorem yields
\[
 0<s_*-s\le2(\alpha_*-L)+\delta
                       \le5\delta=5\eta/16<\eta/2.
\]
This proves \eqref{eq:scaling-factor-conditions}.

Both the degree and the number of bisection steps are
$O(\log(1/\eta)+1)$.  Bisection retains only dyadic endpoints;
polynomial evaluations select the next interval and do not become
later arguments.  Thus all operands have polynomial encoding length,
and the construction uses polynomially many rational operations.

\subsection{Running time of Rounding~1}
\label{sec:running-time}

Every iteration opens a previously unopened facility, so there are
at most $N$ iterations.  The remaining issue is the size of the
rational numbers used to update the opening variables.  Every opening
variable is initially a multiple of $\tau$ and remains so; hence its
denominator divides $1/\tau$.  Each retained assignment variable is a fixed
rational multiple of the corresponding opening variable.

The approximation algorithm solves the original rational LP once.  A
polynomial-time LP algorithm returns an optimum of polynomial encoding
length.  Discard facilities with zero opening variable; the remaining
$N\le m$ facilities retain their original opening and assignment
variables.  Each fixed ratio $\theta_{ij}$ is a quotient of two entries
of this solution.  Let $\ell$ be the total input length of the
original instance and its fractional solution; then $\ell$ is polynomial
in the original input length and $m,n,N=O(\ell)$.  For a standalone call
to the rounding algorithm, use the encoding length of its supplied data.  The
same argument applies without the LP computation.

For this analysis, scale the original opening costs and distances by a common
positive denominator, making them integers of magnitude $2^{O(\ell)}$.
This changes neither the algorithm's comparisons nor its guarantee.
In the approximation algorithm, the supplied opening costs are these integers
multiplied by the rational scaling factor $s$ constructed in
\cref{sec:scaling-factor-choice}.
For the integer $D$ used above,
take the product of the
numerators and denominators of the positive opening variables and the
denominators of the assignment variables, with each input number included
once.  Indeed, the denominator of
$\theta_{ij}=p_{ij}/p_i$ divides the denominator of $p_{ij}$ times
the numerator of $p_i$.  Consequently
\begin{equation}
            \log D=O(\ell),\qquad \log(1/\eta)=O(\ell).
 \label{eq:runtime-initial-encoding}
\end{equation}
The integer $1/\tau$ satisfies
\[
             \frac1\tau<\frac{8N^2}{\eta\xi}+D,
                  \qquad \log(1/\tau)=O(\ell).
\]
Thus $\eta$ and $\tau$ have numerator and denominator of length $O(\ell)$.

\subsection{Rational variables and function evaluations}

The opening variables have denominator dividing $1/\tau$; multiplying by
$\theta_{ij}$ introduces at most one factor of $D$.
Hence each assignment variable, each $y_j,V_j$, and $\sum_jV_j$ has
denominator dividing $D/\tau$, while $\sum_jy_jV_j$ has denominator
dividing $(D/\tau)^2$.
Their magnitudes are bounded by $2^{O(\ell)}$, so the common weighted
mean $\ybar$ has numerator and denominator of length $O(\ell)$ whenever
$\sum_jV_j>0$; otherwise $\ybar=0$.
The opening variables are rounded down before the sets $S_j$ are examined,
so the same bounds apply to every set considered by the algorithm.

The odd partial sum defining $\widehat P$ has degree $O(\ell)$.
For an argument $a/d$ with numerator and denominator of length $O(\ell)$, terms
through degree $r$ share a denominator dividing $d^r(r+2)!$.
Hence $\widehat P$, $\widehat A$, and the upper tangent coefficients
$a^+(y)$ have representations of length $O(\ell^2)$.
The scaling factor $s$ is computed in the same way from a polynomial of
degree $O(\ell)$ at rational bisection endpoints of length $O(\ell)$, and also
has representation of length $O(\ell^2)$.  Bisection carries only its endpoints
between iterations; values of the polynomial do not become new
arguments.

\subsection{Computing the coefficients and choosing the sets}

The coefficients $q_{hi}$ use only facilities in $\mathcal U$.
Fix a facility $i\in\mathcal U$.  The formula for $q_{hi}$ changes
only at radii $d(i,h)$ and
$d(i,h)/(1-\widehat P)$.  Between consecutive breakpoints, let $S_0$
be the sum of $z_h$ over facilities $h\in\mathcal U$ at distance at most
$(1-\widehat P)r$ from $i$, let $S_1$ be the sum of $z_h$ at distances
strictly between $(1-\widehat P)r$ and $r$, and let $S_2$ be the sum
of $z_hd(i,h)$ over the same facilities as $S_1$.
These three sums are fixed on the interval.
The normalization $\sum_hz_hq_{hi}=1$ becomes
\[
              S_0+\frac{S_1-S_2/r}{\widehat P}=1,
              \qquad
              r=\frac{S_2}{S_1-\widehat P(1-S_0)}.
\]
An interior solution has positive denominator.  An equality at a
breakpoint, including the start of a flat interval, is handled
directly.  Sweep the sorted breakpoints, processing equal radii together.
At $d(i,h)$, facility $h$ enters the sums $S_1,S_2$; at
$d(i,h)/(1-\widehat P)$, it leaves them and enters $S_0$.
Each event updates these sums in constant time; facilities at distance
zero from $i$ are included in $S_0$ initially.
If their opening variables sum to at least one, the prescribed choice
of $q_{hi}$ at radius zero uses rational arithmetic.
The three sums have denominator dividing $1/\tau$ and
numerators of length $O(\ell)$, so every radius and coefficient $q_{hi}$ has
polynomial encoding length.  The updated opening variables are rounded
down to multiples of $\tau$ before being stored or used to compute
the sums for a set $S_j$.

For each client, sort the facilities with positive original assignment
by distance once, breaking ties by facility order.  The client stores
the largest position retained in this order; the algorithm also stores
the set $O$ of open facilities.  A facility can belong to $S_j$ only if
its opening variable is positive, it lies outside $O$, and its position
does not exceed the stored position.  An open facility, or a facility
with zero opening variable, cannot satisfy these conditions later.
Choosing $S_j$ only decreases the stored position.  After rounding down
the opening variables for a facility $h$ under consideration, scan the
remaining facilities in order and maintain the sum of assignment
variables and the fractional connection cost.  Evaluate the upper bound
on the contribution of that client for the set consisting of the first $k$
facilities, varying $k$ from zero to the number of remaining facilities.
Each evaluation requires $O(1)$ rational operations, so the scan uses
$O(N)$ operations per client.

The auxiliary variables $b_j$ are minima of original distances and
$5R_j$.  All upper bounds on the client contributions use the coefficients
from the single function evaluation in that iteration; the assignment
sums and fractional connection costs have the common denominators just
described.  Including the supplied opening costs, the quantity minimized
over the choices of $h$ therefore has representation of length $O(\ell^2)$.
Sorting and searching the breakpoints takes
$O(N\log(N+1))$ rational operations per facility $i$.
For each of at most $N$ choices of facility $h$, updating the opening
variables and scanning the ordered facilities for all clients take
$O(nN)$ operations.
The scalar evaluation takes $O(\ell)$ operations once per iteration.
Over at most $N$ iterations, rounding uses at most
\[
              O\bigl(N^3\{n+\log(N+1)\}+N\ell\bigr)
\]
rational operations on polynomial-length operands.  Computing the
scaling factor $s$ uses $O(\ell^2)$ further rational operations.  The initial
orders for the clients, the at most $N+1$ solutions considered in the final
step, and greedy augmentation
also take polynomial time.  Together with the initial LP computation,
this proves polynomial running time in the original input length.

\subsection{Sampling and scalar evaluations in the randomized algorithms}
\label{sec:sampling-details}

A Bernoulli variable with any rational parameter $r\in[0,1]$ uses
at most two fair bits and a constant number of rational arithmetic
operations in expectation.  Compare $r$ with $1/2$ and reveal the first
bit of a uniform binary expansion.  If that half lies entirely below
or above $r$, return the corresponding result.  Otherwise restrict
to the half containing $r$, rescale its parameter to $2r$ or $2r-1$,
and repeat.  Conditional on continuation, the unrevealed bits are still
independent fair bits.  Each stage continues with probability at most
$1/2$, so the procedure terminates almost surely and has the stated
expectation.  Equality with $1/2$ is decided by the next bit immediately.

To choose a current opening location, scan the locations and use
successive rational coins with parameters $z_h$ divided by the total
mass not yet passed.  This samples $z_h/Y$ exactly, using $O(m)$
expected arithmetic operations and fair bits.  The final randomized
choice is sampled in the same way, including a final outcome with
mass $1-Y$.

The lower estimate to $P$ can be evaluated from its alternating series
\[
                  P(\ybar)=\sum_{k\ge0}\frac{(-\ybar)^k}{k!(k+2)}.
\]
An odd partial sum of degree $O(\log(1/\delta_P))$ is a rational
lower bound with error at most $\delta_P$.  Take its maximum with
$1/4$.  Since $P(1)>1/4$ and $P(\ybar)\le1/2$,
\eqref{eq:quadratic-P-precision} holds.  The exact rational $\ybar$ and
its Taylor partial sum have polynomial-size representations.  Scalar
evaluations cost $O(m\log(m+1))$ arithmetic operations over all opening
iterations, which is $O(m^2)$.

\paragraph{Elementary bounds and the augmentation parameter.}
The elementary bounds $8/3<e<11/4$ imply
$1/4<P(\ybar)\le1/2$ and $5/3<\beta<7/4$.
Also $e^{3/2}>4$ from its first four Taylor terms, while
$e^{3/2}<(11/4)(7/4)=77/16$.
For the increasing function $v\mapsto ve^{1+v}$ these inequalities
give
\[
       \frac{41}{100}<t_*<\frac12.
\]
Indeed $(41/100)e^{141/100}<(41/100)(77/16)<2$, whereas
$(1/2)e^{3/2}>2$.
It follows that the rational parameter $t$ of
\cref{thm:grouped-augmentation} lies in $[2/5,1/2]$.
It is computed as follows.  Certified rational bisection on
$u e^{u+1}=2$, initially on $[2/5,1/2]$, locates $t_*$ in a rational
interval $[\ell,u]$ of width at most $1/(172800L)$.
Set $t=\ell-1/(86400L)$.  At a midpoint $v$, use Taylor bounds to
enclose $v e^{v+1}-2$ in a rational interval of width at most
$1/(10^6L)$.  They require $O(\log(L+1))$ terms on the specified
interval.  If this interval contains zero, the inequality
$(v e^{v+1})'>1$ localizes the root to
$[v-1/(10^6L),v+1/(10^6L)]$, whose width is less than
$1/(172800L)$, and the search stops.
Otherwise its certified sign determines the next bisection step.
Thus no arbitrarily precise sign comparison is required, and
$O(\log^2(L+1))$ rational arithmetic operations suffice.  This is
within the asserted quadratic bound.

\section{The integrality-gap instances}
\label{sec:matching-lp-lower-bound}

We prove the lower bound in \cref{thm:main-finite-gap}.  The instances
are the classical ones of Guha and Khuller~\cite{GK98,GK99}; what is
added here is an estimate of their integrality gap that is uniform in
the number of open facilities and quantitative in the number of
facilities.

\subsection{The family and its exact costs}

Fix an integer $k\ge2$ and a rational $t\in(0,2)$.  There are
$m=k^2$ facilities and one client $j_A$ for each $k$-element subset
$A\subseteq[m]$.  Put
\[
               n=\binom mk,\qquad
               f_i=t\binom{m-1}{k-1}=\frac{tn}{k}.
\]
Client $j_A$ has distance one to facilities in $A$ and distance three
to the others.  Distances between distinct facilities or distinct
clients are two.  These distances satisfy the triangle inequality:
a path through a third vertex from a facility to a client has length
at least $2+1$, and such a path between two distinct facilities or
two distinct clients has length at least two.
Denote the instance by $I_{k,t}$.

The fractional solution $y_i=1/k$ and
$x_{iA}=\mathbf1_{\{i\in A\}}/k$ has cost $(1+t)n$.
To see that it is optimal, assign every client the dual value
$\alpha_A=1+t$.  For a fixed facility $i$, precisely
$\binom{m-1}{k-1}$ clients are at distance one.  Each contributes $t$
to its dual constraint; all other clients contribute zero because
$1+t<3$.  Thus
\[
 \sum_A(\alpha_A-d(i,j_A))_+
       =t\binom{m-1}{k-1}=f_i.
\]
The dual objective is $(1+t)n$, equal to the displayed fractional
cost.  Weak duality therefore gives
\begin{equation}
                      \operatorname{LP}(I_{k,t})=(1+t)n.
 \label{eq:lower-exact-lp}
\end{equation}

If exactly $s$ facilities open, precisely $\binom{m-s}{k}$ clients
have no open facility at distance one.  Hence every set of that cardinality
has cost
\begin{equation}
 n\bigl(1+tz+2q_k(s)\bigr),\qquad
 z=\frac{s}{k},\qquad
 q_k(s)=\frac{\binom{k^2-s}{k}}{\binom{k^2}{k}}.
 \label{eq:lower-exact-integral}
\end{equation}
A binomial coefficient is zero when its upper argument is below $k$.
The integral optimum minimizes this expression over all $1\le s\le k^2$.

\subsection{A bound for every number of open facilities}

On any bounded range $0\le z\le z_0$, the fraction of clients with no
open facility at distance one is
uniformly close to $e^{-z}$.  Specifically, for $k\ge2(z_0+1)$,
\begin{equation}
 e^{-z}-\frac{2z_0(z_0+1)}{k}
                      \le q_k(s)\le e^{-z}.
 \label{eq:lower-uniform-product}
\end{equation}
For $0\le r<k$, the denominator $k^2-r$ lies between
$k(k-1)$ and $k^2$.  Writing the ratio as a product and comparing its factors gives
\[
 \left(1-\frac{z}{k-1}\right)^k
 \le q_k(s)=\prod_{r=0}^{k-1}
             \left(1-\frac{s}{k^2-r}\right)
 \le\left(1-\frac zk\right)^k\le e^{-z}.
\]
All factors are positive because $k\ge2(z_0+1)$.
Using $\log(1-u)\ge-u/(1-u)$ for $0\le u<1$, we obtain
\[
 \log q_k(s)
 \ge-\frac{kz}{k-1-z}
 =-z-\frac{z(1+z)}{k-1-z}
 \ge-z-\frac{2z_0(z_0+1)}k.
\]
The last step uses $k-1-z\ge k/2$.  Now $e^{-a}\ge1-a$
gives \eqref{eq:lower-uniform-product}.

Define
\[
                   R(t)=1+\frac{t\log(2/t)}{1+t}.
\]
The continuous expression $tz+2e^{-z}$ is minimized at
$z_t=\log(2/t)$, with minimum $t+t\log(2/t)$.
Choose a fixed interval $J$ around $t_*:=\alpha_*-1$ and a constant
$z_0$ such that, for every $t\in J$,
\begin{equation}
        z_t<z_0,\qquad
        \frac{1+tz_0}{1+t}>R(t),\qquad |R'(t)|<1.
 \label{eq:lower-parameter-conditions}
\end{equation}
\Cref{sec:lower-constants} supplies explicit choices.
For $z\le z_0$, the uniform estimate and continuous minimum give
\[
 \frac{1+tz+2q_k(s)}{1+t}
       \ge R(t)-\frac{4z_0(z_0+1)}{k(1+t)}.
\]
For $z>z_0$, the opening cost and the connection cost of at least one per client give
a ratio greater than $R(t)$.
Thus every possible number of openings obeys the same lower bound:
\begin{equation}
 \frac{\operatorname{OPT}(I_{k,t})}{\operatorname{LP}(I_{k,t})}
       \ge R(t)-\frac{4z_0(z_0+1)}{k(1+t)}.
 \label{eq:lower-finite-ratio}
\end{equation}

The bound also identifies the classical limiting curve.  Taking
$s=\lceil z_tk\rceil$ makes $z\to z_t$, and
\eqref{eq:lower-uniform-product} gives $q_k(s)\to e^{-z_t}$.
This set of open facilities has ratio tending to $R(t)$.  Together with
\eqref{eq:lower-finite-ratio}, the optimum ratio therefore converges
to $R(t)$.

\subsection{Finite instances approaching the lower bound}

The derivative
\[
                       R'(t)=\frac{\log(2/t)-1-t}{(1+t)^2}
\]
changes from positive to negative at its unique zero
$t_*=\alpha_*-1$.  Indeed its numerator is strictly decreasing,
and vanishes precisely when $t e^{1+t}=2$.
At that point
\[
                \log(2/t_*)=1+t_*,\qquad R(t_*)=\alpha_*.
\]

Given any $\varepsilon>0$, choose a rational $t\in J$ with
$|t-t_*|<\varepsilon/2$.  The derivative bound gives
$R(t)>\alpha_*-\varepsilon/2$.  Choose an integer $k$ satisfying
\begin{equation}
 k\ge2(z_0+1),\qquad
 k>\frac{8z_0(z_0+1)}{\varepsilon(1+t)}.
 \label{eq:lower-finite-choices}
\end{equation}
Then \eqref{eq:lower-finite-ratio} gives the desired strict ratio
$\operatorname{OPT}(I_{k,t})/\operatorname{LP}(I_{k,t})>
\alpha_*-\varepsilon$.  Every input number is rational and the number
$\binom{k^2}{k}$ of clients is finite.  Thus the integrality gap of the standard LP is arbitrarily close to
$\alpha_*$.

\subsection{The dependence on the number of facilities}
\label{sec:finite-gap-order}

\begin{proof}[Proof of the lower bound in \cref{thm:main-finite-gap}]
Put $k=\lfloor\sqrt m\rfloor\ge8$.
Choose a rational $t\in[2/5,1/2]$ with $|t-t_*|<1/k$ and use
the preceding instance with $k^2$ facilities.
The bounds in \cref{sec:lower-constants} give $|R'(t)|<1$ and
$z_0=3$.  Thus \eqref{eq:lower-finite-ratio} yields
\[
 \frac{\OPT(I_{k,t})}{\LP(I_{k,t})}
 \ge R(t)-\frac{48}{k(1+t)}
 >\alpha_*-\frac{1+240/7}{k}
 >\alpha_*-\frac{36}{k}.
\]
There are at most $m$ facilities, and
$k\ge\sqrt m/2$.  The ratio is therefore greater than
$\alpha_*-72/\sqrt m$.  This is the lower bound of the theorem.
\end{proof}

\subsection{Constants for the integrality-gap instances}
\label{sec:lower-constants}

Take
\[
                         J=[2/5,1/2],\qquad z_0=3.
\]
Since $e^{7/5}=e\,e^{2/5}<3/(1-2/5)=5$, the threshold equation
gives $\alpha_*>7/5$.  The inequality $\alpha_*<3/2$ follows from $e^{3/2}>4$; together these bounds place $t_*=\alpha_*-1$ in the interior of $J$.  For $t\in J$,
\[
                         1<\log(2/t)\le\log5<2.
\]
Indeed $2/t\ge4>e$, while the first three terms of the positive series
for $e^2$ sum to five.  Thus
$-1/2<\log(2/t)-1-t<3/5$, implying $|R'(t)|<1$.
Also $z_t=\log(2/t)<2<z_0$, and
\[
           \frac{1+3t}{1+t}-R(t)
             =\frac{t}{1+t}\bigl(2-\log(2/t)\bigr)>0.
\]
All three conditions in \eqref{eq:lower-parameter-conditions} follow.
The uniform estimate is therefore
\[
            e^{-z}-\frac{24}{k}\le q_k(s)\le e^{-z}
                   \qquad(0\le z\le3,\ k\ge8).
\]

\section{Verification of the scalar inequalities}
\label[appendix]{sec:scalar-verification}

\subsection{The parameters}
\label{sec:comparison-parameter-values}

For $0\le \ybar\le1$ and $0\le w\le1-\ybar$, put $y=\ybar+w$ and
$z=P(\ybar)w^2$.  The choices in
\cref{lem:general-scalar-choices} are
\begin{equation}
\begin{gathered}
 \lambda=\frac12-\frac \ybar3,\qquad \nu=\frac32+\frac \ybar3,\qquad
 k(P)=\frac{5P}{20-11P},\qquad k_s=(1-s)k(P),\\
 d_0=\frac{13}{25},\qquad d_1=\frac25,\qquad
 H_0=\frac25-\frac45\ybar^2(1-\ybar)^2,\qquad H_1=\frac4{15}.
\end{gathered}
\label{eq:general-parameter-values}
\end{equation}
For the correction in \cref{lem:general-charging}, take
$\varepsilon(r)=\lambda z$ and $B=\nu z$.
These choices satisfy the four scalar inequalities in
\cref{sec:general-comparison-proof}.  The verification consists of
rational inequalities and one polynomial coefficient calculation;
we give all details in \cref{sec:scalar-verification}.

\subsection{Elementary bounds}
\label{sec:general-scalar-choices}

We now verify Lemma~\ref{lem:general-scalar-choices}.  Throughout this
subsection, $0\le \ybar\le1$, $0\le w\le1-\ybar$, $y=\ybar+w$, and $z=Pw^2$.
The remaining notation $P,A,\mu_0,a,C_s,b_s,L_s,K_s,G$ is as in
\cref{sec:general-comparison-proof}.
Use the explicit values in \eqref{eq:general-parameter-values}.

The parameters $\lambda,\nu$ determine the two terms of the correction
$\varepsilon D+B$.  The parameters $d_s,H_s$ appear in the bounds
for indices in $\mathcal T$ and its complement, while $k$ bounds
$\kappa_s(q)$.  These parameters are used only in the proof of the
comparison inequality.  The rounding algorithm evaluates $P$ once per
iteration, as specified in
\cref{sec:quadratic-rounding}.
For the common correction in the charging lemma, take
$\varepsilon=\lambda z$ and $B=\nu z$.
Then $0\le\varepsilon\le B$.
Since $\ybar(1-\ybar)\le1/4$, we have
\[
 \frac16\le\lambda\le\frac12,\quad \lambda+\nu=2,\qquad
 \frac4{15}<\frac7{20}\le H_0\le\frac25
                                      =d_1<d_0.
\]
Thus the ordering in \eqref{eq:parameter-order} holds.  We prove
\eqref{eq:scalar-inequality-one}, \eqref{eq:scalar-inequality-two}, and
\eqref{eq:scalar-inequality-four} here.  Inequality~\eqref{eq:scalar-inequality-three}
also has a short proof for $s=1$; its proof for $s=0$ appears in the
next subsection.

\paragraph{Bounds for the analytic coefficients.}
The even Taylor bound for the exponential, integrated against $t$, gives
\begin{equation}
 P(\ybar)\le U(\ybar):=\frac12-\frac \ybar3+\frac{\ybar^2}{8}
                           -\frac{\ybar^3}{30}+\frac{\ybar^4}{144}
              \le\frac3{\ybar^2+4\ybar+6}.
 \label{eq:general-rational-upper}
\end{equation}
Indeed, integrate
$e^{-\ybar t}\le1-\ybar t+(\ybar t)^2/2-(\ybar t)^3/6+(\ybar t)^4/24$
for $0\le t\le1$ after multiplying by $t$.
For the second inequality, put $d=\ybar^2+4\ybar+6$ and observe that
\[
 3-dU(\ybar)=\frac{\ybar^2}{720}
       \{60-\ybar^4+24\ybar(1-\ybar)+4\ybar^3(1-\ybar)\}\ge0.
\]

The function $P$ decreases from $P(0)=1/2$, and
$P(1)=1-2/e>1/4$ because $e>1+1+1/2+1/6=8/3$.
Thus
\[
                       \frac14<P(\ybar)\le\frac12.
\]
Directly from the integral definitions, $A\ge1$, $A-1\le \ybar$,
and $\mu_0\ge0$.
The identity $(1+\ybar)A+2\ybar P=1+3\ybar$ gives
\begin{equation}
 A+P=2-\frac{(1-\ybar)(1-P)}{1+\ybar}\le2.
\label{eq:general-coefficient-range}
\end{equation}
Two consequences of the coarse upper bound in
\eqref{eq:general-rational-upper} are
\begin{equation}
 g:=\frac{1-P}{P(1+\ybar)}\ge1+\frac \ybar3,\qquad
                         1-P\ge\frac12+\frac \ybar6.
\label{eq:general-g-bound}
\end{equation}
Indeed, the first is equivalent to $P\le3/(\ybar^2+4\ybar+6)$; for the
second, subtract $3/(\ybar^2+4\ybar+6)$ from $(3-\ybar)/6$.

\paragraph{Bounding $\kappa_s(q)$.}
On $1/4\le P\le1/2$,
\begin{equation}
 \frac{k(P)}P\le\frac{10}{29},\qquad
 k(P)\le\frac5{29}.
\label{eq:general-k-bounds}
\end{equation}
Moreover,
\[
 4k-P(1+k)^2=
                  \frac{4P^2(5-9P)}{(20-11P)^2}>0.
\]
Thus \eqref{eq:kappa-upper-bound} applies and gives
$\kappa_s(q)\le k_s$ for $s=0$ and $s=1$.

\paragraph{The first inequality.}
Since $a-1=A-1+2Pw\le \ybar+w=y$, the identity
\[
                  \mu_0+b_s=C_s(a-1)+sP(a-1-y)
\]
gives
\begin{equation}
             \frac{\mu_0+b_s}{C_s}\le a-1\le y\le1.
\label{eq:common-linear-coefficient}
\end{equation}
Writing $X=1-\ybar$ and $t=1-\ybar/3$, the coefficient identity gives
\[
 a-1=1-P+P(2w-gX)
          \le1-P+Ptw,
\]
where we used $w\le X$ and \eqref{eq:general-g-bound}.
Using \eqref{eq:common-linear-coefficient}, $C_s\le1$, and
$(1-Pr)^{-1}\ge1$, we obtain
\begin{align*}
 L_s(r)-\frac{C_sPr(1-r)}{1-Pr}
 &\ge C_sP\left\{\frac{(1-P)r^2}{1-Pr}-twr+\lambda w^2\right\}\\
 &\ge C_sP\{(1-P)r^2-twr+\lambda w^2\}.
\end{align*}
The quadratic on the right is nonnegative by the
arithmetic--geometric mean inequality, since
\[
 4\lambda(1-P)
 \ge4\left(\frac12-\frac \ybar3\right)
         \left(\frac12+\frac \ybar6\right)
 =t^2+\frac{\ybar(1-\ybar)}3\ge t^2.
\]
This proves \eqref{eq:scalar-inequality-one}, including $r=0$.

\paragraph{A lower bound on $K_s(r)$.}
We next show
\begin{equation}
                        K_s(r)>H_1=\frac4{15}
                  \qquad(0\le r\le1).
\label{eq:positive-constant}
\end{equation}
Put $d=\ybar^2+4\ybar+6<15$ and
\[
 T_0=2\ybar+2w-\frac32w^2+\frac{10}{29},\qquad
 T_1=1+\ybar+w-\frac32w^2.
\]
Both are nonnegative.  Since $K_s$ decreases in $r$ and
$\nu\ge3/2$, the bounds on $P$ and $k/P$ imply
$K_s(r)\ge K_s(1)\ge1-3T_s/d$.
Using $\ybar\le1-w$, we obtain
\begin{align*}
 d-4T_0&\ge7\left(w-\frac37\right)^2+\frac{68}{203}
                                                        >\frac13,\\
 d-4T_1&=\ybar^2+6\left(w-\frac13\right)^2+\frac43
                                                        >\frac13.
\end{align*}
For the first inequality, expand $d-4T_0$ as
$\ybar^2-4\ybar+6w^2-8w+134/29$, which decreases in $\ybar\in[0,1]$,
and substitute $\ybar=1-w$.
It follows that
\[
 K_s(r)\ge\frac14+\frac{3(d-4T_s)}{4d}
                    >\frac14+\frac1{60}=\frac4{15}.
\]
Since $k_s-\kappa_s(q)\ge0$, this also proves
\eqref{eq:scalar-inequality-four} with $t=1$.

\paragraph{The second inequality.}
Both $K_s(r)$ and $G(r)=a(1-r)$ are affine in $r$.
At $r=1$, inequality~\eqref{eq:scalar-inequality-two} follows from $G(1)=0$ and
$K_s(1)>0$.  It remains to check $r=0$.
The bound $g\ge1$ and $w\le1-\ybar$ give
\[
                     a\le2-P+Pw.
\]
For $s=0$, using $\nu\ge3/2$, we obtain
\begin{align*}
 K_0(0)-d_0a
 &\ge-\frac1{25}
      +P\left(\frac{13}{25}-\frac{13}{25}w+\frac32w^2\right)-k(P)\\
 &\ge-\frac1{25}+\frac{47}{100}P-k(P)>0.
\end{align*}
The second inequality follows from
\[
 150w^2-52w+5=150(w-13/75)^2+37/75>0.
\]
For the last, $k$ is convex, so the expression is concave in $P$;
its values at $P=1/4,1/2$ are $139/27600$ and $131/5800$, respectively.

For $s=1$, it suffices to show
\[
                     1-P-\frac25a+\frac32Pw^2>0.
\]
Substitute $A=(1+3\ybar-2\ybar P)/(1+\ybar)$ and regard this expression as affine
in $P\in[0,3/d]$, where $d=\ybar^2+4\ybar+6$.
At $P=0$, it is $(3-\ybar)/[5(1+\ybar)]>0$.
At $P=3/d$, it equals
\[
 \frac{45w^2-24w-2\ybar^2+6}{10d}
 \ge\frac{43(w-10/43)^2+72/43}{10d}>0,
\]
using $\ybar\le1-w$.
Affineness in $P$ and then in $r$ proves
\eqref{eq:scalar-inequality-two} for both values of $s$.

\paragraph{The third inequality for $s=1$.}
Put $C=1-P$, $b=Py$, $\ell=C-\mu_0-b\ge0$, and
$T(r)=(CH_1-b)r+\nu z$.
The needed margin is
\begin{equation}
 S_1=\frac4{15}C-b+\frac25\ell+\frac{17}{10}z>0.
 \label{eq:scalar-margin-s1}
\end{equation}
It is affine in $P\in[0,3/d]$.
At $P=0$, it is at least $4/15$; at $P=3/d$, it equals
\[
 \frac{153w^2-126w-4\ybar^2-46\ybar+60}{30d}
 \ge\frac{149(w-36/149)^2+194/149}{30d}>0,
\]
again using $\ybar\le1-w$.
Since $\nu+d_1\lambda=17/10+\ybar/5\ge17/10$, we have
\[
 T(r)+d_1L_1(r)
 =r\{CH_1-b+d_1\ell+(\nu+d_1\lambda)z\}
                 +(1-r)(\nu+d_1\lambda)z\ge0.
\]
Inequality~\eqref{eq:scalar-inequality-two} and $L_1,G\ge0$ now give
\[
 K_1L_1+TG
       =L_1(K_1-d_1G)+G(T+d_1L_1)\ge0.
\]
This is \eqref{eq:scalar-inequality-three} for $s=1$.

It remains to prove \eqref{eq:scalar-inequality-four} for $t=0$.
For both values of $s$, we use the following positive quantity.  Put
\[
 d=\ybar^2+4\ybar+6,\qquad \omega=\ybar(1-\ybar),\qquad
 M=1-H_0-P(\ybar+7/6).
\]
Since $P\le3/d$ and $d\ge6$,
\begin{align*}
 dM
 &\ge d(1-H_0)-3(\ybar+7/6)
   =\frac1{10}-\frac35\omega+\frac45d\omega^2\\
 &\ge3(\omega-1/10)^2+\frac7{100}>0.
\end{align*}

\paragraph{The fourth inequality for $s=1,t=0$.}
Here $k_1=0$ and $\kappa_1(q)\le0$.
Dropping the nonnegative term $-d_0\kappa_1(q)$ and using $q\le1$
gives
\[
 K_1(q)+d_0\{-\kappa_1(q)\}
 \ge1-P\{1+\ybar+w-\nu w^2\}
 \ge1-P(\ybar+7/6)=H_0+M>H_0.
\]
The second inequality uses
$w-\nu w^2\le1/(4\nu)\le1/6$.

\paragraph{The fourth inequality for $s=t=0$.}
It suffices to prove
\begin{equation}
 1-\frac{12}{25}k
  -P\left[2\ybar+\max_{0\le v\le1-\ybar}(2v-\nu v^2)\right]
                                                    \ge H_0.
\label{eq:remaining-scalar-inequality}
\end{equation}
For $0\le q\le1$,
\[
 \kappa_0(q)\le\frac{P(1-q^2)}{2(1-P)};
\]
after clearing denominators, this follows from
$(1-q)(1+Pq)\ge0$.
Put $\chi=d_0/[2(1-P)]\le13/25$, so
$\nu\chi\le143/150<1$.
We therefore have
\[
 K_0(q)+d_0\{k-\kappa_0(q)\}
                  \ge1-\frac{12}{25}k-Pf(q),
\]
where
\[
 f(q)=2\ybar q+\max_{0\le v\le1-\ybar}(2vq-\nu v^2)+\chi(1-q^2).
\]
The maximum occurs at $v=\min(q/\nu,1-\ybar)$.
On the first branch, $f'(q)=2\ybar+2q(1/\nu-\chi)\ge0$;
on the second, $f'(q)=2(1-\chi q)>0$.
Thus the largest loss occurs at $q=1$, proving the reduction.

To prove \eqref{eq:remaining-scalar-inequality}, put $X=1-\ybar$ and consider
the quadratic
\[
 J(w)=1-\frac{12}{25}k-P(2\ybar+2w-\nu w^2)-H_0,
                    \qquad 0\le w\le X.
\]
Its middle coefficient on $[0,X]$ is
\[
 J_{\rm mid}=1-\frac{12}{25}k-P(1+\ybar)-H_0,
\]
so
\[
 J(Xu)=(1-u)^2J(0)+2u(1-u)J_{\rm mid}+u^2J(X)
                         \qquad(0\le u\le1).
\]
Since $k\le10P/29$ and $24/145<1/6$,
\[
 J_{\rm mid}\ge1-H_0-P(1+\ybar+24/145)>M>0.
\]
Also $J(0)=J_{\rm mid}+PX>0$.  Only the endpoint $J(X)$ remains.

The quantity $J(X)$ decreases in $P$: both $k(P)$ and
$P\{2-\nu(1-\ybar)^2\}$ increase.  We may therefore replace $P$ by
$U(\ybar)$ from \eqref{eq:general-rational-upper}.
Define
\begin{equation}
 \begin{split}
 T(\ybar)={}&\{20-11U(\ybar)\}
 \bigl[1-H_0(\ybar)-U(\ybar)\{2-\nu(1-\ybar)^2\}\bigr]\\
 &-\tfrac{12}{5}U(\ybar).
 \end{split}
 \label{eq:scalar-polynomial}
\end{equation}
This polynomial is $J(X)$ after substituting $P=U(\ybar)$ and multiplying by
$20-11U(\ybar)>0$, with the denominator in $k(U(\ybar))$ canceled.
It has degree eleven.  It remains to show $T>0$ on $[0,1]$.

One Taylor expansion suffices.  Write
\[
                       T(3/4+t)=\sum_{i=0}^{11}c_it^i.
\]
Direct expansion of \eqref{eq:scalar-polynomial} gives
\[
\begin{gathered}
 c_0>\frac1{160},\qquad -\frac38<c_1<0,\qquad c_2>6,\\
 0<c_3<11,\qquad c_4>14,\qquad
                         \sum_{i=5}^{11}|c_i|<3.
\end{gathered}
\]
These bounds are specified entirely by $T$: if
$T(\ybar)=\sum_{j=0}^{11}a_j\ybar^j$, then
$c_i=\sum_{j=i}^{11}\binom ji a_j(3/4)^{j-i}$.
For $|t|\le3/4$, the higher terms satisfy the single bound
\[
 \sum_{i=4}^{11}c_it^i
 \ge t^4\left(c_4-\frac34\sum_{i=5}^{11}|c_i|\right)
 \ge11t^4.
\]
If $0\le t\le1/4$, dropping this nonnegative remainder and the cubic term gives
\[
 T(3/4+t)>
 \frac1{160}-\frac38t+6t^2
       =6(t-1/32)^2+\frac1{2560}>0.
\]
If $-3/4\le t\le0$, then $c_1t\ge0$ and $c_3t\ge11t$, so
\begin{align*}
 T(3/4+t)
 &\ge c_0+t^2(6+11t+11t^2)\\
 &=c_0+t^2\{11(t+1/2)^2+13/4\}>0.
\end{align*}
Thus $T$ is positive throughout $[0,1]$, and hence $J(X)>0$.
All three coefficients in the displayed expansion of $J(Xu)$ are
positive.  This proves \eqref{eq:remaining-scalar-inequality} and hence
\eqref{eq:scalar-inequality-four} for every $s,t\in\{0,1\}$.

\subsection{The third inequality for \texorpdfstring{$s=0$}{s=0}}
\label{sec:third-inequality-s0}

We verify \eqref{eq:scalar-inequality-three} for $s=0$.  Fix $\ybar\in[0,1]$,
put $X=1-\ybar$, and retain the choices of
Section~\ref{sec:general-scalar-choices}.  Write
\[
 \zeta=1-k,\qquad H=H_0,\qquad
 a=A+2Pw,\qquad b=2P(\ybar+w),\qquad z=Pw^2.
\]
For $0\le w\le X$ and $0\le r\le1$, the relevant functions are
\[
 L=(2-a)r+\lambda z,\qquad K=\zeta-br+\nu z,\qquad
 G=a(1-r),\qquad T=(H-b)r+\nu z.
\]
Thus the left side of \eqref{eq:scalar-inequality-three} is a quadratic in $r$:
\begin{equation}
 Q(w,r)=KL+TG=B_0(w)+B_1(w)r+B_2(w)r^2,
 \label{eq:quadratic-in-r}
\end{equation}
where
\begin{align}
 B_0&=z(\nu a+\lambda\zeta)+\lambda\nu z^2,\nonumber\\
 B_1&=(H-b)a+\zeta(2-a)
                  +z\{2\nu(1-a)-\lambda b\},
                  \label{eq:quadratic-coefficients}\\
 B_2&=2b(a-1)-aH.\nonumber
\end{align}
We prove $Q\ge0$ on the whole rectangle.  Its endpoint values are
already nonnegative:
\[
 Q(w,0)=B_0(w)\ge0,\qquad Q(w,1)=K(1)L(1)\ge0.
\]
Here $\zeta=1-k>0$, and the preceding subsection proved $K,L\ge0$.
A split of the $r$ interval reduces the interior to two polynomial
inequalities, verified together below.

\paragraph{Splitting the quadratic.}
We use a split proportional to $w$, so that the left middle
coefficient has a factor $w$.  Choose
\[
 \gamma(\ybar)=(1+\ybar)\left(1+\frac{\ybar^2}{6}\right),\qquad
                         t=w\gamma(\ybar).
\]
Then $0\le t\le X\gamma(\ybar)\le1$, since
\[
                  1-X\gamma(\ybar)=\frac{\ybar^2(5+\ybar^2)}6\ge0.
\]
Put
\[
 M_\ell=B_0+\frac t2B_1,\qquad
 M_r=B_0+\frac{1+t}{2}B_1+tB_2.
\]
We will show $M_\ell\ge0$ and $M_r>0$.
These are the middle coefficients of the quadratic on $[0,t]$ and
$[t,1]$.  To see why they suffice, first note that
\[
                         Q(w,t)=(1-t)M_\ell+tM_r.
\]
For $0\le\theta\le1$, the two interval expansions are
\begin{align*}
 Q(w,t\theta)
   &=(1-\theta)^2Q(w,0)
             +2\theta(1-\theta)M_\ell+\theta^2Q(w,t),\\
 Q(w,t+(1-t)\theta)
   &=(1-\theta)^2Q(w,t)
             +2\theta(1-\theta)M_r+\theta^2Q(w,1).
\end{align*}
All weights are nonnegative.  The identities remain valid when
$t=0$ or $t=1$, so no endpoint case is lost.

\paragraph{One coefficient calculation for both middle terms.}
For this calculation replace $P$ by an independent variable $p$ in
\eqref{eq:quadratic-coefficients}, and write
\[
 A=\frac{1+3\ybar-2\ybar p}{1+\ybar},\qquad k=\frac{5p}{20-11p}.
\]
Use the polynomial upper bound $U(\ybar)$ from
\eqref{eq:general-rational-upper}.  Parameterize
$1/4\le p\le U(\ybar)$ and $0\le w\le X$ by
\[
 p(\ybar,v)=\frac14+\left(U(\ybar)-\frac14\right)v,\qquad
 w=Xu,\qquad 0\le u,v\le1.
\]
Every admissible pair $(P(\ybar),w)$ has this form.
Put $\Omega=(1+\ybar)(20-11p)$ and define
\begin{equation}
 \begin{split}
 F_\ell(\ybar,u,v)
   &=\left.\Omega\frac{M_\ell}{w}
                              \right|_{p=p(\ybar,v),\ w=Xu},\\
 F_r(\ybar,u,v)
   &=\left.\Omega M_r
                              \right|_{p=p(\ybar,v),\ w=Xu}.
 \end{split}
 \label{eq:coefficient-polynomials}
\end{equation}
The factor $w$ in the first line is canceled before the substitution:
$B_0$ is divisible by $w^2$ and $t=w\gamma(\ybar)$.
Thus the definition is valid also at $w=0$.
The factor $\Omega$ clears the variable denominators.
Thus $F_\ell,F_r$ are polynomials with rational coefficients and degrees at most
$(21,3,3)$ and $(21,4,3)$ in $(\ybar,u,v)$.

For completeness, the one-variable coefficient identity is
\begin{equation}
 \sum_{a=0}^d c_a z^a
  =\sum_{i=0}^d
       \left(\sum_{a=0}^i\binom{d-a}{i-a}c_a\right)
                              z^i(1-z)^{d-i}.
 \label{eq:general-homogeneous-coefficients}
\end{equation}
To verify it, insert $1=(z+(1-z))^{d-a}$ into each monomial $z^a$,
apply the binomial theorem, and then sum with coefficients $c_a$.

\paragraph{Reproducing the coefficient calculation.}
Here is an explicit polynomial form of the calculation.  It avoids
division by $w$ and specifies the coefficients using rational
arithmetic alone.  Before substituting $p=p(\ybar,v)$ and $w=(1-\ybar)u$, put
\[
 D=1+\ybar,\qquad E=20-11p,\qquad Z=20-16p,\qquad
 N=1+3\ybar-2\ybar p+2pwD.
\]
Thus $a=N/D$, $\zeta=Z/E$, and $\Omega=DE$.
With the previously defined $b=2p(\ybar+w)$, let
\begin{align*}
 \widetilde B_0={}&pw^2(\nu EN+\lambda DZ)+\lambda\nu p^2w^4DE,\\
 \widetilde B_1={}&(H-b)EN+Z(2D-N)
                 +pw^2\{2\nu E(D-N)-\lambda bDE\},\\
 \widetilde B_2={}&2bE(N-D)-HEN.
\end{align*}
These are exactly $\Omega B_0$, $\Omega B_1$, and $\Omega B_2$.
Consequently the two polynomials are obtained by the substitutions
$p=p(\ybar,v)$ and $w=(1-\ybar)u$ in
\begin{align*}
 F_\ell={}&pw(\nu EN+\lambda DZ)+\lambda\nu p^2w^3DE
                                             +\frac\gamma2\widetilde B_1,\\
 F_r={}&\widetilde B_0+\frac{1+w\gamma}{2}\widetilde B_1+w\gamma \widetilde B_2.
\end{align*}
Every expression on the right is a polynomial; in particular, the
formula for $F_\ell$ is valid at $w=0$.

For either resulting polynomial, write
\[
 F(\ybar,u,v)=\sum_{a=0}^{d}\sum_{i=0}^{q}\sum_{j=0}^{3}
                         c_{aij}\ybar^au^iv^j,
\]
where coefficients outside the actual degree are zero.  The coefficient
of $\ybar^a(1-\ybar)^{d-a}u^i(1-u)^{q-i}v^j(1-v)^{3-j}$ is
\begin{equation}
 h_{aij}=
 \sum_{b=0}^{a}\sum_{k=0}^{i}\sum_{\ell=0}^{j}
 \binom{d-b}{a-b}\binom{q-k}{i-k}\binom{3-\ell}{j-\ell}
                                      c_{bk\ell}.
 \label{eq:explicit-coefficient-conversion}
\end{equation}
This is the displayed one-variable identity applied in each variable.
Expansion of the explicit polynomials and this formula give
\[
\begin{array}{c|c|c|c}
 & (d,q,3)&\min h_{aij}&\text{an index attaining the minimum}\\\hline
 F_\ell&(21,3,3)&406176701/559872000&(21,0,3)\\
 F_r&(21,4,3)&58025243/186624000&(21,0,3)
\end{array}
\]
There are $22\cdot4\cdot4=352$ coefficients in the first line and
$22\cdot5\cdot4=440$ in the second.  The minima exceed $1/2$ and
$1/4$, respectively.  The formulas above specify each coefficient and
every entry of the table, so no additional data or program is needed
to reproduce the calculation.  This coefficient identity proves
positivity throughout the continuous cube: all basis factors are
nonnegative there, and at every point at least one is positive.

For the actual coefficient $p=P(\ybar)$, the factor $\Omega$ is
positive.  The coefficient table therefore gives
$M_\ell/w>0$ after polynomial cancellation and $M_r>0$.
Hence $M_\ell\ge0$, with equality when $w=0$.
The two interval expansions prove $Q(w,r)\ge0$ throughout the
rectangle, establishing \eqref{eq:scalar-inequality-three} for $s=0$.
Together with the preceding subsection, this completes the verification
of Lemma~\ref{lem:general-scalar-choices}.

\section{A flow formulation of the fractional problem}
\label{sec:flow-formulation}

The fractional problem has a direct formulation by flow requirements.
The relevant capacities are reused separately for each client.  This is
the capacity model used for minimum-cost multicast with network
coding~\cite{LunEtAl06}.

Construct a directed graph with a root $r$, one vertex for every
facility, and one vertex for every client.  Its arcs are
$r\to i$ for every facility $i$ and $i\to j$ for every facility--client
pair.  Purchasing capacity $u_{ri}$ costs $f_i u_{ri}$, and purchasing
capacity $u_{ij}$ costs $d(i,j)u_{ij}$.  For every client $j$, require
a flow of value one from $r$ to $j$ under the purchased capacities.
Each requirement is tested separately: the flows of different clients
are not added on a shared arc.

\begin{lemma}
\label{lem:separate-flow-equivalence}
The minimum capacity-purchasing cost in this graph equals the standard
fractional facility-location optimum.  Feasible solutions can be
converted in either direction in $O(mn)$ time.  For fixed capacities,
all flow requirements can also be tested in $O(mn)$ time.
\end{lemma}
\begin{proof}
Given a feasible fractional solution $(x,y)$, purchase $u_{ri}=y_i$
and $u_{ij}=x_{ij}$.  For the separate test of client $j$, send
$x_{ij}$ units along each path $r,i,j$.  The value is one, the
capacities are respected because $x_{ij}\le y_i$, and the total
purchasing cost is unchanged.

Conversely, the paths from $r$ to a fixed client $j$ have the form
$r,i,j$ and are otherwise arc-disjoint.  The maximum flow value is
therefore exactly
\[
                         \sum_i\min\{u_{ri},u_{ij}\}.
\]
When this sum is at least one, a scan chooses
$0\le x_{ij}\le\min\{u_{ri},u_{ij}\}$ with $\sum_i x_{ij}=1$.
Set $y_i=\min\{1,u_{ri}\}$.  Since each $x_{ij}\le1$, the standard
assignment bounds $x_{ij}\le y_i$ hold.  The resulting fractional
cost is no greater than the purchased-capacity cost.  The scans also
establish the running-time claims.
\end{proof}

Equivalently, the capacity constraints are
\[
 \sum_{i\notin S}u_{ri}+\sum_{i\in S}u_{ij}\ge1
 \qquad(j\in\mathcal C,\ S\subseteq\mathcal F).
\]
For a fixed client, the minimum cut chooses the cheaper of the two
arcs on each facility path.  Thus separation is particularly simple
on this graph.  Optimizing the capacities remains a coupled problem:
the capacity purchased on $r\to i$ can be used in every client's
separate flow test.  A feasible flow computation for fixed capacities
does not perform this capacity optimization.  Requiring a simultaneous
flow from the root to all clients would instead sum the client flows
on $r\to i$ and would describe a different relaxation.


\end{document}